\def\prd{Phys. Rev. D}
\def\prl{Phys. Rev. Lett.}
\def\apj{Astrophys. J.}

\def\apjs{Astrophys. J.Suppl.}
\def\mnras{Mon. Not. R. Astr. Soc.}
\def\aap{Astr. Astrophys.}

\def\aj{Astr. J.}

\def\jcap{JCAP}

\def\procspie{Proceedings of SPIE}
\def\pasp{Publications of the Astronomical Society of the Pacific}
\def \<{\langle}
\def \>{\rangle}

\newcommand{\ra}{\;\raise1.0pt\hbox{$'$}\hskip-6pt\partial\;}
\newcommand{\lo}{\;\overline{\raise1.0pt\hbox{$'$}\hskip-6pt\partial}\;}

\newcommand{\degree}{^\circ}

\newcommand{\red}{\textcolor[rgb]{1.00,0.00,0.00}}
\newcommand{\exptt}[1]{\langle#1\rangle}

\newcommand{\Abs}[1]{\begin{abstract} #1 \end{abstract}}
\newcommand{\Ack}[1]{\begin{acknowledgments} #1 \end{acknowledgments}}
\newcommand{\mktt}{\maketitle}

\documentclass[onecolumn,amsmath,amssymb,floatfix,superscriptaddress,showkeys,prd,nofootinbib]{revtex4-2}

\usepackage{graphicx,epsfig,natbib,color,times,bm,amsmath,multirow,hyperref,harpoon,amssymb,mathtools,mathrsfs,extarrows}
\usepackage[ddmmyyyy,hhmmss]{datetime}

\usepackage{xcolor, academicons}
\def\orcid#1{\kern
.08em\href{https://orcid.org/#1}{\includegraphics[width=1.0em]{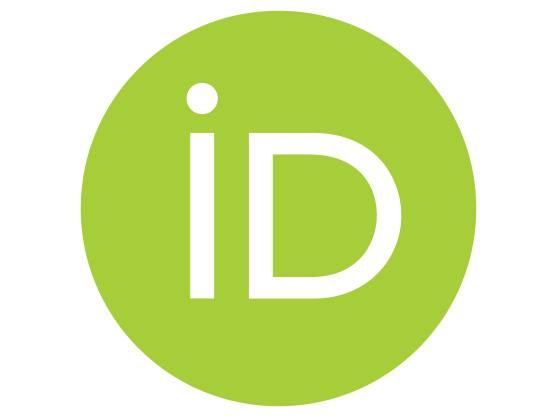}}}

\begin{document}

\title{The polarization quaternion and its applications: a joint representation
of the Q/U Stokes parameters and E/B mode polarizations}

\author{Hao Liu\orcid{0000-0003-4410-5827}}\email[]{ustc\_liuhao@163.com}
\affiliation{School of Physics and optoelectronics engineering, Anhui University, 111 Jiulong Road, Hefei, Anhui, China 230601.}

\author{James Creswell}\email[]{creswelljames@gmail.com}
\affiliation{Arnold Sommerfeld Center for Theoretical Physics, Ludwig Maximilian University of Munich, Theresienstr.\ 37, 80333 Munich, Germany.}

\author{Chao-Wei Tsai\orcid{0000-0002-9390-9672}}\email[]{cwtsai@nao.cas.cn}
\affiliation{National Astronomical Observatories, Chinese Academy of Sciences, 20A Datun Road, Chaoyang District, Beijing, 100012, People's Republic of China.s}

\author{Pavel Naselsky\orcid{0000-0002-8891-0273}}\email[]{naselsky@nbi.dk}
\affiliation{The Niels Bohr Institute, Blegdamsvej 17, DK-2100 Copenhagen, Denmark.}

\Abs{

In this work, we point out that the Q/U Stokes parameters and E/B mode
polarizations are the four components of a unique quaternion, which describes at
the same time the directions and the parity states of spherical linear
polarizations. We then point out that, with this polarization quaternion, the
mathematical form of all Q/U and E/B transforms are greatly simplified, to an
extent that requires only one quaternion multiplication for each transform. A
preliminary application of the polarization quaternion is shown as an example to
detect peculiar pixel domain patterns within the E- and B-families, which are
the former and latter halves of the polarization quaternion.

}

\mktt
\tableofcontents

\section{Introduction}
\label{sec:intro}

In modern cosmology, one of the working hypotheses holds that the large-scale
structure of matter is related to the evolution of quantum fluctuations during
the inflationary expansion of the Universe. The cosmic microwave background and
its anisotropy and polarization are the strongest observational evidences of
this theory. The B-mode of the primordial CMB polarization encodes the tensor to
scalar ratio $r$, which measures the power of gravitational waves in the early
Universe that are related to the inflationary potential. Planck observations
limit $r$ to about $r \le 0.036-0.1$~\citep{2013ApJS..208...19H,
2018arXiv180706209P, 2015PhRvL.114j1301B, 2018PhRvL.121v1301B,
PhysRevLett.127.151301}. Unfortunately, the theoretical predictions of the
$r$-parameter are model-dependent and range from $10^{-2}$ to $10^{-4}$. In the
next decade, there will be new efforts to study the CMB polarization in order to
improve constraints on cosmological parameters and the physics of inflation. The
next generation of CMB experiments include LiteBIRD~\citep{2012SPIE.8442E..19H},
CMB-S4~\citep{2016arXiv161002743A}, the Simons Observatory
\citep{2019JCAP...02..056A}, POLARBEAR \citep{2011arXiv1110.2101K}, and AliCPT
\citep{doi.10.1093.nsr.nwy019}. These experiments are targetting a sensitivity
of $r=0.001$. However, the lesson after BICEP2 is that measuring only the power
spectrum of the B-mode polarization, which in the simplest theoretical
simulation is zero in the absence of cosmological gravitational waves (GW), is
insufficient.

Additionally to the power spectrum estimation, special attention should be paid
to the study of the statistical properties of the B-mode sky map in order to
isolate potential non-Gaussianities indicating its contamination by foreground
remnants and systematic effects. The obvious importance of this problem for
understanding the physics of inflation and GWs requires the development of
methods complementary to the standard decomposition of the $Q$/$U$ Stokes
parameters into E/B components. If we are sure that in the absence of
cosmological GWs we have zero B-mode, but the measured B-mode is non-zero and
Gaussian, this signal would be an important indicator of the properties of
inflation. But already the experience of Planck shows that at $r\sim 0.05-0.10$
the B-mode polarization is different from zero, and is associated with
non-cosmological (and non-Gaussian) signals. Here one should also add the effect
of lensing of the primary E-mode, leading to the appearance of a nonzero B-mode
even in the absence of GWs.

Thus, in addition to the standard transition $(Q,U)\rightarrow(E,B)$, where the
Stokes parameters $(Q,U)$ are treated as components of a pseudo-vector and
convolved into a combination of scalar E and pseudoscalar B, it is important to
analyze also the possibility of transforming Q/U into two pseudo-vectors
$(Q_E,U_E)$ and $(Q_B,U_B)$ called the E- and
B-families~\cite{2018JCAP...05..059L}, where the B-family $(Q_B,U_B)$ is still
zero in the absence of GWs and lensing. Among other interesting properties, this
conception of the E/B decomposition leads to useful approaches in polarized
foreground analysis~\cite{2018JCAP...05..059L,2018arXiv180711940R}.

To achieve this goal, we will in this paper augment the previous work
\cite{2018JCAP...05..059L} and exploit the concept of quaternion decomposition
of the Stokes parameters into $(Q_E,U_E)$ and $(Q_B,U_B)$, which leads to two
pairs of maps with morphology different from the morphology of the scalar E and
B maps.

The quaternion concept was introduced by the Irish mathematician William Rowan
Hamilton in 1843, and the German mathematician Frobenius in 1877 proved that
every finite-dimensional associative division algebra over the real numbers is
isomorphic to one of the three: real numbers, complex numbers and the
quaternions. Here we present implementation of that concept for the CMB
polarization from the recent Planck data release. The idea is to represent the
polarization data as a quaternion with $(Q_E,U_E)$ and $(Q_B,U_B)$ in separate
quaternionic components. Then the E and B modes, in either harmonic
($a_{E/B,\ell m}$) or real-space ($P_{E/B}$) representations, are related by
convolution with an appropriate kernel. Again exploiting the dimensions of the
quaternion to combine the E and B modes into a single mathematic object, the
equations can be written compactly, and new mathematical properties of the E/B
transform are revealed. Furthermore, we also consider generalizations such as
the quaternionic spin eigensystem for arbitrary spins and the general relation
between the real space maps and the parity space representation.

The outline of the paper is the follows: In Section \ref{sec:review} the
formalism of the Stokes-space EB family decomposition is reviewed, and Section
\ref{sec:quat} presents the new theory of the polarization quaternion. In
Section \ref{sec:eigen} the quaternionic eigenproblem of a spin system is
discussed, and Section \ref{sec:application} is devoted to application of the
EB-families to foreground analysis, focusing especially on the Planck 30 GHz and
353 GHz frequency maps, in comparison with other astrophysical datasets. The
results reveal local features associated with B-mode emission. Lastly a brief
discussion is given in Section \ref{sec:disscuss}.

\section{Review of the basis}\label{sec:review}

We start from the well known forward-backward spin-2 spherical harmonics
transforms of the Stokes parameter:
\begin{align}\label{equ:spin 2 sht}
	a_{\pm2,\ell m} &= \int (Q \pm U\bm{i}) [_2 Y^*_{\ell m}] d\bm{n}
	\\ \nonumber
	Q \pm U\bm{i} &= \sum_{\ell m} a_{\pm2,\ell m} [_{\pm 2} Y_{\ell m}],
\end{align}
note that $Q=Q(\bm{n})$, $U=U(\bm{n})$ and $_{\pm 2} Y_{\ell m}=_{\pm 2} Y_{\ell
m}(\bm{n})$ are all functions of the unit pointing vector $\bm{n}$, which is
omitted in the above equation for convenience. The E- and B-mode harmonic
coefficients are defined as \footnote{\label{foo:convention}~Note that the
definition here is different to the definition in~\cite{PhysRevD.55.1830} by
factor $-1$ for both the E and B mode coefficients. The reason is that we choose
to attach $\bm{i}$ to $U$ and $a_{B,\ell m}$ as $U\bm{i}$ and $a_{B,\ell
m}\bm{i}$ to simplify the equations (which is also important for the quaternion
form to be presented later). For the same reason, the definition in
eq.~(\ref{equ:define F}) is also different to our previous
work~\citep{2018JCAP...05..059L} by a constant factor.}:
\begin{align}\label{equ: def of almEB}
	a_{E,\ell m} &= \frac{a_{2,\ell m} + a_{-2,\ell m}}{2}
	\\ \nonumber
	a_{B,\ell m} \bm{i} &= \frac{a_{2,\ell m} - a_{-2,\ell m}}{2}.
\end{align}
From eq.~(\ref{equ:spin 2 sht}) we get
\begin{align} \label{equ:qu from spin alm}
    Q &= \frac{1}{2} \left( \sum_{\ell m} a_{2,\ell m} [{}_{2}Y_{\ell m}] +
    \sum_{\ell m} a_{-2,\ell m}\;[_{-2}Y_{\ell m}] \right),
    \nonumber \\
    U \bm{i} &= \frac{1}{2}\left( \sum_{\ell m} a_{2,\ell m} [{}_{2}Y_{\ell m}]
    - \sum_{\ell m} a_{-2,\ell m}\;[_{-2}Y_{\ell m}]
    \right),
\end{align}
and by assuming  $a_{E,\ell m}=0$ and $a_{B,\ell m}=0$ respectively, we get
\begin{align}\label{equ:eb only sh domain}
	\left\{
	\begin{matrix*}[l]
		a_{B,\ell_m}=0 & \Longrightarrow & a_{2,\ell m} = a_{-2,\ell m} =
		a_{E,\ell m} \\ a_{E,\ell_m}=0 & \Longrightarrow & a_{2,\ell m} = -
		a_{-2,\ell m} = a_{B,\ell m} \bm{i}
	\end{matrix*}
	\right. 
	\\ \nonumber
\end{align}
Then we define two functions with even and odd parities respectively:
\begin{align}\label{equ:define F}
    F_{+,lm} &= \frac{1}{2} \left({}_{2}Y_{lm} + {}_{-2}Y_{lm} \right)
    \quad
    F_{-,lm} = \frac{1}{2} \left({}_{2}Y_{lm} - {}_{-2}Y_{lm} \right),
\end{align}
which gives
\begin{align}\label{equ: QU eb solution E}
	a_{B,\ell_m} &= 0 \Longrightarrow
	\left\{
	\begin{matrix*}[r]
		Q_E = \sum_{\ell m} a_{E,\ell m} F_{+,\ell m} \\		
		U_E\bm{i} = \sum_{\ell m} a_{E,\ell m} F_{-,\ell m}
	\end{matrix*}
	\right.
	\\ \nonumber
	a_{E,\ell_m} &= 0 \Longrightarrow
	\left\{
	\begin{matrix*}[l]
		Q_B = \sum_{\ell m} (a_{B,\ell m}\bm{i}) F_{-,\ell m} \\		
		U_B\bm{i} = \sum_{\ell m} (a_{B,\ell m}\bm{i}) F_{+,\ell m}
	\end{matrix*}
	\right. .
\end{align}
Combining eq.~(\ref{equ:spin 2 sht}) and eq.~(\ref{equ:define F}) gives
\begin{align}\label{equ:aeab in form of qu and fpfm}
	a_{E,\ell m} &= \frac{a_{2,\ell m}+a_{-2,\ell m}}{2} 
	= \frac{1}{2}\int \left\{ (Q+U\bm{i})[ _2Y^*_{\ell m}] +
	(Q-U\bm{i})[ _-2Y^*_{\ell m}]\right\} d\bm{n}
	\\ \nonumber
	&= \int \left(Q F^*_{+,\ell m} + (U\bm{i}) F^*_{-,\ell m} \right)
	d\bm{n}
	\\ \nonumber
	a_{B,\ell m}\bm{i} &= \frac{a_{2,\ell m}-a_{-2,\ell m}}{2} 
	= \frac{1}{2}\int \left\{ (Q+U\bm{i})[ _2Y^*_{\ell m}] -
	(Q-U\bm{i})[ _-2Y^*_{\ell m}]\right\} d\bm{n}
	\\ \nonumber
	&= \int \left(Q F^*_{-,\ell m} + (U\bm{i}) F^*_{+,\ell m} \right)
	d\bm{n};
\end{align} 

Thus, we have
\begin{align}\label{equ:QE init}
	& Q_E(\bm{n}) = \sum_{\ell m} a_{E,\ell m} F_{+,\ell m}(\bm{n}) = 
	\\ \nonumber
	& \int d\bm{n}' \left\{ Q(\bm{n}') \sum_{\ell m}\left[F_{+,\ell
	m}(\bm{n})F^*_{+,\ell m}(\bm{n}')\right] + U(\bm{n}')\bm{i}
	\sum_{\ell m} \left[ F_{+,\ell m}(\bm{n})F^*_{-,\ell m}(\bm{n}')
	\right] \right\}.
\end{align}
Use eq.~(7) of~\cite{1997PhRvD..56..596H} to compute the $F_{+}F^*_{+}$ and
$F_{+}F^*_{-}$ terms, we get
\begin{align}
	\sum_{\ell m} F_{+,\ell m}(\bm{n})F^*_{+,\ell m}(\bm{n}') &=
	\frac{1}{4} \sum_\ell \left( {}_{2}Y_{\ell,-2} + {}_{2}Y_{\ell ,2} +
	{}_{-2}Y_{\ell,-2} + {}_{-2}Y_{\ell,2} \right)(\beta,\alpha)
	\\ \nonumber 
	&= \sum_\ell \mathrm{Re}\left( \sqrt{\frac{2\ell+1}{4\pi}} \,
	\frac{_2Y_{\ell, -2}(\beta,\alpha) + {}_2Y_{\ell,
	2}(\beta,\alpha)}{2} \right)
	\\ \nonumber
	\sum_{\ell m} F_{+,\ell m}(\bm{n})F^*_{-,\ell m}(\bm{n}') &=
	\frac{1}{4} \sum_\ell \left( {}_{2}Y_{\ell,-2} - {}_{2}Y_{\ell ,2} +
	{}_{-2}Y_{\ell,-2} - {}_{-2}Y_{\ell,2} \right)(\beta,\alpha)
	\\ \nonumber 
	&= \sum_\ell \bm{i}\,\mathrm{Im}\left(\sqrt{\frac{2\ell+1}{4\pi}} \,
	\frac{_2Y_{\ell, -2}(\beta,\alpha) - {}_2Y_{\ell,
	2}(\beta,\alpha)}{2} \right),
\end{align}
where $(\beta,\alpha)$ is the Euler angles of rotation from $\bm{n}$ to
$\bm{n}'$. Continue with the other two combinations gives
\begin{align}\label{equ:fpfm all}
	\sum_{\ell m} F_{+,\ell m}(\bm{n})F^*_{+,\ell m}(\bm{n}') &=
	\mathrm{Re}(\mathscr{F}_\mathrm{+}) ;\;\;
	\sum_{\ell m} F_{+,\ell m}(\bm{n})F^*_{-,\ell m}(\bm{n}') =
	\bm{i}\,\mathrm{Im}(\mathscr{F}_\mathrm{-})
	\\ \nonumber
	\sum_{\ell m} F_{-,\ell m}(\bm{n})F^*_{-,\ell m}(\bm{n}') &=
	\mathrm{Re}(\mathscr{F}_\mathrm{-}) ;\;\;
	\sum_{\ell m} F_{-,\ell m}(\bm{n})F^*_{+,\ell m}(\bm{n}') =
	\bm{i}\,\mathrm{Im}(\mathscr{F}_\mathrm{+})
\end{align}
where $\mathscr{F_\pm}$ are defined as
\begin{align}\label{equ:define ypm}
	\mathscr{F}_+(\bm{n},\bm{n}') &= \sum_\ell \sqrt{\frac{2\ell+1}{4\pi}} \,
	\frac{_2Y_{\ell, -2}(\beta,\alpha) + {}_2Y_{\ell,
	2}(\beta,\alpha)}{2}
	\\ \nonumber
	\mathscr{F}_-(\bm{n},\bm{n}') &= \sum_\ell \sqrt{\frac{2\ell+1}{4\pi}} \,
	\frac{_2Y_{\ell, -2}(\beta,\alpha) - {}_2Y_{\ell,
	2}(\beta,\alpha)}{2}.
\end{align}
Substitute the above equation back to eq.~(\ref{equ:QE init}) gives
\begin{align}\label{equ:compute QE}
	Q_E(\bm{n}) = \int \left(
	\mathrm{Re}[\mathscr{F}_+(\bm{n},\bm{n}')]Q(\bm{n}') -
	\mathrm{Im}[\mathscr{F}_-(\bm{n},\bm{n}')]U(\bm{n}') \right)
	d\bm{n}'.
\end{align}
Continue with $U_E$, $Q_B$, $U_B$ and with similar processes we get
\begin{align}\label{equ:QUEB final}
	\begin{pmatrix}
		Q_E \\ U_E
	\end{pmatrix} &= \int 
	\begin{pmatrix*}[r]
		\mathrm{Re}(\mathscr{F}_+) & -\mathrm{Im}(\mathscr{F}_-) \\ 
		\mathrm{Im}(\mathscr{F}_+) & \mathrm{Re}(\mathscr{F}_-)  
	\end{pmatrix*}
	\begin{pmatrix}
		Q \\ U
	\end{pmatrix} d\bm{n}'
	\\ \nonumber
	\begin{pmatrix}
		Q_B \\ U_B
	\end{pmatrix} &= \int 
	\begin{pmatrix*}[r]
		\mathrm{Re}(\mathscr{F}_-) & -\mathrm{Im}(\mathscr{F}_+) \\ 
		\mathrm{Im}(\mathscr{F}_-) & \mathrm{Re}(\mathscr{F}_+)  
	\end{pmatrix*}
	\begin{pmatrix}
		Q \\ U
	\end{pmatrix} d\bm{n}'.
\end{align}
This equation is equivalent to eq. (2.23) of~\cite{2018JCAP...05..059L},
but is further simplified.

\section{Quaternion representation of the E and B families}
\label{sec:quat}

In this section, we rewrite all EB-decompositions in quaternionic forms. Some
necessary introduction of rules and conventions of quaternion conjugate and
multiplication can be found in Appendix~\ref{app:sys of quat mul}, and for
reader's convenience, a direct summary of the main results is given in
Appendix~\ref{app:eb family quick reference}.

\subsection{The pixel-to-harmonic domain transform}

By definition, each quaternion has two equivalent forms: either consisting of
four real numbers or two complex numbers:
\begin{align}
	z_1 + z_2 \bm{j} &= (a+b\bm{i})+(c+d\bm{i})\bm{j} =
	a+b\bm{i}+c\bm{j}+d\bm{k}.
\end{align}
Therefore, with the fact that all the following quantities are ordinary complex
numbers: $a_E$, $a_B\bm{i}$, $F_{\pm,\ell m}$, $\mathscr{F}_\pm$, $P_E=Q_E + U_E
\bm{i}$, $P_B=Q_B + U_B \bm{i}$; we can define five quaternions
straightforwardly\footnote{Note that we use math calligraphy characters for
quaternions, bold letters for matrices and vectors, and follow footnote
\ref{foo:convention} to attach $\bm{i}$ with $U$ and $a_{B}$, }:
\begin{align}\label{equ:quat form base}
	\mathcal{F}_{\ell m} &= F_{+,\ell m} + F_{-,\ell m} \bm{j}
	\\\nonumber
	\mathcal{G} &= \mathscr{F}_+ + \mathscr{F}_- \bm{j}
	\\\nonumber
	\mathcal{D} &= P_E + P_B \bm{j}
	\\ \nonumber
	\mathcal{A}_{\ell m} &= a_{E,\ell m} + (a_{B,\ell m}\bm{i})\bm{j} =
	a_{E,\ell m} + a_{B,\ell m}\bm{k}
	\\\nonumber
	\mathcal{P} &= Q + (U\bm{i})\bm{j} = Q + U\bm{k},
\end{align}
where $\mathcal{F}_{\ell m}$ is the quaternion version of the spin-spherical
harmonic function; $\mathcal{G}$ is the pixel domain quaternionic convolution
kernel; $\mathcal{D}$ is the pixel domain polarization quaternion that directly
contains the EB-families; $\mathcal{A}$ is the harmonic domain counterpart of
$\mathcal{D}$ and is a simple combination of the E- and B-mode harmonic
coefficient; and $\mathcal{P}$ is a reduced form of $\mathcal{D}$ that contains
only the pixel domain Q and U Stokes parameters.

With these definitions, the complete forward EB-transform (pixel domain to
harmonic domain) is given by a simple quaternionic integration:
\begin{align}\label{equ:harm quat forward}
	\mathcal{A}_{\ell m} = \int \mathcal{F}^{*_{101}}_{\ell m} \,
	\mathcal{P},
\end{align}
where $*_{101}$ refers to the type-101 quaternion conjugate
(Appendix~\ref{sub:quat conj}), which is introduced to match the normal E- and
B-mode definitions (more possibilities of conjugate will be introduced below).
The above equation gives the E- and B-mode harmonic coefficients at the same
time as components of $\mathcal{A}$. It is also easy to see that the backward
transform is (note that we need eq.~(\ref{equ:quat conj odd even}) to get the
following equation):
\begin{align}\label{equ:harm quat backward}
	\mathcal{P} &= 
	\sum_{\ell m} \mathcal{A}^{*_{010}}_{\ell m} \, \mathcal{F}_{\ell
	m} = 
	\sum_{\ell m} \mathcal{F}^{*_{010}}_{\ell m} \, \mathcal{A}_{\ell
	m},
\end{align}
and the self-consistency of the forward/backward transforms is verified in the
following proof (also pay attention to eq.~(\ref{equ:quat conj odd even}), and
note that $\mathcal{P}^{*_{010}} = \mathcal{P}$):
\begin{align}
	\mathcal{P}
	=& \sum_{\ell m} \mathcal{A}_{\ell m}^{*_{010}} \, \mathcal{F}_{\ell
	m} =
	\sum_{\ell m} \left(\int \mathcal{F}_{\ell m}^{*_{101}} \,
	\mathcal{P}\right)^{*_{010}} \, \mathcal{F}_{\ell m}
	\\ \nonumber =& \sum_{\ell m} \left(\int \mathcal{P}^{*_{010}} \,
	(\mathcal{F}_{\ell m}^{*_{101}})^{*_{010}}\right) \,
	\mathcal{F}_{\ell m} = \sum_{\ell m} \left(\int
	\mathcal{P}^{*_{010}} \,
	\mathcal{F}_{\ell m}^{*_{111}}\right) \, \mathcal{F}_{\ell m}
	\\ \nonumber =& \sum_{\ell m} \int \mathcal{P} \,
	\left(\mathcal{F}^{*}_{\ell m} \mathcal{F}_{\ell m}\right) 
	\\ \nonumber
	=&  \int d\bm{n}'\mathcal{P}(\bm{n}') \, \sum_{\ell m}
	\mathcal{F}^{*}_{\ell m}(\bm{n}') \mathcal{F}_{\ell m}(\bm{n}) 
	\\ \nonumber
	=& \,\mathcal{P} .
\end{align}

If we write $\mathcal{F}_{\ell m}$ into a quaternion matrix $\bm{\mathcal{F}}$
\footnote{For the matrix form $\bm{\mathcal{F}}$, we no longer need the
subscripts $\ell m$, because they are now the column indices of the matrix.},
with the columns being the value of $\mathcal{F}_{\ell m}$ at the spherical
directions $\bm{n}$ (preferably sorted in the HEALPix~\cite{2005ApJ...622..759G}
ring ordering) and rows being the value of $\mathcal{F}_{\ell m}$ at combination
of $\ell$ and $m$, sorted with an $\ell$-first dictionary order, and
correspondingly convert $\mathcal{P}$ and $\mathcal{A}$ into column vectors of
consistent orders, then eqs.~(\ref{equ:harm quat forward}--\ref{equ:harm quat
backward}) can be further simplified as
\begin{align}\label{equ:pixel-to-harmo quat}
	\bm{\mathcal{A}} &= \bm{\mathcal{F}}^{H_{101}}
	\bm{\mathcal{P}}
	\\ \nonumber
	\bm{\mathcal{P}} &= \bm{\mathcal{F}}^{*_{010}}
	\bm{\mathcal{A}},
\end{align}
where $H_{101}$ is analog to the conjugate transpose of matrices, but the
conjugate part must be of type-101, and $^{*_{010}}$ means a type-010 conjugate
without transposing. This is quite different to a complex matrix because a
quaternion matrix has seven conjugates rather than one. The above equations are
self-consistent because we can easily prove $\bm{\mathcal{F}}^{*_{010}}
\bm{\mathcal{F}}^{H_{101}} = \bm{I}$, i.e., the $\bm{\mathcal{F}}$ matrix is unitary.

\subsection{The pixel-to-pixel domain transform}

The pixel-to-pixle domain EB-family decomposition is
\begin{align}
	\mathcal{D} = \int \mathcal{G}(\bm{n},\bm{n}') \mathcal{P}(\bm{n}')\,d\bm{n}' =
	\mathcal{G} \ast \mathcal{P}.
\end{align}
Like above, this can be further simplified to a quaternion matrix form as
\begin{align}
	\bm{\mathcal{D}} = \bm{\mathcal{G}} \bm{\mathcal{P}},
\end{align}
where $\bm{\mathcal{G}}$ is a quaternion matrix with the columns and rows being
the values of $\mathcal{G}$ at $(\bm{n},\bm{n}')$, respectively. Note that
because quaternion multiplication is non-commutative, we must be very careful in
changing the order of all quaternion equations, especially the ones in
combination with a matrix multiplication. This is further discussed in
Appendix~\ref{app:sys of quat mul}.

\subsection{Transform the polarization quaternion to the harmonic domain and
back}

It is also easy to prove that we can get $\mathcal{D}$ directly from
$\mathcal{A}$ by using only the spin-2 spherical harmonic function:
\begin{align}
	\mathcal{D} &= \sum_{\ell m} (F_{+,_{\ell m}} + F_{-,_{\ell
	m}})\mathcal{A}_{\ell m} = \sum_{\ell m} {}_{2} Y_{\ell m}
	\mathcal{A}_{\ell m}
	\\ \nonumber
	\mathcal{A}_{\ell m} &= \int {}_{2} Y^*_{\ell m} \mathcal{D},
\end{align}
Let $\bm{\mathcal{Y}}_2$ be a matrix form with the columns being the values of
${}_{2} Y_{\ell m}(\bm{n})$ at various $\bm{n}$, and the rows being its values
at different combinations of $\ell$ and $m$; then we get the matrix forms of the
above equations:
\begin{align}\label{equ:d4 to a4}
	\bm{\mathcal{D}} &= \bm{\mathcal{Y}}_2 \bm{\mathcal{A}}
	\\ \nonumber
	\bm{\mathcal{A}} &= \bm{\mathcal{Y}}^{H}_2
	\bm{\mathcal{D}},
\end{align}
where $H \equiv H_{111}$ means to take the full conjugate transpose of the
quaternion matrix. Therefore, by using the E and B families, the spherical
harmonic transform of polarizations can be done even without a minus-spin
spherical harmonics.

\subsection{Alternative forms}

Due to quaternion's complexity, the form of transform is not unique. In
this section, we discuss the possibilities of various forms of the
transform.

\subsubsection{Different choices of the quaternion conjugates}

The forward and backward EB-transforms can actually be expressed using the other
types of quaternion conjugates (see Appendix~\ref{sub:quat conj}), provided that
they are paired correctly. The pairs are type-1/type-6, type-3/type-4, in
addition to the type-2/type-5 used above. Depending on the type of conjugation
used, the specific representation of $a_E$ and $a_B$ in a quaternion form can
vary from $\mathcal{A}_4$, but in all cases $a_E$ and $a_B$ are well-defined,
and it always corresponds to the same E and B power spectra.

Using type-0 conjugation,
\begin{equation}
    a_{E,\ell m}^* - a_{B,\ell m}^* \bm{k} = \int \mathcal{F}_{\ell m}
    (Q - U \bm{k}),
\end{equation} Using type-3 conjugation (011),
\begin{equation}
    a_{E,\ell m}^* + a_{B,\ell m}^* \bm{k} = \int \mathcal{F}_{\ell
    m}^{*_{011}} (Q + U \bm{k}),
\end{equation}
Using type-5 conjugation (101), which was already stated above,
\begin{equation}
    a_{E,\ell m} + a_{B,\ell m} \bm{k} = \int \mathcal{F}_{\ell
    m}^{*_{101}} (Q + U \bm{k}),
\end{equation}
Using the type-6 conjugation (110) for the forward transform, 
\begin{equation}
    a_{E,\ell m} - a_{B,\ell m} \bm{k} = \int \mathcal{F}_{\ell
    m}^{*_{110}} (Q - U \bm{k}),
\end{equation}
The other conjugation types produce uneven conjugates between $F_+$ and $F_-$
that require slightly different representations. Suppose that $$a_{E, \ell m} =
x_{E, \ell m} + y_{E, \ell m} \bm{i},$$ where $x_{E,
\ell m}$ and $y_{E, \ell m}$ are both real. Then we define the modified
$a_{E, \ell m}$ as
\begin{equation}
    \overline{a}_{E,\ell m} = x_{E, \ell m} + (-1)^{m} y_{E, \ell m} \bm{i}.
\end{equation}
The key is, that this kind of modification does not change the power
spectrum:
\begin{equation}
    \left< |\overline{a}_{E,\ell m}|^2 \right> =  \left< |a_{E,\ell m}|^2
    \right>.
\end{equation}



\begin{table}[!h]
    \centering
    \begin{tabular}{c|c|l|c}
        type & output & kernel & input \\
        \hline
        0 & $a_E^* - a_B^* \bm{k}$ & $\mathcal{F}_{\ell m} $ & $Q - U \bm{k}$ \\ 
        1 & $\overline{a}_E^* + \overline{a}_B^* \bm{k}$ & $\mathcal{F}_{\ell m}^{*_{001}}$ & $Q + U \bm{k}$ \\
        2 & $\overline{a}_E^* - \overline{a}_B^* \bm{k}$ & $\mathcal{F}_{\ell m}^{*_{010}}$ & $Q - U \bm{k}$ \\
        3 & $a_E^* + a_B^* \bm{k}$ & $\mathcal{F}_{\ell m}^{*_{011}}$ & $Q + U \bm{k}$\\
        4 & $ \overline{a}_E - \overline{a}_B \bm{k}$ & $\mathcal{F}_{\ell m}^{*_{100}}$ & $Q - U \bm{k}$ \\
        5 & $a_E + a_B \bm{k}$ & $\mathcal{F}_{\ell m}^{*_{101}}$ & $Q + U \bm{k}$ \\
        6 & $a_E - a_B \bm{k}$ & $\mathcal{F}_{\ell m}^{*_{110}}$ & $Q - U \bm{k}$ \\
        7 & $\overline{a}_E + \overline{a}_B \bm{k}$ & $\mathcal{F}_{\ell m}^{*_{111}}$ & $Q + U \bm{k}$\\
    \end{tabular}
    \caption{Summary of all forward transform types. }
    \label{tab:transform-types}
\end{table}


\subsubsection{Different definitions of the basic quaternions}

We also point out that the basic quaternion definitions in eq.~(\ref{equ:quat
form base}) are not unique. For example, assume we have an alternative form as
follows:
\begin{align}\label{equ:quat form alt}
	\mathcal{F}_4 &= F_{+} + F_{-} \bm{x}
	\\ \nonumber
	\mathcal{A}_4 &= a_{E} + a_{B} \bm{y} 
	\\ \nonumber
	\mathcal{P}_2 &= Q + U\bm{z},
\end{align}
and we still want eq.~(\ref{equ:harm quat forward}) to be correct. Then by
computing the quaternion multiplication and integration and comparing the
results with eq.~(\ref{equ:aeab in form of qu and fpfm}), we get the following
constraints for $x$, $y$ and $z$:
\begin{align}
	x z = \bm{i}, \quad y=z, \quad x = -\bm{i}y.
\end{align}
There are several possible solutions to these three equations, including:
\begin{align}
	1:\left\{
	\begin{matrix*}[l]
		x = 1 \\ 
		y = z = \bm{i}
	\end{matrix*}
	\right. \quad
	2:\left\{
	\begin{matrix*}[l]
		x = -1 \\ 
		y = z = -\bm{i}
	\end{matrix*}
	\right. \quad
	3:\left\{
	\begin{matrix*}[l]
		x = \bm{j} \\ 
		y = z = \bm{k}
	\end{matrix*}
	\right. \quad
	4:\left\{
	\begin{matrix*}[l]
		x = -\bm{k} \\ 
		y = z = \bm{j}.
	\end{matrix*}
	\right. \quad
\end{align}
It is easy to prove that solutions 1 and 2 correspond to the basic spin-2
spherical harmonic transforms (eq.~\ref{equ:spin 2 sht}), and solutions 3 and 4
give two variations of the same E and B decomposition.

\section{The eigen-problem of a quaternionic system and the parity
space}\label{sec:eigen}

\subsection{The eigen-problem of a quaternionic system}

According to eq.~(\ref{equ:d4 to a4}), if the column vector $\bm{\mathcal{A}}$
contains only one non-zero element, then the matrix multiplication
$\bm{\mathcal{Y}}_2\bm{\mathcal{A}}$ is nothing but one column of
$\bm{\mathcal{Y}}_2$ right-multiplied by this non-zero element (as a single
quaternion). A more general case is: if a linear spin system can be diagonalized
in the space of spin spherical harmonics, then it can be represented by the
following equation:
\begin{align}\label{equ:quat linear system}
	\bm{\mathcal{M}} = 
	\bm{\mathcal{Y}}_2 \begin{pmatrix}
		\mathcal{A}_1 & 0 & 0 & \cdots \\
		0 & \mathcal{A}_2 & 0 & \cdots \\
		\cdots & \cdots & \cdots & \cdots \\
		0 & 0 & \cdots & \mathcal{A}_n \\
	\end{pmatrix}\bm{\mathcal{Y}}_2^H,
\end{align}
where $\mathcal{A}_i$ represents the $i$-th quaternionic eigenvalue of the
$\bm{\mathcal{M}}$-system, and the eigenvectors (eigen-modes) of the linear
system are nothing but columns of $\bm{\mathcal{Y}}_2$ which, with the
quaternion multiplication with $\mathcal{A}_i$, automatically splits into the
odd and even parity states of the spin, as components of the quaternionic
eigen-state.

Mathematically, eq.~(\ref{equ:quat linear system}) is true if all eigen-modes of
the system are represented by the spin-spherical harmonics $_sY_{\ell m}$.
Therefore, it is the general form of every measurable system that is based on
spins.

Eqs.~(\ref{equ:d4 to a4}, \ref{equ:quat linear system}) together indicate that
spins and parities are probably the two faces of one thing: only that one is
more obvious in the pixel domain, whereas the other is more obvious in the
harmonic domain.

\subsection{From real space to parity space}

Exploration of the quaternion transforms in section~\ref{sec:quat} further
reveals a new space called the parity space, which is the dual of the real
space. First, with Eq.~(\ref{equ:pixel-to-harmo quat}) we obtain
\begin{align}
	\bm{\mathcal{A}} &= \bm{\mathcal{F}}^{H_{101}}
	\bm{\mathcal{P}} = \bm{a}_{E,\ell m} + \bm{a}_{B,\ell m}\bm{k},
\end{align}
which means:
\begin{align}
	\bm{\mathcal{Y}}_0 \bm{\mathcal{A}} = \bm{E} + \bm{B}\bm{k},
\end{align}
where $\bm{\mathcal{Y}}_0$ is the spin-zero spherical harmonic matrix whose
columns consist of the values of $Y_{\ell m}(\bm{n})={}_0Y_{\ell m}(\bm{n})$ at
different directions $\bm{n}$. Therefore, we have the following:
\begin{align}
	(\bm{E} + \bm{B}\bm{k}) = \left[\bm{\mathcal{Y}}_0
	\bm{\mathcal{F}}^{H_{101}}\right] (\bm{Q}+\bm{U}\bm{k}).
\end{align}
Apparently, $(\bm{Q}+\bm{U}\bm{k})$ contains all information of polarization in
the real space, whereas $(\bm{E} + \bm{B}\bm{k})$ contains the same amount of
information, but in a new space called the parity space. Because for
polarization, the transform from the real space to the parity space is done by
the spin-0 and spin-2 spherical harmonics, we rewrite the above equation and
explicitly show spin-$s$ as subscripts:
\begin{align}
	(\bm{E} + \bm{B}\bm{k})_s = \left[\bm{\mathcal{Y}}_0
	\bm{\mathcal{F}}^{H_{101}}_s\right] (\bm{Q}+\bm{U}\bm{k})_s,
\end{align}
whose eigen-system problem (with focus of derivative operators) is fully
compatible with the well known angular momentum operators $\bm{\hat{L}}_z$ and
$\bm{\hat{L}}^{(2)}$:
\begin{align}
	\bm{\hat{L}}_z(\bm{E}_{\ell m}) &= m\bm{E}_{\ell m} \\ \nonumber
	\bm{\hat{L}}^{(2)}(\bm{E}_{\ell m}) &= \ell(\ell+1)\bm{E}_{\ell m} \\ \nonumber
	\bm{\hat{L}}_z(\bm{B}_{\ell m}\bm{k}) &= m\bm{B}_{\ell m}\bm{k} \\ \nonumber
	\bm{\hat{L}}^{(2)}(\bm{B}_{\ell m}\bm{k}) &= \ell(\ell+1)\bm{B}_{\ell m}\bm{k}
\end{align}
For simplicity, we write $\bm{\rho}_s=(\bm{E} + \bm{B}\bm{k})_s$ for the parity
space; and $\bm{P}_s=(\bm{Q}+\bm{U}\bm{k})_s$ for the real space, then the
relationship between the real and parity spaces is nothing but
\begin{align}
	\bm{\rho}_s &= \left[\bm{\mathcal{Y}}_0
	\bm{\mathcal{F}}^{H_{101}}_s\right] \bm{P}_s.
\end{align}
A special case is $s=0$, (e.g., temperature anisotropy), which shows
$\bm{\rho}_0=\bm{P}_0=\bm{T}$, i.e., the real and parity spaces are identical
for $s=0$. This also indicates that both real and parity spaces belong to the
pixel domain.

An interesting fact is: although we can measure all four Stokes parameters: $I$,
$Q$, $U$ and $V$ in the real world, they are actually distributed in two different
spaces: $I$ and $V$ are in the parity space; whereas $Q$ and $U$ are in the real space.
Therefore, the four Stokes parameters cannot be properly included in one
quaternion state. Another fact is: it is possible to transform $\bm{\rho}_s$
back to $\bm{P}_s$ using another spin $s'\ne s$, which is related to the
possible coupling of two systems with different intrinsic spins, i.e., the
parity space is probably a convenient bridge between different spins.

\section{Example of application}\label{sec:application}

\subsection{Morphology of the E- and B-families}

In this section we provide some examples of the application of the E and B
families. The primary estimators in use are the polarization intensities of the
E and B families, which can be written using the normal complex or quaternion
modulus
\begin{align}
    ||P_E|| &= ||Q_E + U_E \bm{i}|| = \sqrt{Q_E^2 + U_E^2} \\
    ||P_B|| &= ||Q_B + U_B \bm{i}|| = \sqrt{Q_B^2 + U_B^2} ,
\end{align}
and the corresponding orientation of the Stokes fields. The E and B intensities
show directly the amount of emission in E and B, and they can be used for the
identification and characterization of local features, which is obscured in the
normal E and B mode maps. Note that if the EB angular power spectrum is
negligible compared to the EE and BB angular power spectrum, then statistically,
we get $\exptt{P_E^2+P_B^2} = \exptt{P^2}$.
 
\begin{figure*}[!htb]
  \centering
  \includegraphics[width=0.32\textwidth]{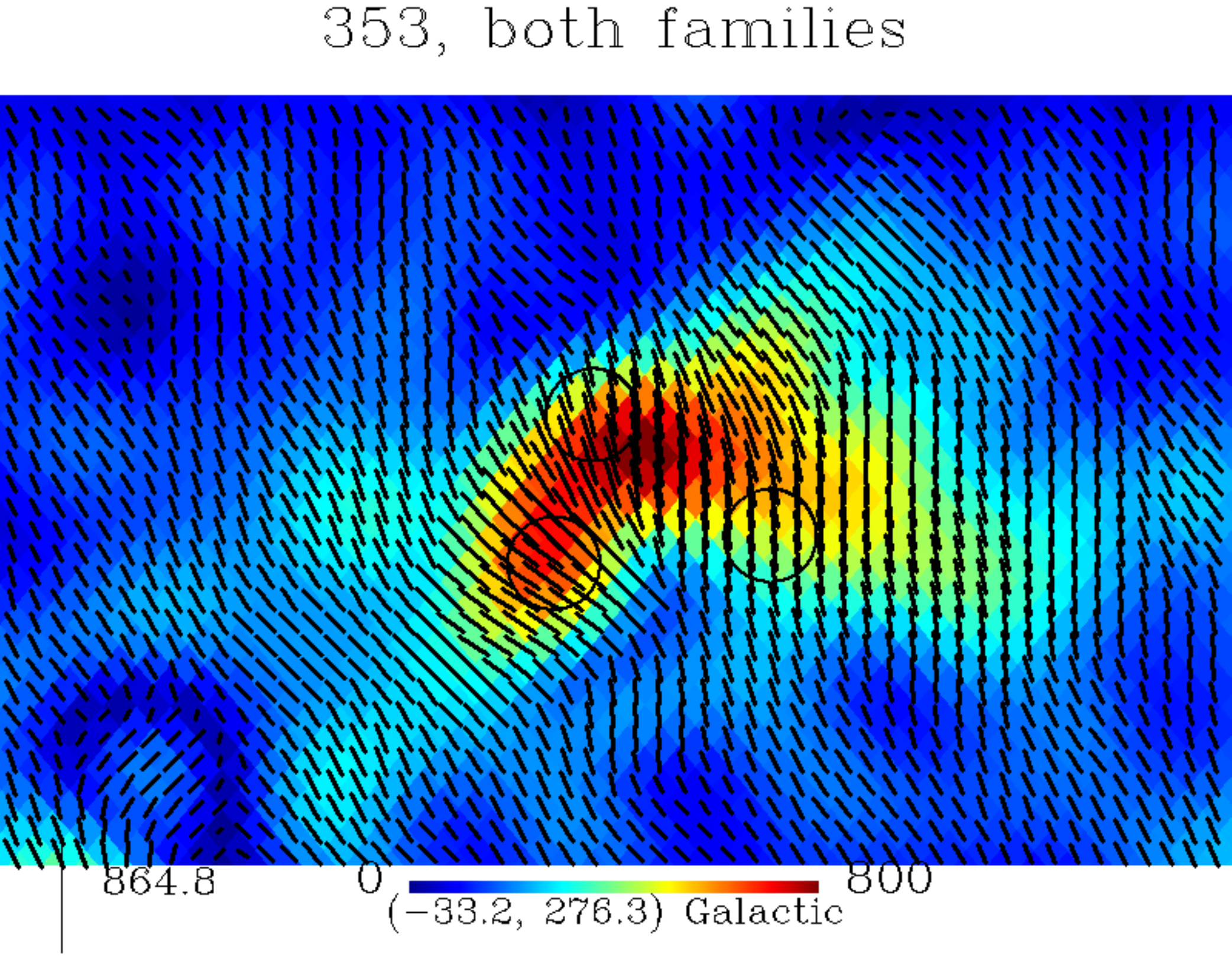}
  \includegraphics[width=0.32\textwidth]{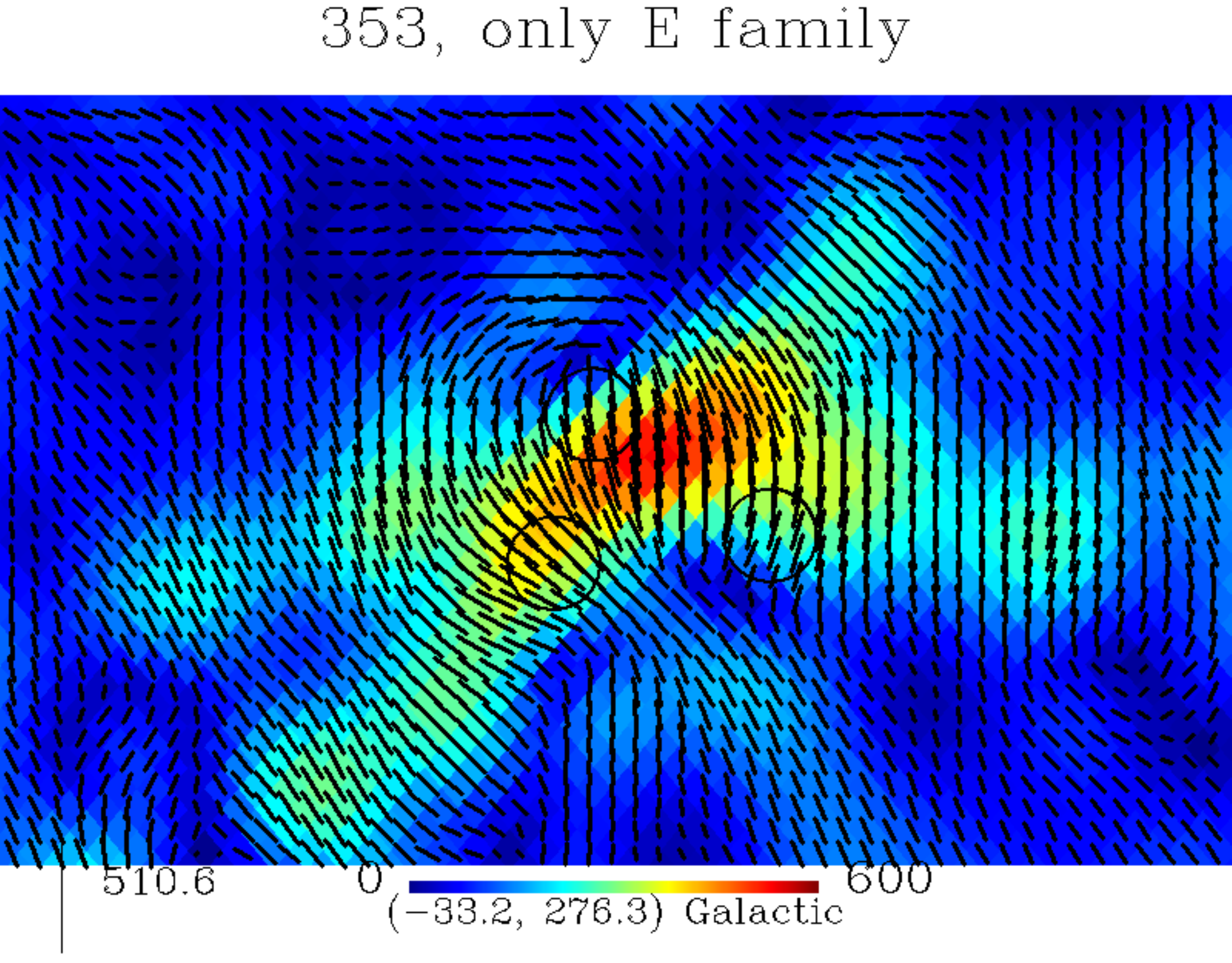}
  \includegraphics[width=0.32\textwidth]{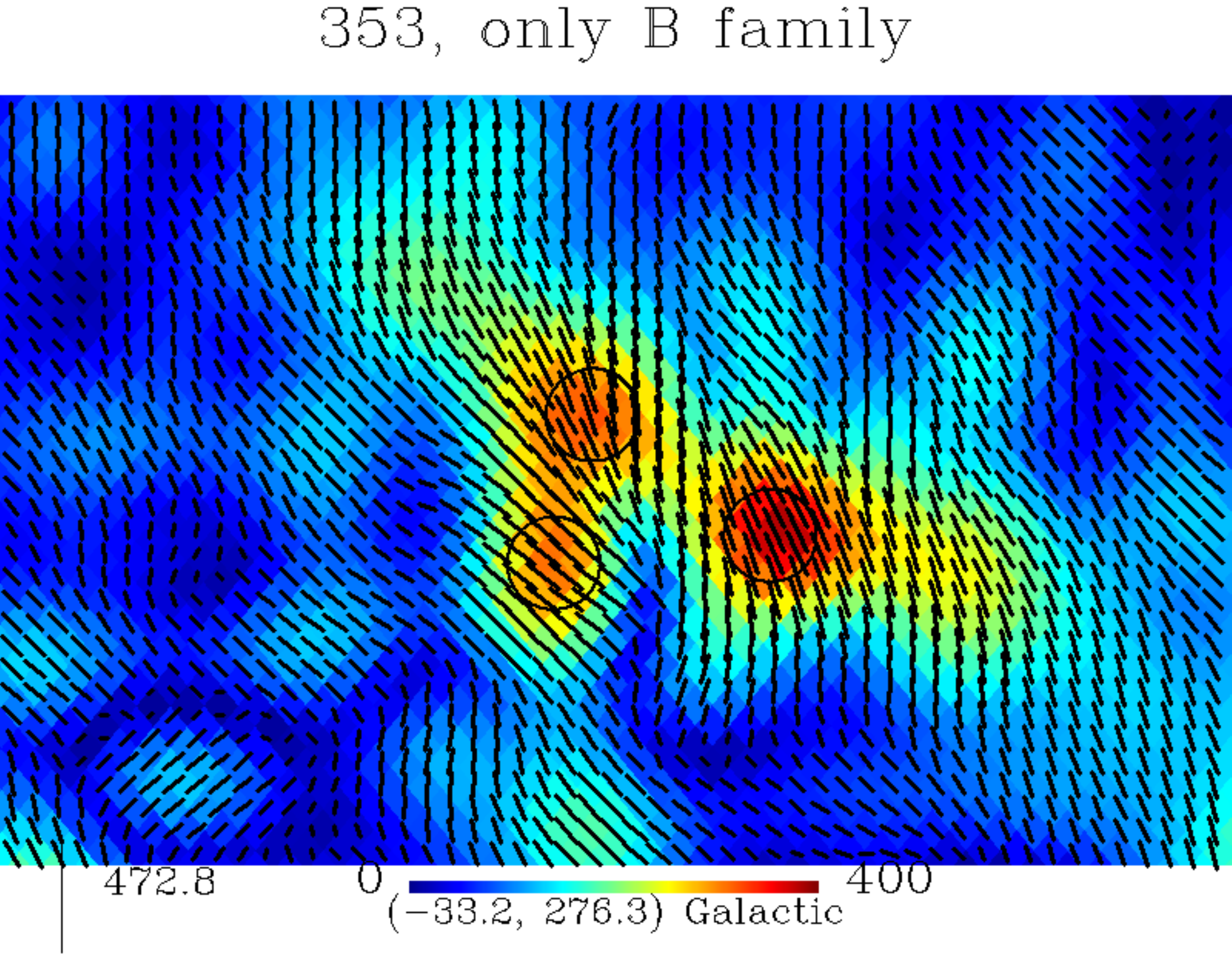}

  \includegraphics[width=0.32\textwidth]{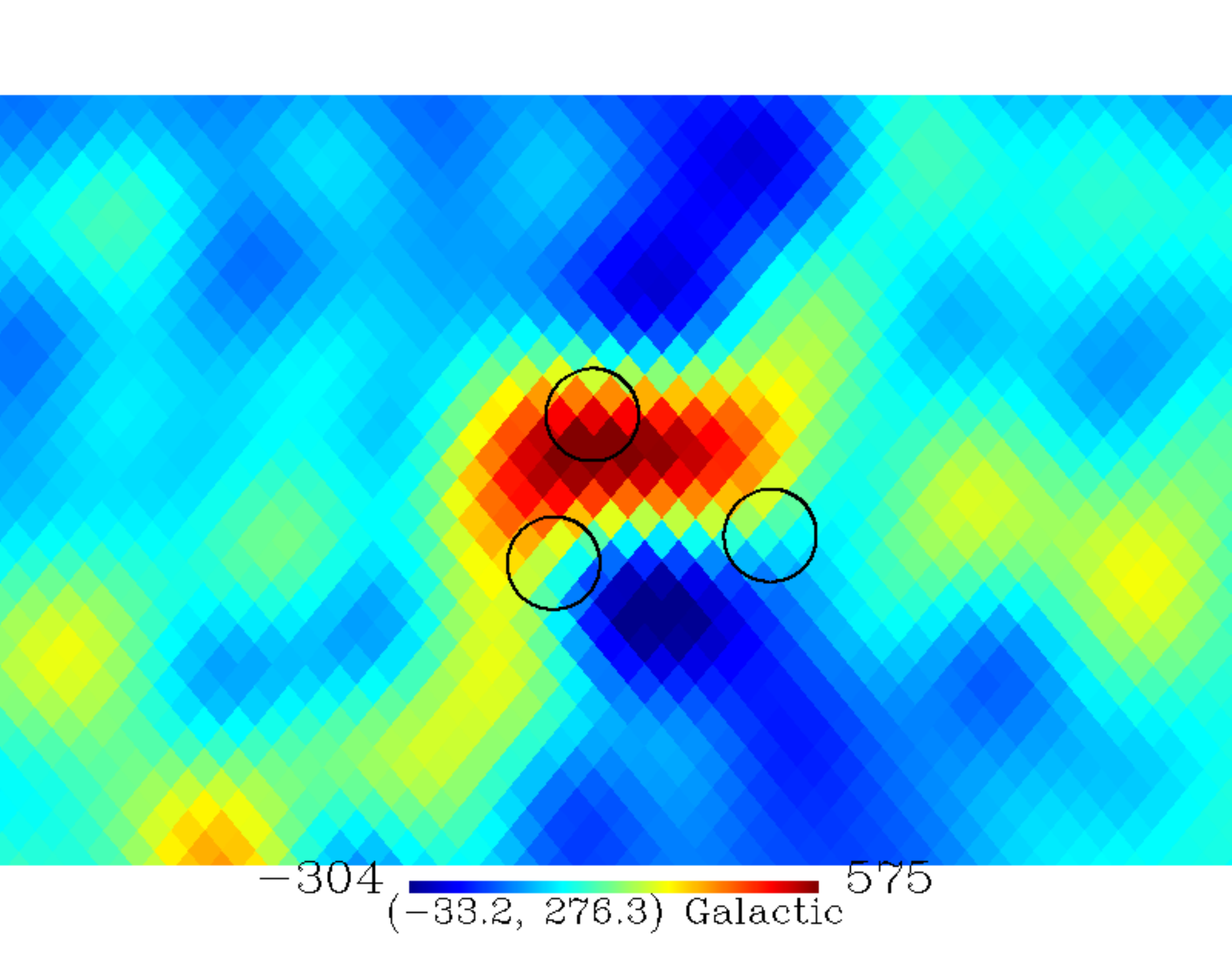}
  \includegraphics[width=0.32\textwidth]{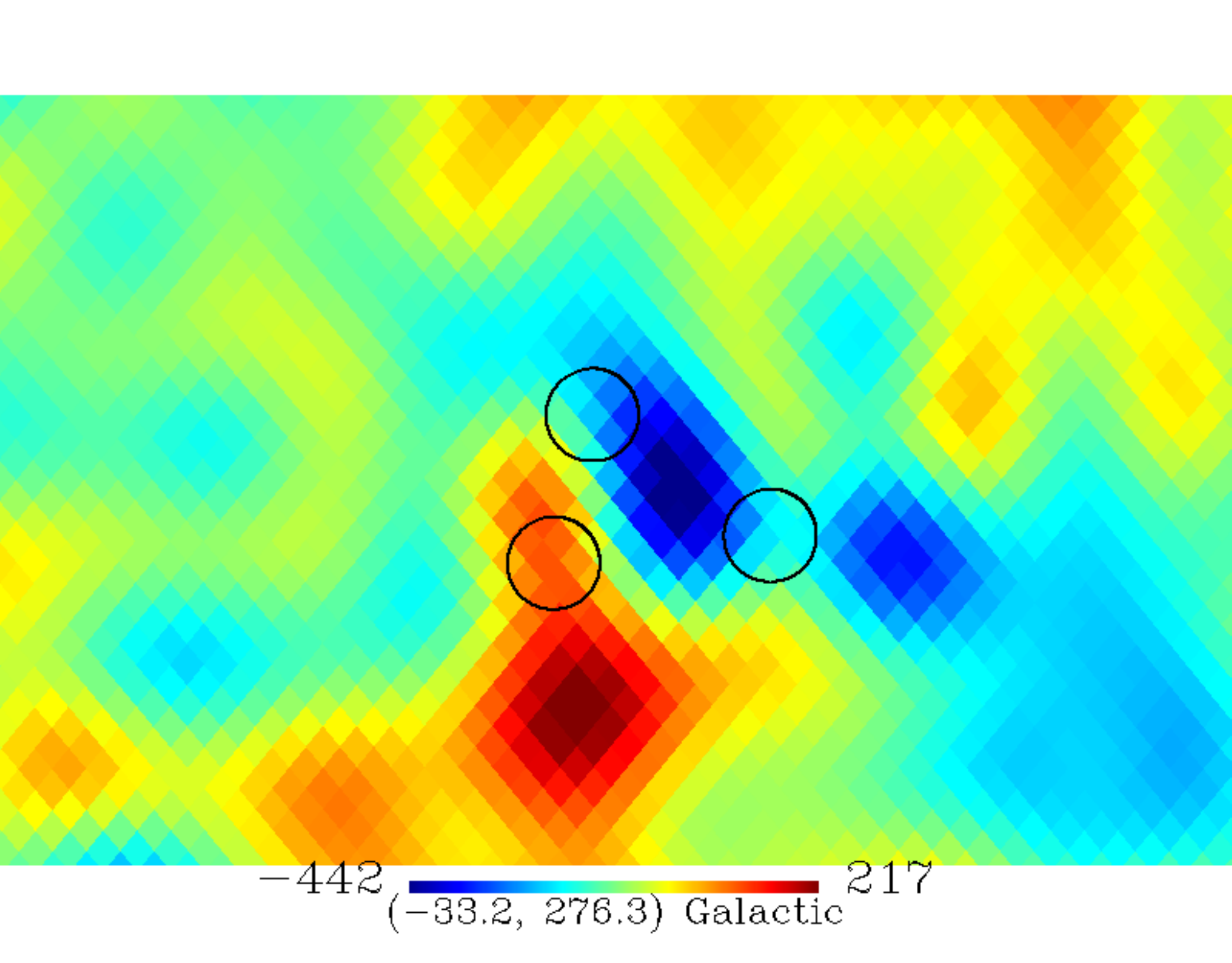}
  \caption{The Planck 353 GHz polarization map around the LMC region at
  $N_{\mathrm{side}}=2048$. \emph{Upper}, from left to right: the original
  polarization (including both the E and B families) and the E, B families.
  \emph{Lower}, from left to right: the normal scalar E, B mode maps. At least
  three point source like structures can be seen in the B-family, and an
  onion-like polarization pattern can be seen in the E-family. However, both of
  them are hard to see in the original polarization map or any of the normal E
  or B maps. The diameter of black circle is 6 arcmin.}
  \label{fig:LMC example A}
\end{figure*}

First we check the Planck 353 GHz dust polarization map, which has a relatively
high resolution. We use the original resolution and choose the well known Large
Magellanic Cloud (LMC) region as an example to plot three polarization maps of
the original polarization and the E, B families respectively in
figure~\ref{fig:LMC example A}. We can see an onion-like polarization pattern in
the E-family, and at least three point source like structures in the B-family;
however, both are hard to see in the original polarization map. We also show the
normal scalar E and B mode maps in the same region, which does not show any
special structure, and most importantly, we cannot get the polarization
direction from a scalar E or B mode map. Thus, it is apparent that the E and B
families can help to detect special structures in a polarization map.

\begin{figure*}[!htb]
  \centering
  \includegraphics[width=0.32\textwidth]{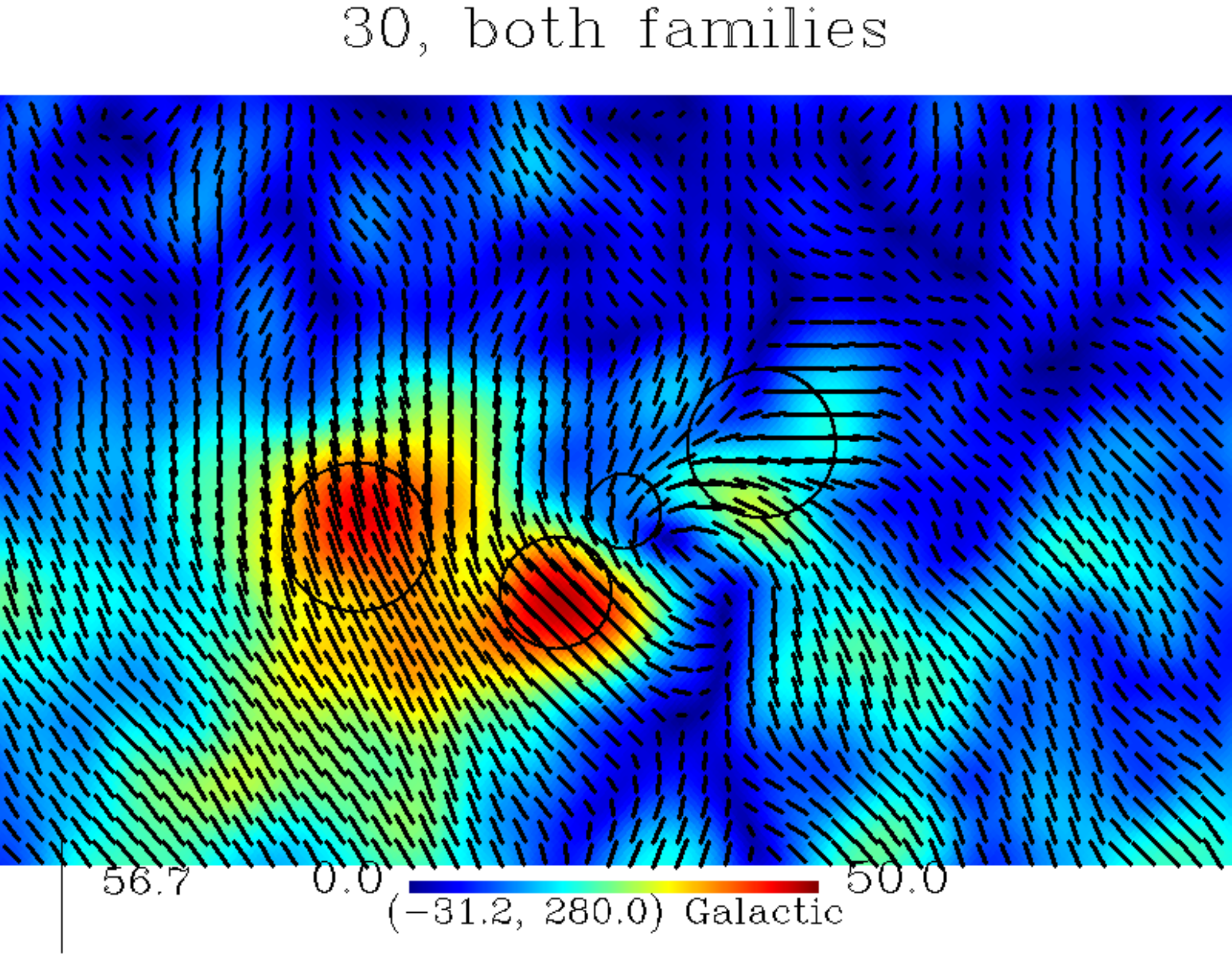}
  \includegraphics[width=0.32\textwidth]{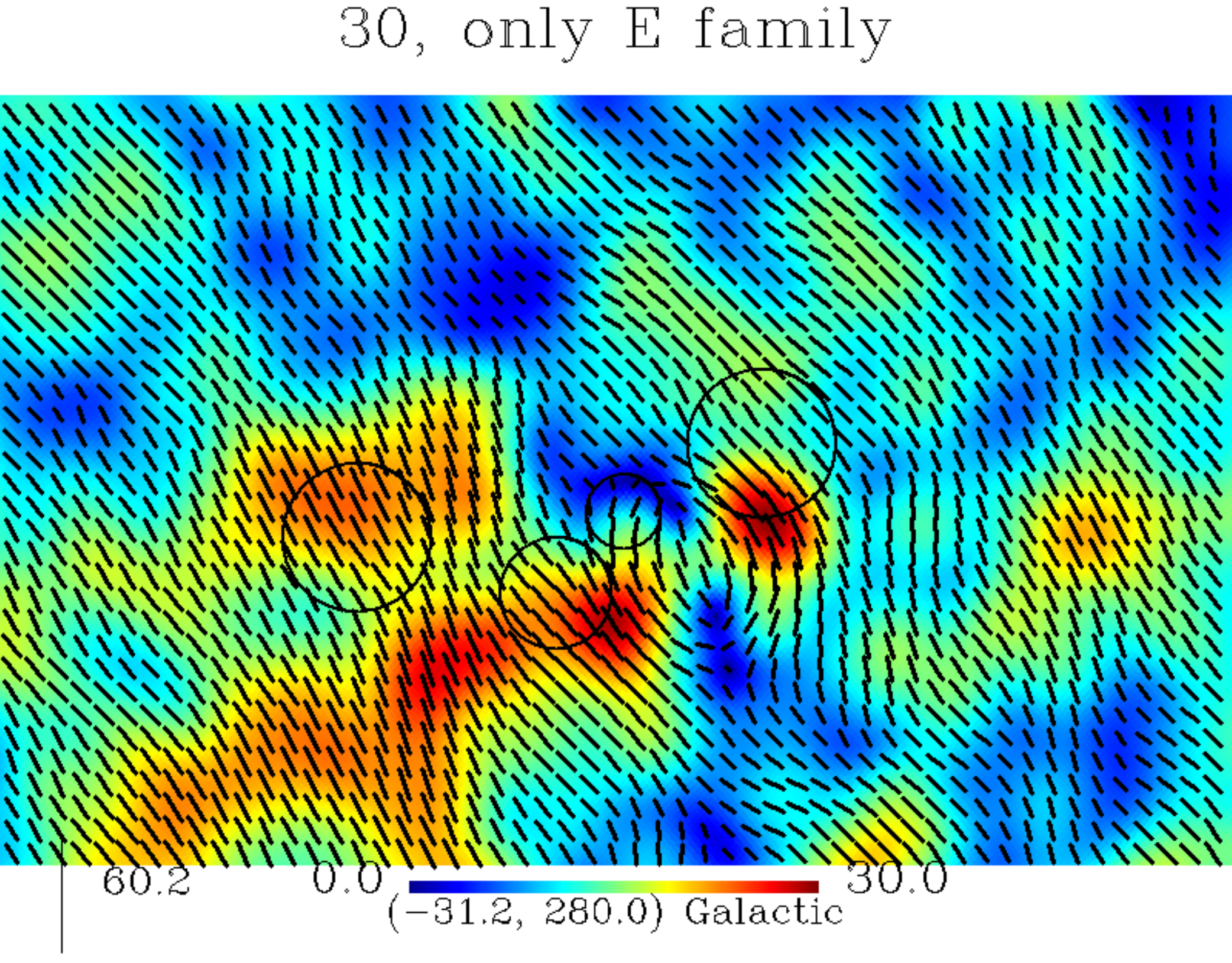}
  \includegraphics[width=0.32\textwidth]{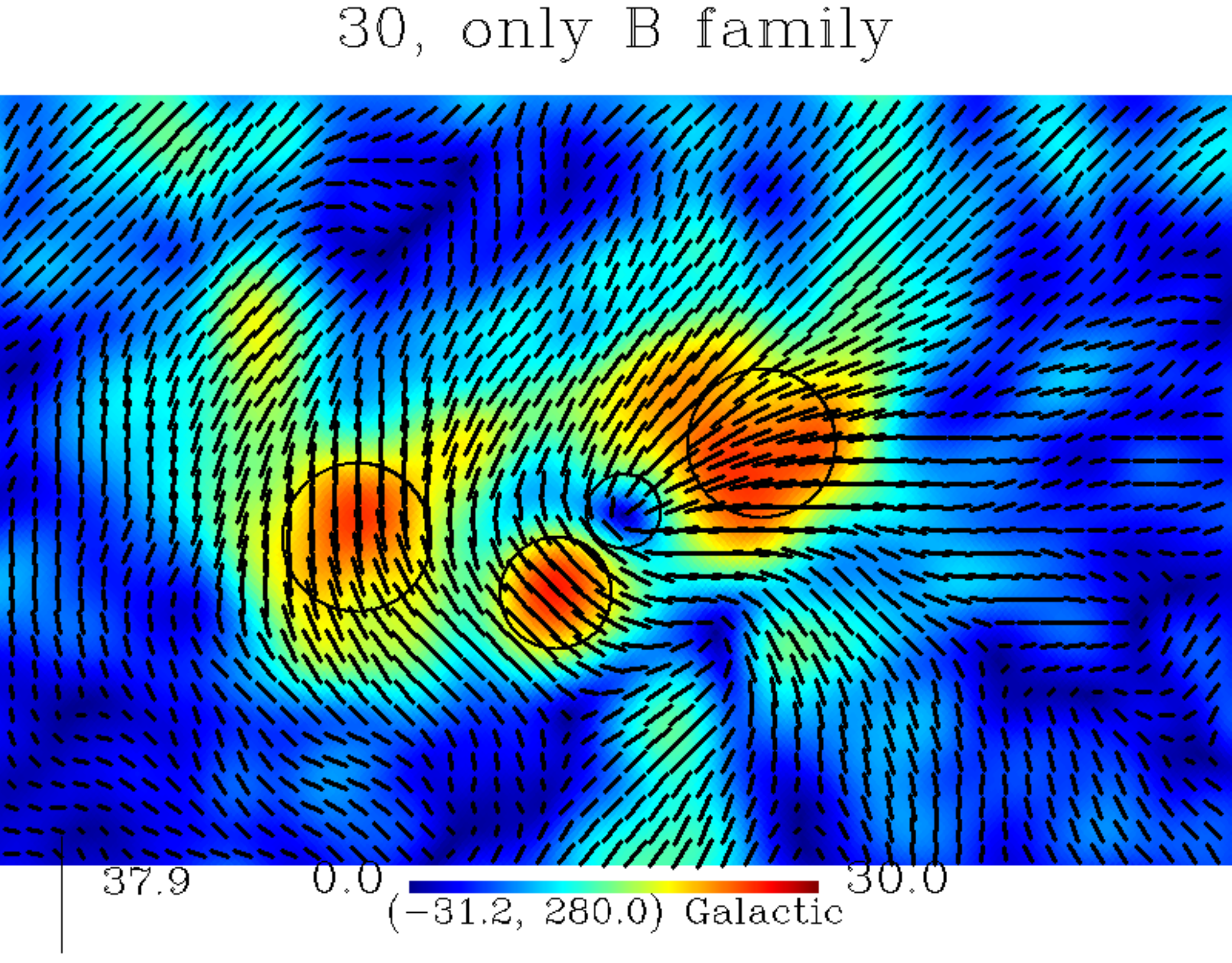}
  \caption{The Planck 30 GHz polarization map at $N_{\mathrm{side}}=1024$,
  re-beamed from $32.23'$ to $40'$ to suppress the noise, and around
  $(b,l)=(-31.4\degree,280\degree)$ in the LMC region. From left to right: the
  original polarization and the E, B family. An interesting jet-like structure
  can be seen clearly in the B-family, which starts from a small cold spot, and
  ejects towards the north-east direction. This structure is also marginally
  visible in the original polarization map, but is completely missing in the
  E-family. The diameter of the biggest black circle is $0.8\degree$.}
  \label{fig:LMC example B}
\end{figure*}

We then examine the Planck 30 GHz polarization map, which has a original
resolution of $N_{\mathrm{side}}=1024$, and we smooth it to $1\degree$-FWHM to
suppress the noises. Again three maps of the original polarization and the E and
B families are shown in figure~\ref{fig:LMC example B} for the position
$(b,l)=(-31.2\degree,280\degree)$ of the LMC region, respectively. We can see an
interesting jet-like structure in the B-family, starting from the blue (cold)
spot on the upper right corner of the black circle, and ejects towards the
north-east direction. This structure is also marginally visible in the original
polarization map, but is completely missing in the E-family. Therefore, special
structures can be much more visible in the E or B family than in the original
polarization map.

\begin{figure*}[!htb]
  \centering
  \includegraphics[width=0.32\textwidth]{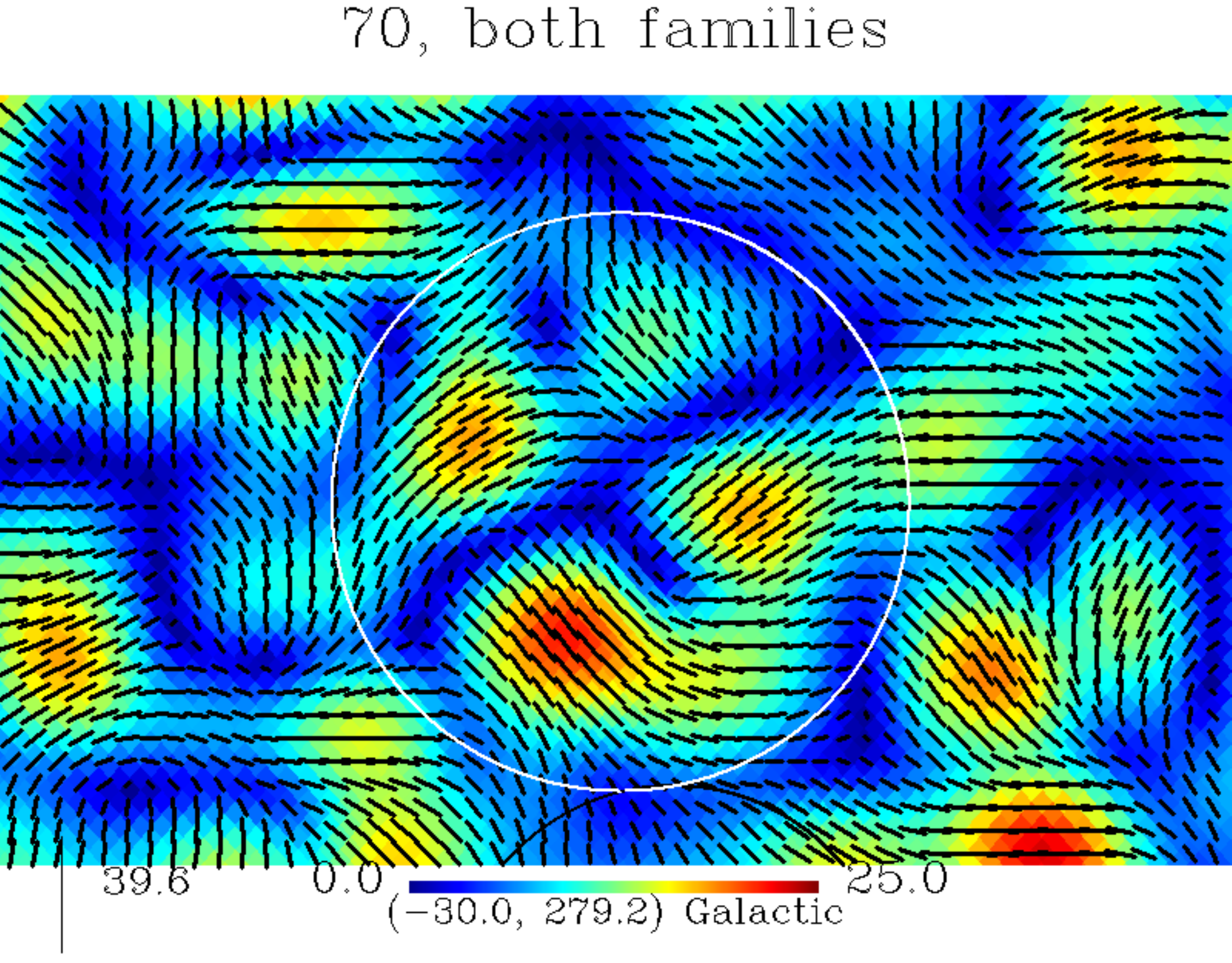}
  \includegraphics[width=0.32\textwidth]{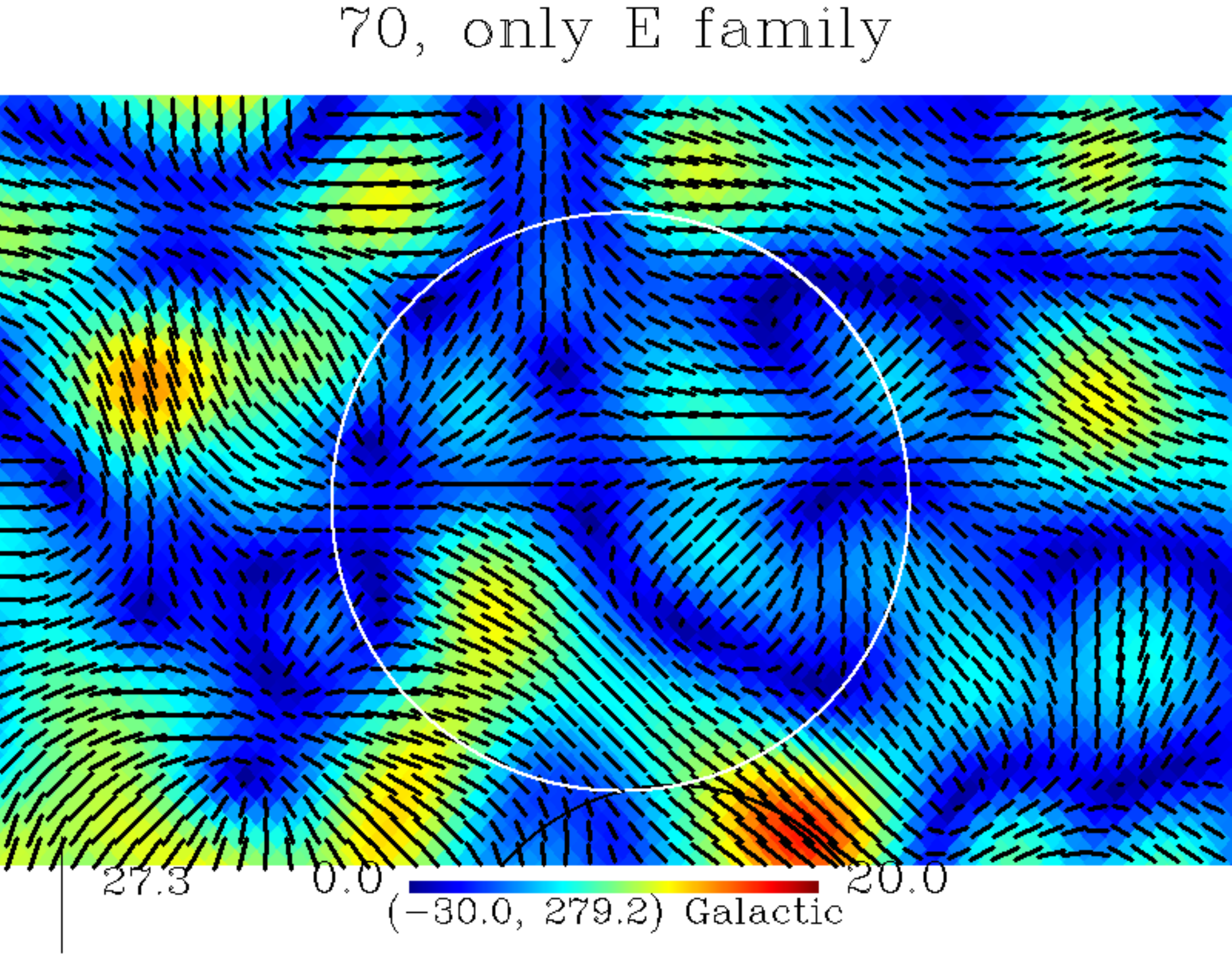}
  \includegraphics[width=0.32\textwidth]{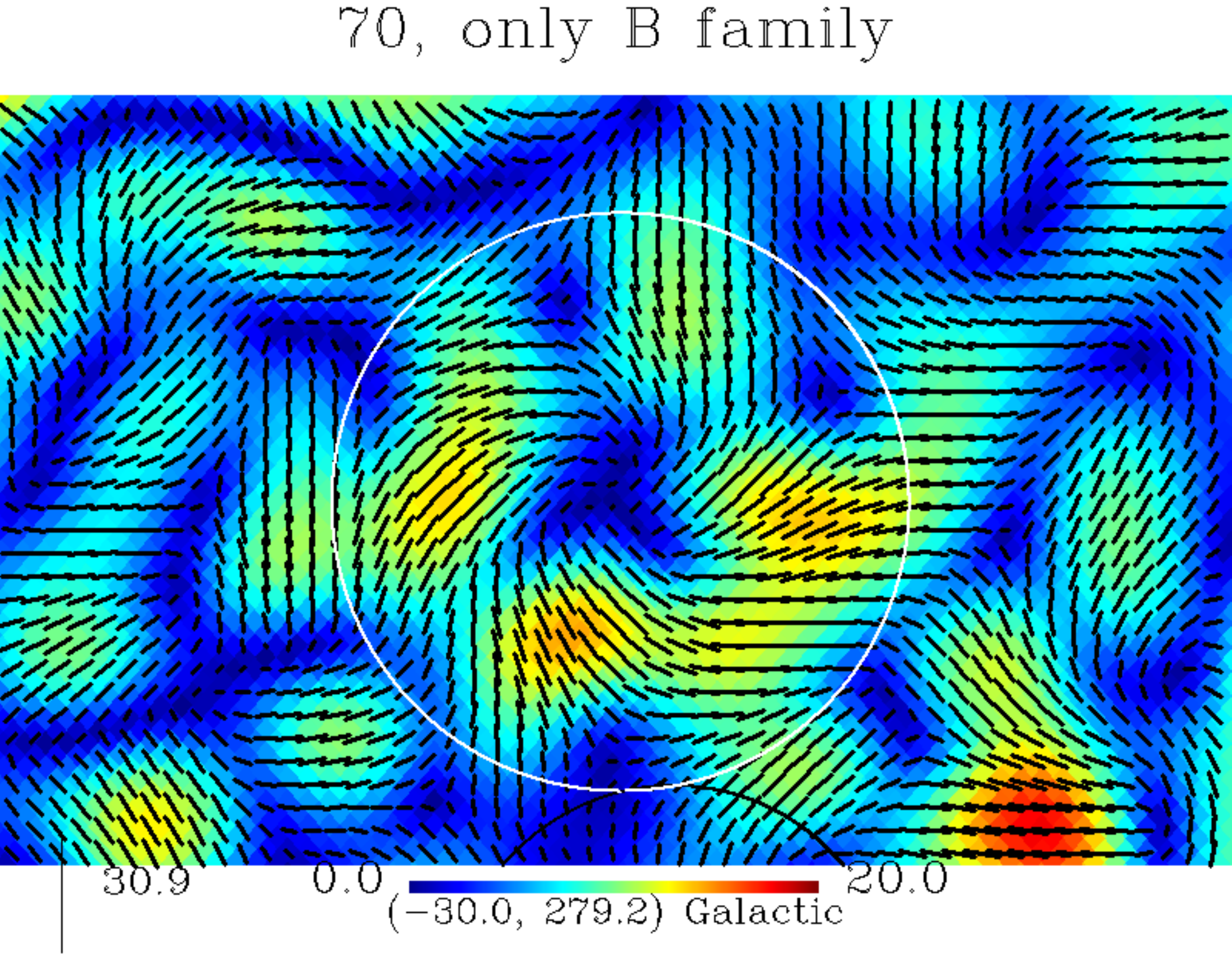}

  \includegraphics[width=0.32\textwidth]{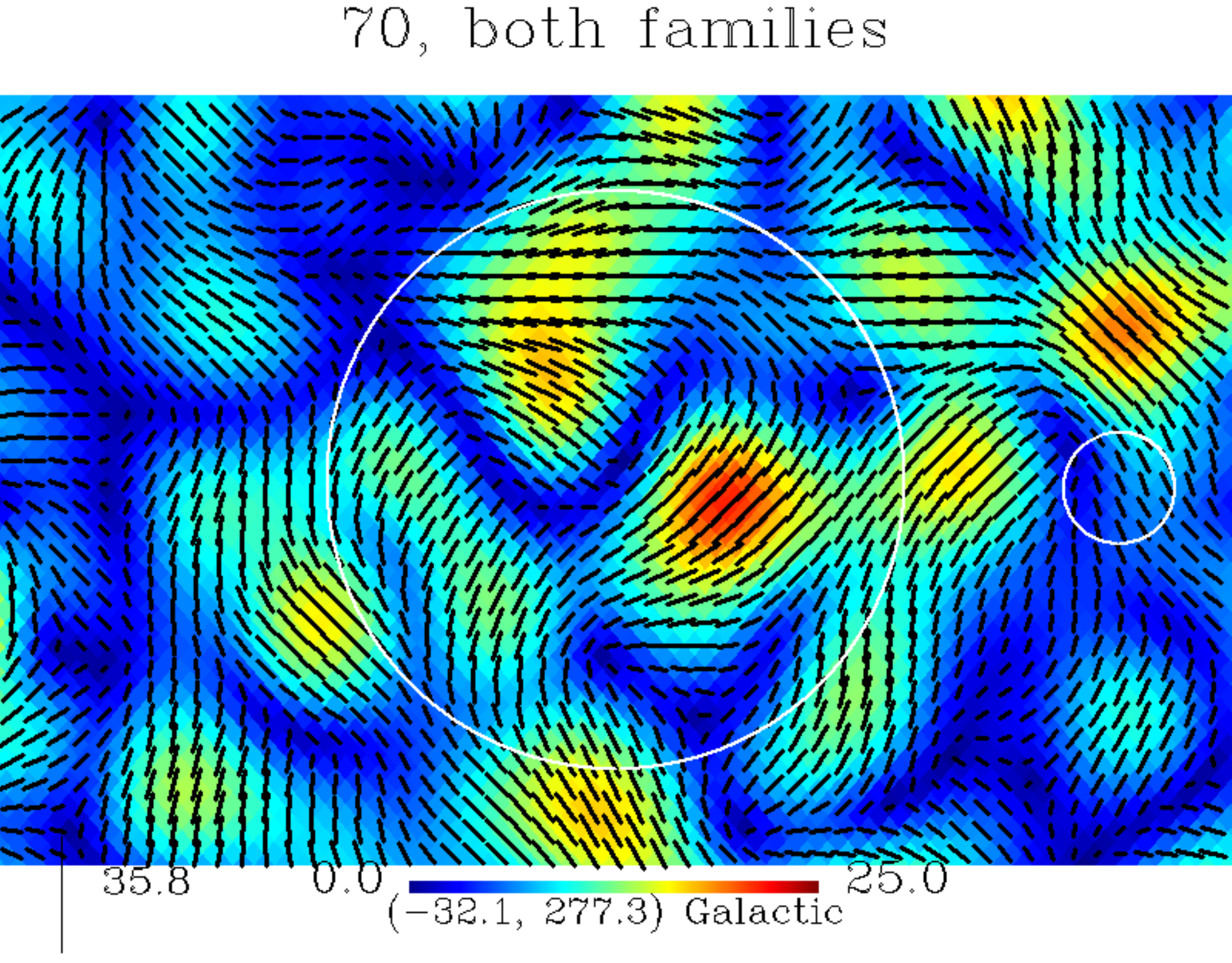}
  \includegraphics[width=0.32\textwidth]{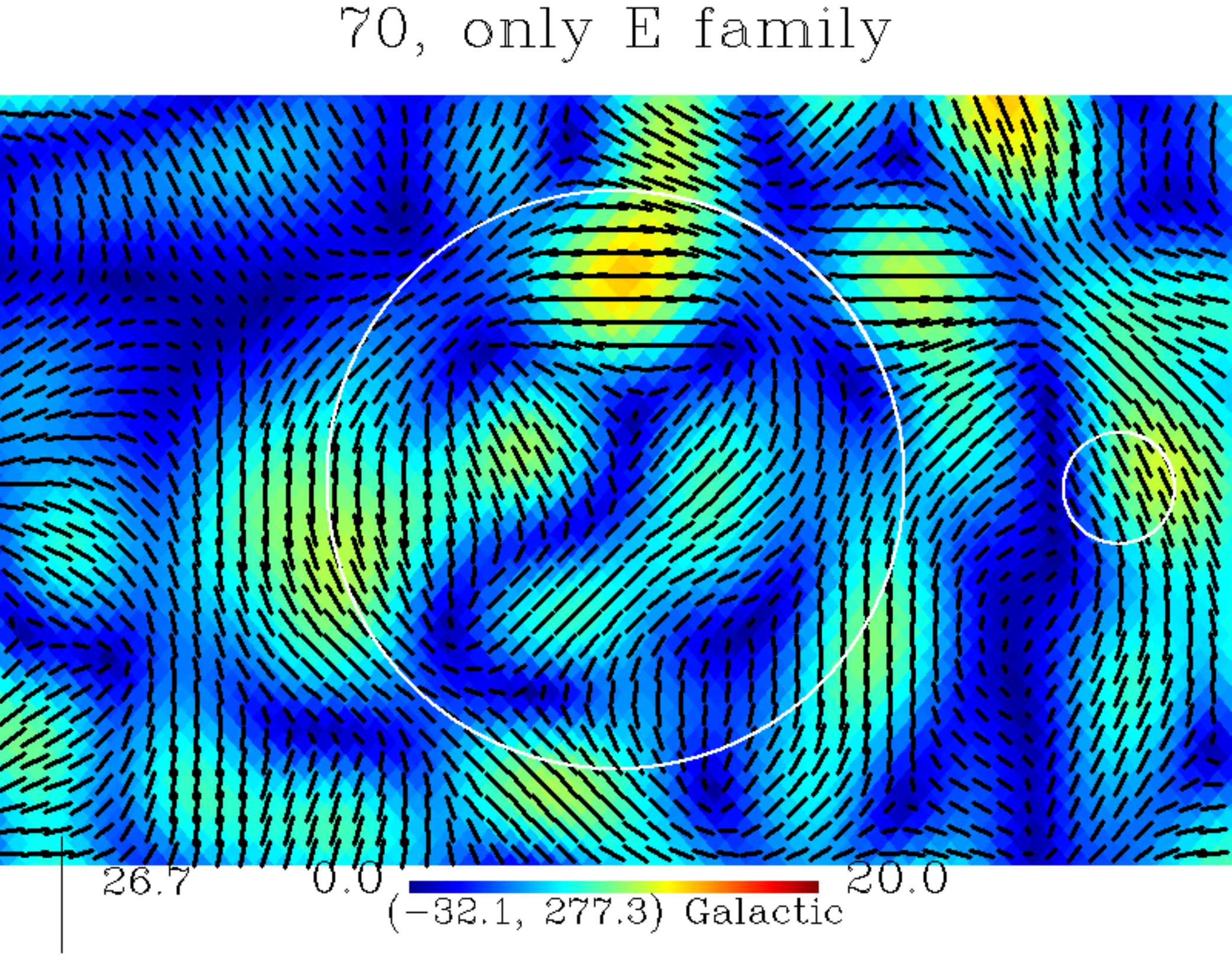}
  \includegraphics[width=0.32\textwidth]{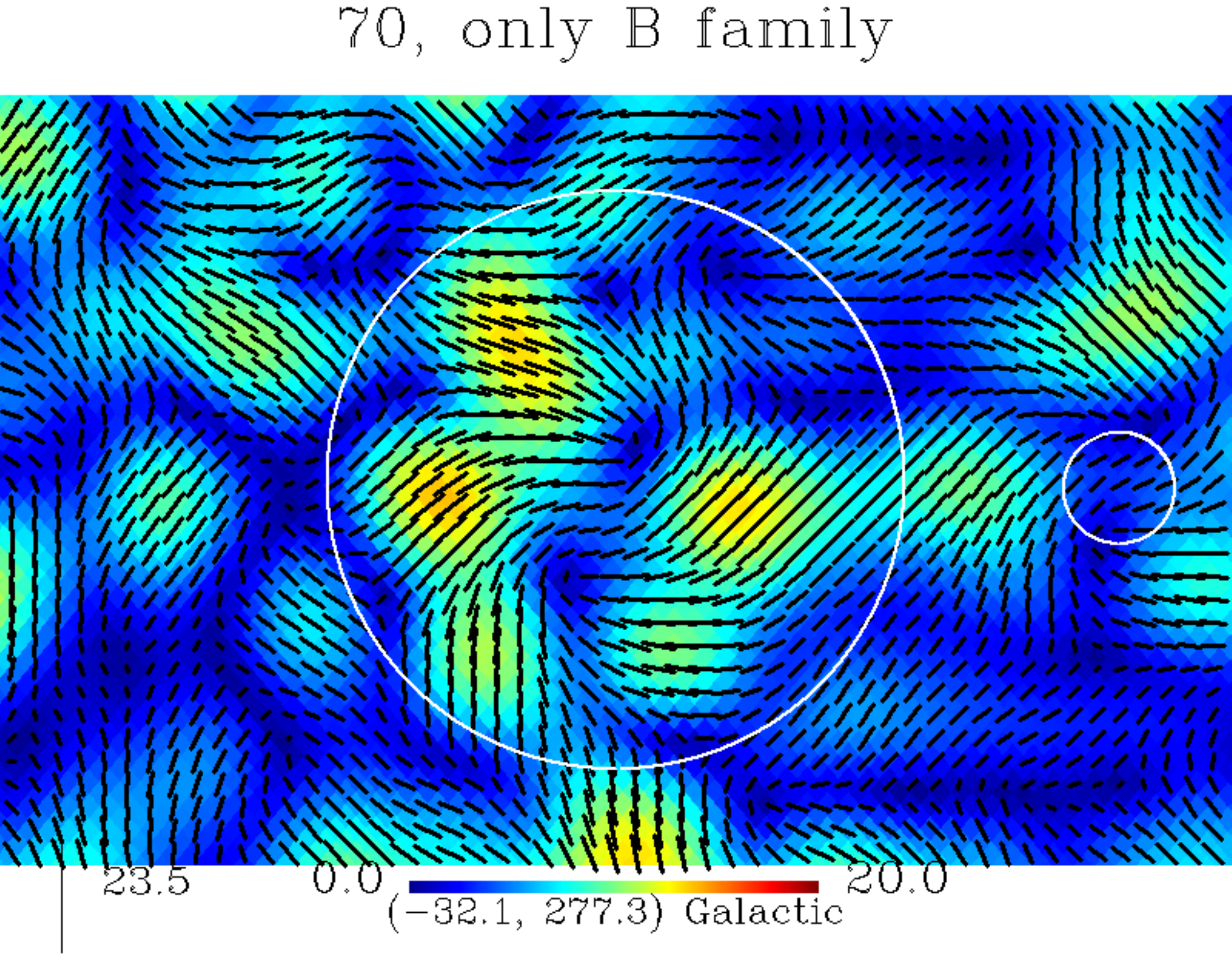}
  \caption{The Planck 70 GHz polarization map at $N_{\mathrm{side}}=1024$ and
  re-beamed from $13.25'$ to $16'$ FWHM to suppress the noise. The two positions
  in the LMC region are presented. One is around
  $(b,l)=(-30.0\degree,279.2\degree)$ (upper), and the other is around
  $(b,l)=(-32.1\degree,277.3\degree)$ (lower). From left to right: the original
  polarization and the E, B family. Two nearly perfect spiral structures
  belonging to the B-mode can be seen in B-family and partially in the original
  polarization, but are of course missing in the E-family. Also note that both
  of them are cold (less polarized) in the center. The diameter of the white
  circle is $0.5\degree$.}
  \label{fig:LMC example CD}
\end{figure*}

Next in figure~\ref{fig:LMC example CD}, we focus on the
$(b,l)=(-30.0\degree,279.2\degree)$ and $(b,l)=(-32.1\degree,277.3\degree)$
positions, which are both in the LMC region and close to the region of
figure~\ref{fig:LMC example B}. In this two regions we see two nearly perfect
spiral structure belonging to the B-family, which does not appear in 30, 44 or
100 GHz, and the amplitudes seem to be above the noise contribution. Therefore,
they are unlikely to be the noise, CMB or synchrotron emission. One possibility
is that they are due to the Anomalous microwave emission (AME). If this is true,
then there should be some hidden mechanism that is able to create nearly pure
B-mode polarization in AME.

\begin{figure*}[!htb]
  \centering
  \includegraphics[width=0.32\textwidth]{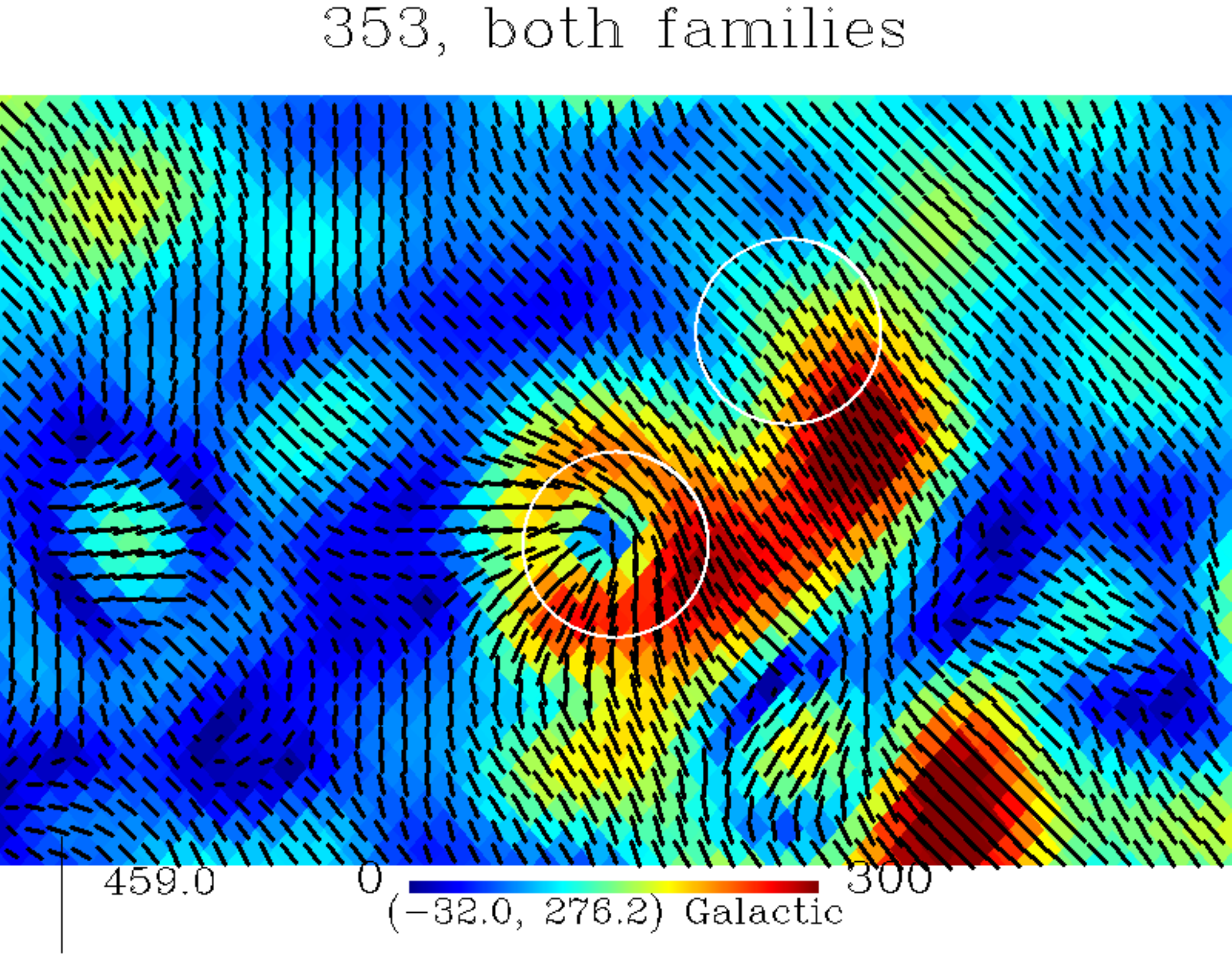}
  \includegraphics[width=0.32\textwidth]{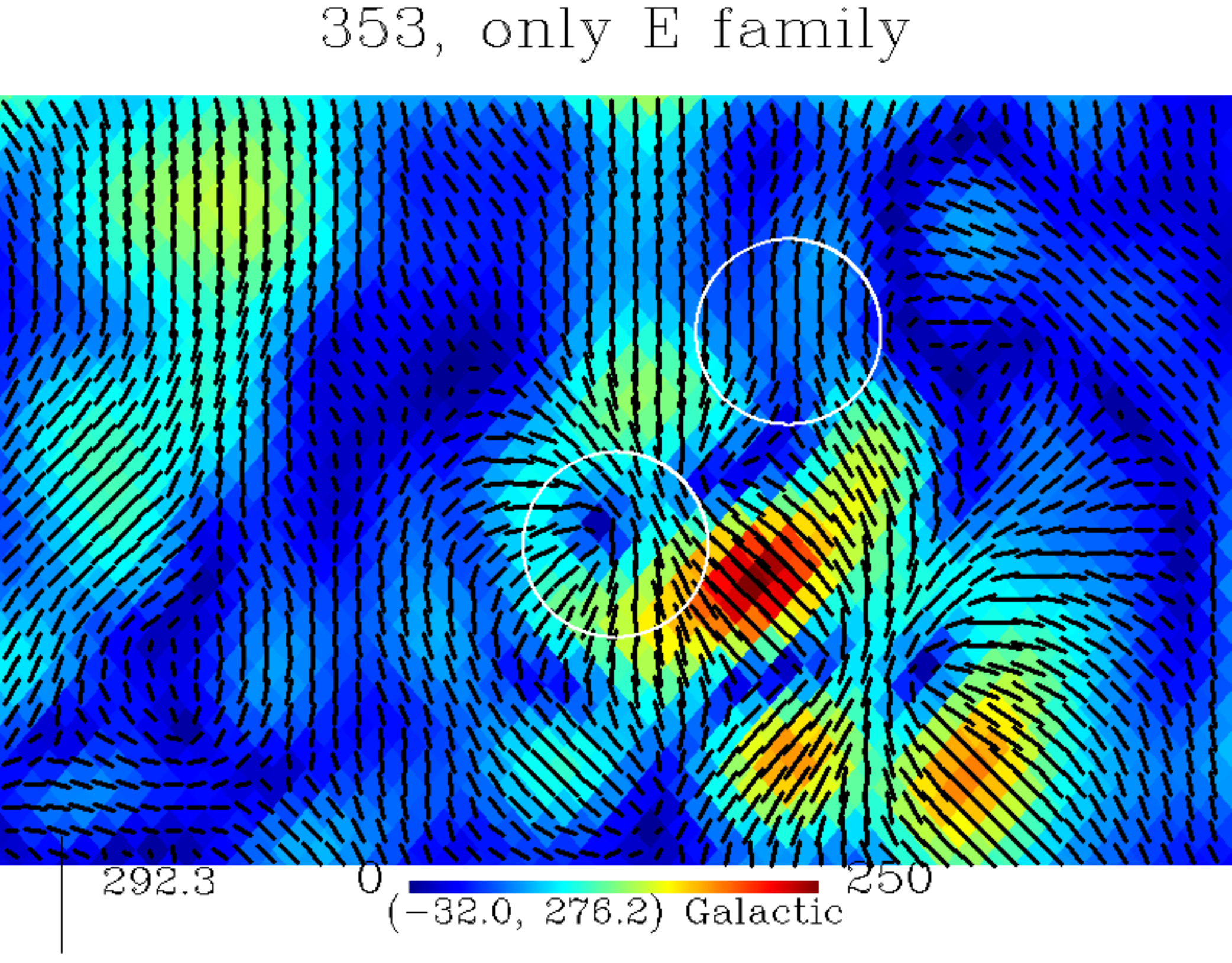}
  \includegraphics[width=0.32\textwidth]{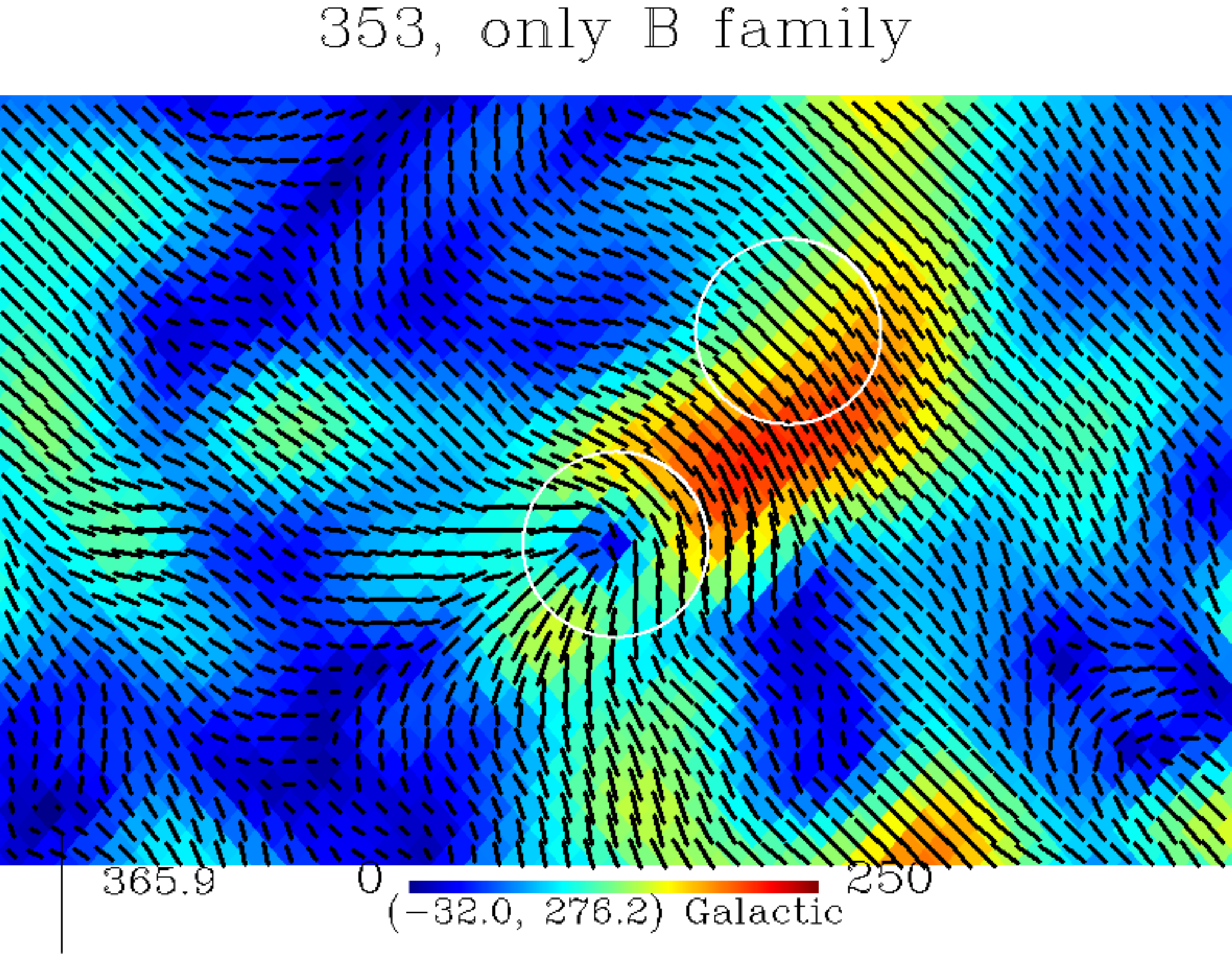}
  \caption{Similar to figure~\ref{fig:LMC example B} but for the Planck 353 GHz
  polarization map at $N_{\mathrm{side}}=2048$ without additional smoothing,
  showing position $(b,l) = (-32.07\degree, 276.24\degree)$ in the LMC region. A
  similar ejecting structure to figure~\ref{fig:LMC example B} can be seen in
  the B-family at a much higher frequency.}
  \label{fig:LMC example E}
\end{figure*}

Next in figure~\ref{fig:LMC example E}, we focus on the
$(b,l)=(-32.0\degree,276.2\degree)$ position in the LMC region, on the right of
the lower part in figure~\ref{fig:LMC example CD}. In this region we see another
ejecting structure that is very similar to figure~\ref{fig:LMC example B};
however, the input map is Planck 353 GHz rather than 30 GHz; thus it is almost
certainly due to the thermal dust emission. According to the similarity of the
structures, a reasonable guess is that the structure in figure~\ref{fig:LMC
example B} is probably also associated to the dust emission. However, dust
polarization in 30 GHz can only be AME (spinning or magnetic dust emissions).
This again requires a mechanism that can create B-family polarization in the
AME.

Then we focus on the ejecting structure in figure~\ref{fig:LMC example B} and
compare it for eight different bands from WMAP and Planck, including the WMAP
K-band (22.8 GHz) and the Planck 30, 44 70, 100, 143, 217 and 353 GHz bands. The
K to 70 GHz bands are re-beamed to $1\degree$ FWHM, and the higher bands are
re-beamed to $30'$. The results are shown in figure~\ref{fig:LMC example B all
bands}. From this figure, we can see that the structure is visible in all
frequency bands from 22 to 353 GHz, no matter WMAP or Planck, which safely
excludes the possibility of CMB, noise or systematics. The amplitude variation
from K-band to 30 GHz is consistent with the synchrotron emission spectrum
($\beta\sim -3$), and the amplitude variation from 217 to 353 GHz is also
consistent with a thermal dust spectrum of $\beta\sim1.6$ and $T_{dust}\sim 16$
K. These facts tell us that this structure includes significant synchrotron and
thermal dust polarization at the same time. Thus, we have the following
conclusions: 1) The B-family nature is most likely coming from a B-type magnetic
field distribution rather than a specific emission mechanism. 2) There
should be a physical mechanism that can create such a B-type magnetic
field distribution, and 3) the possibility of spin or magnetic dust
emission cannot be completely excluded.
\begin{figure*}[!htb]
  \centering
  \includegraphics[width=0.24\textwidth]{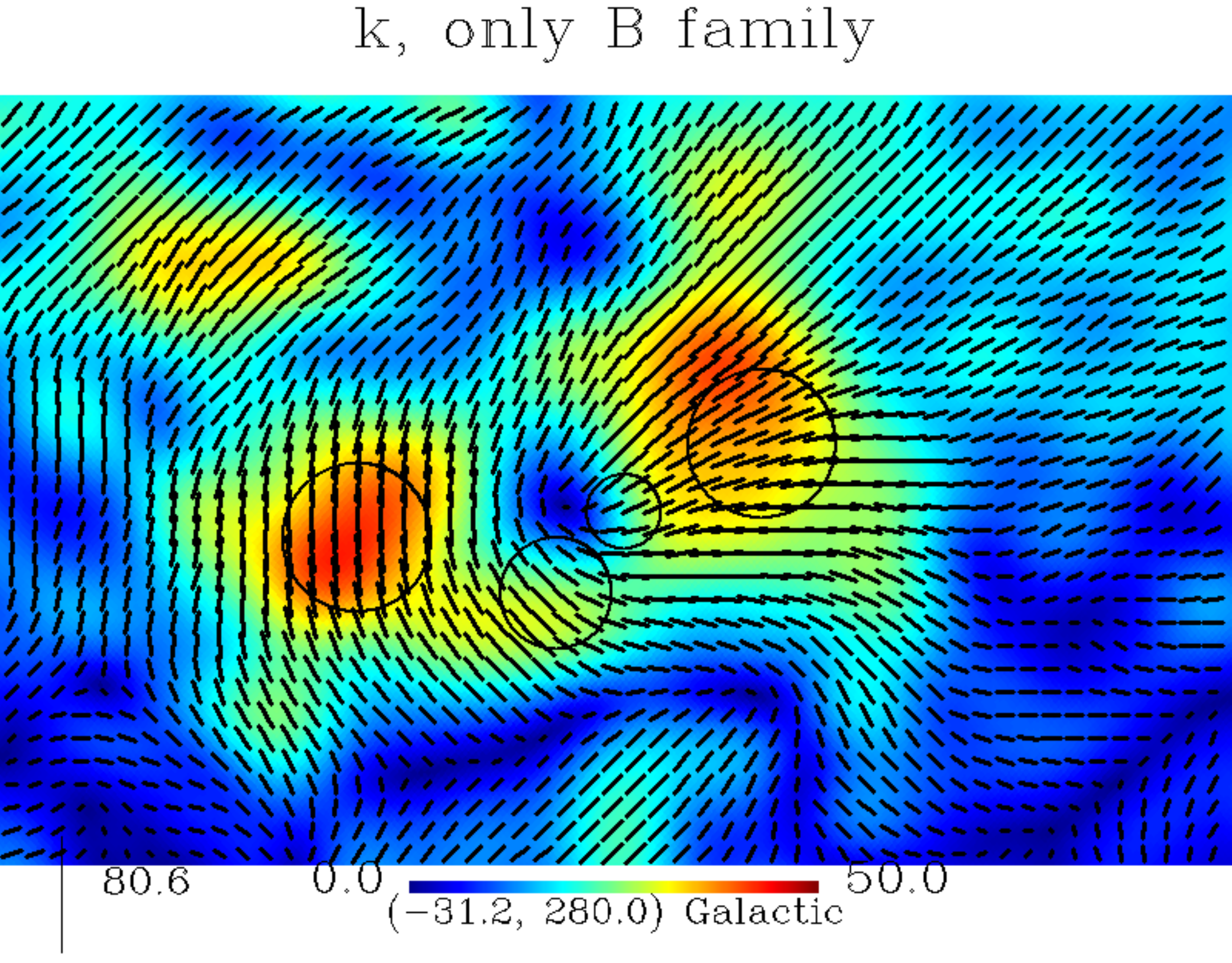}
  \includegraphics[width=0.24\textwidth]{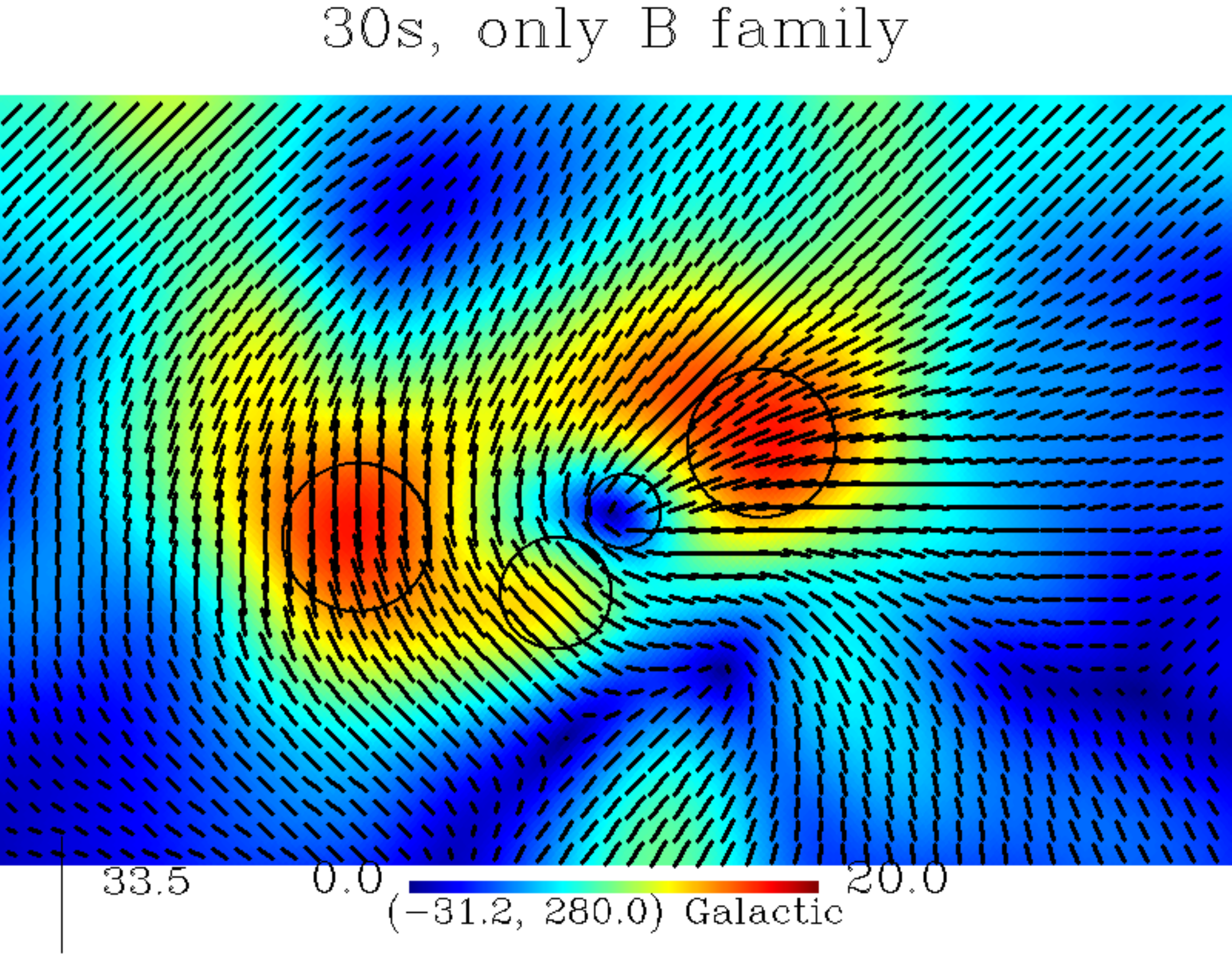}
  \includegraphics[width=0.24\textwidth]{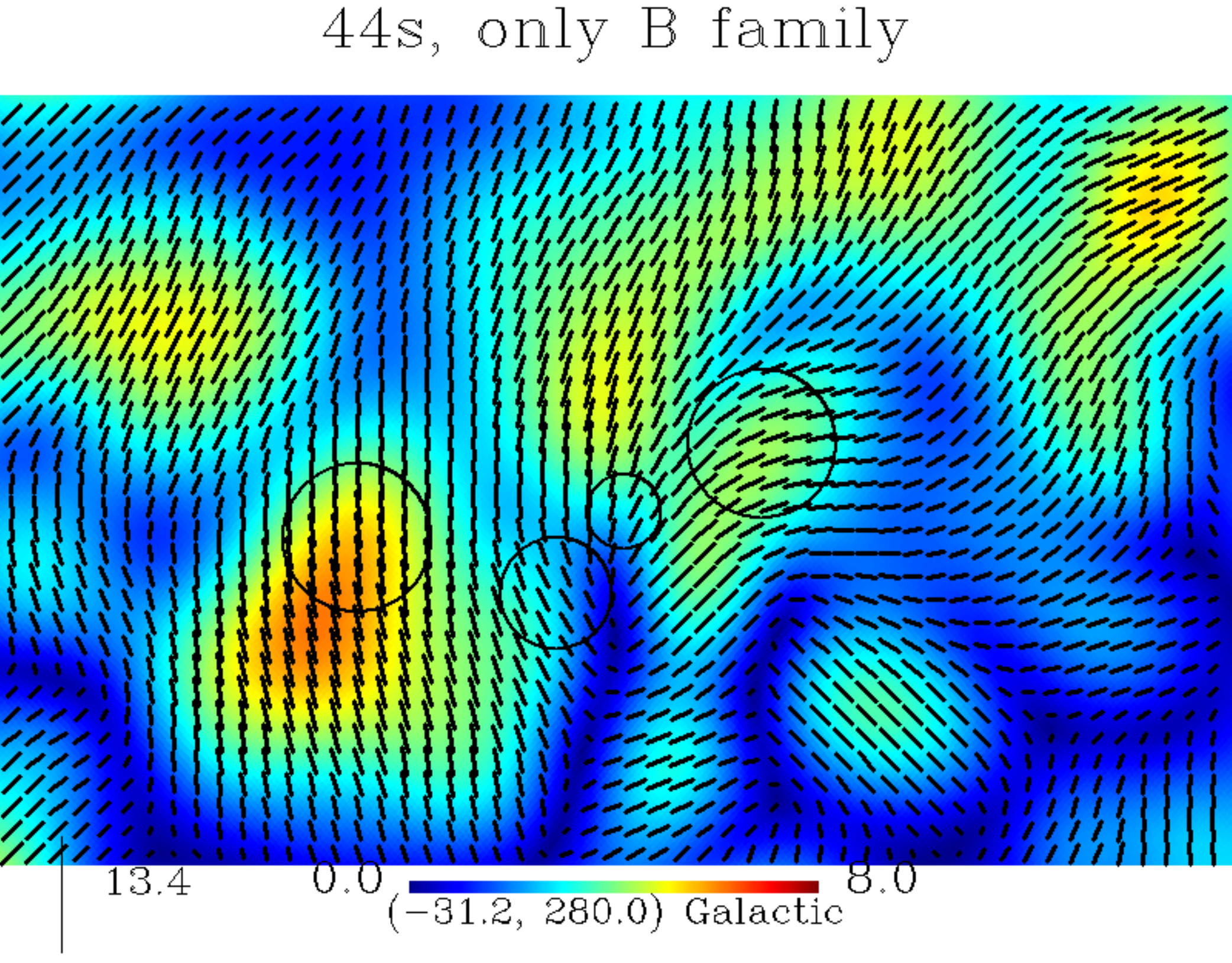}
  \includegraphics[width=0.24\textwidth]{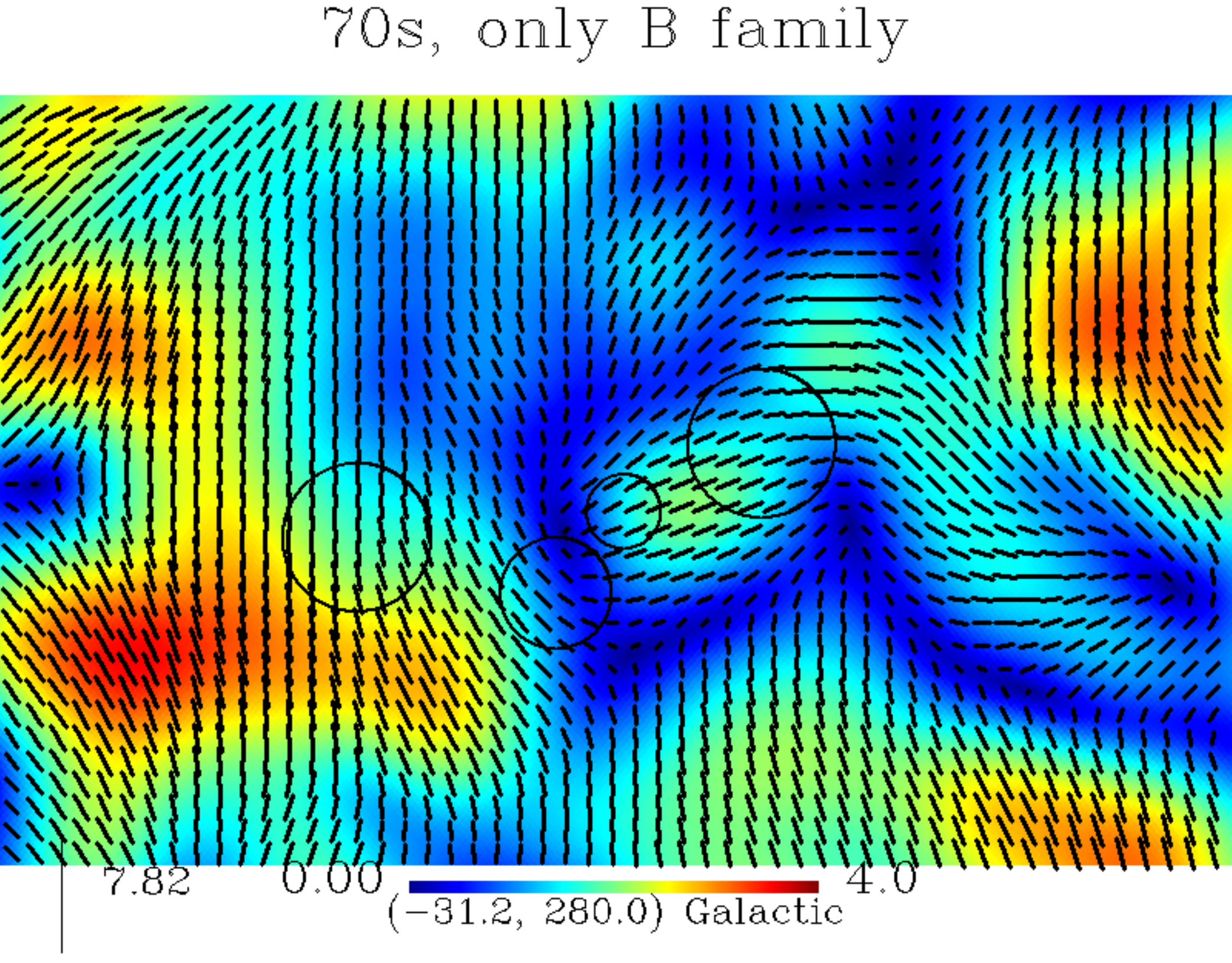}

  \includegraphics[width=0.24\textwidth]{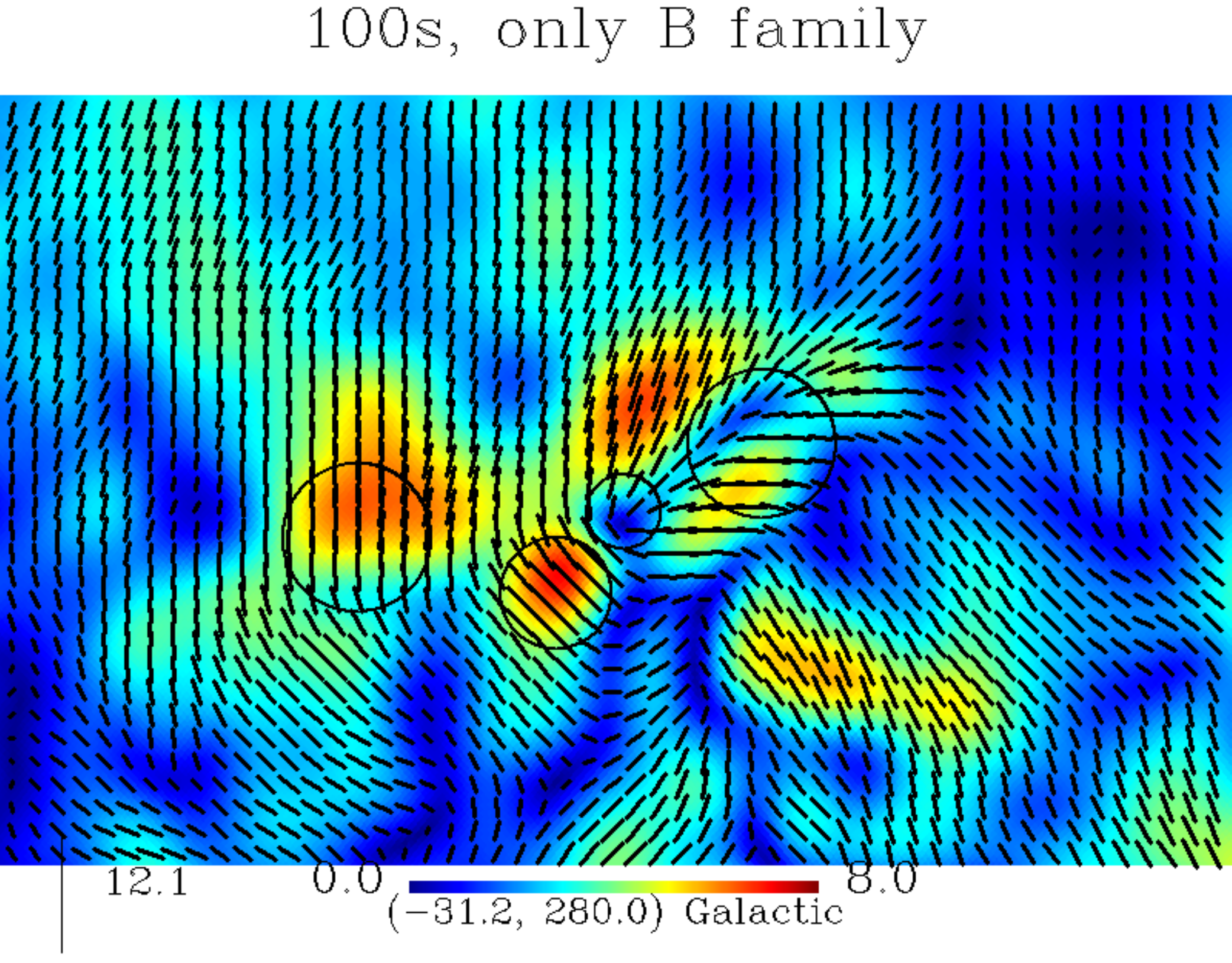}
  \includegraphics[width=0.24\textwidth]{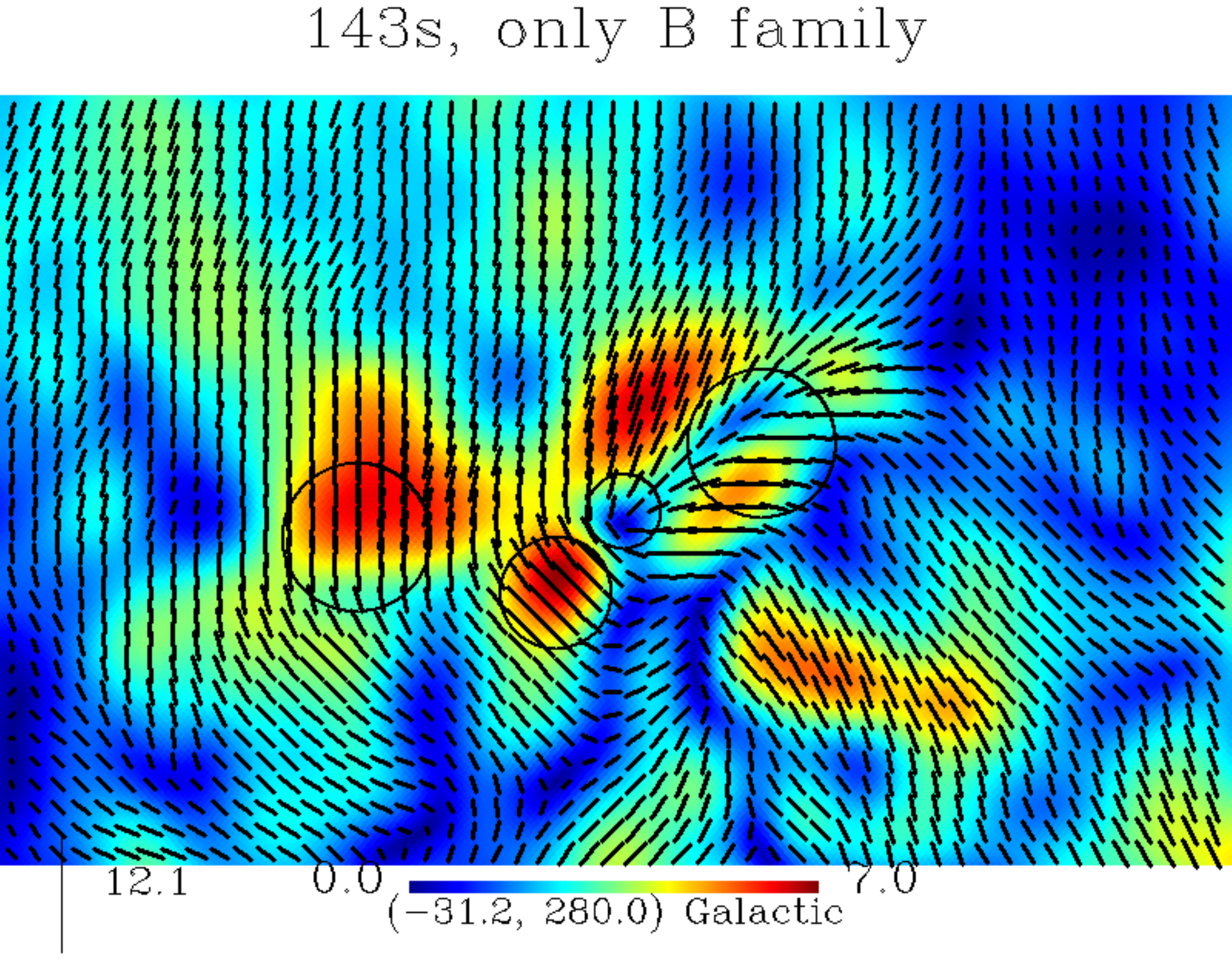}
  \includegraphics[width=0.24\textwidth]{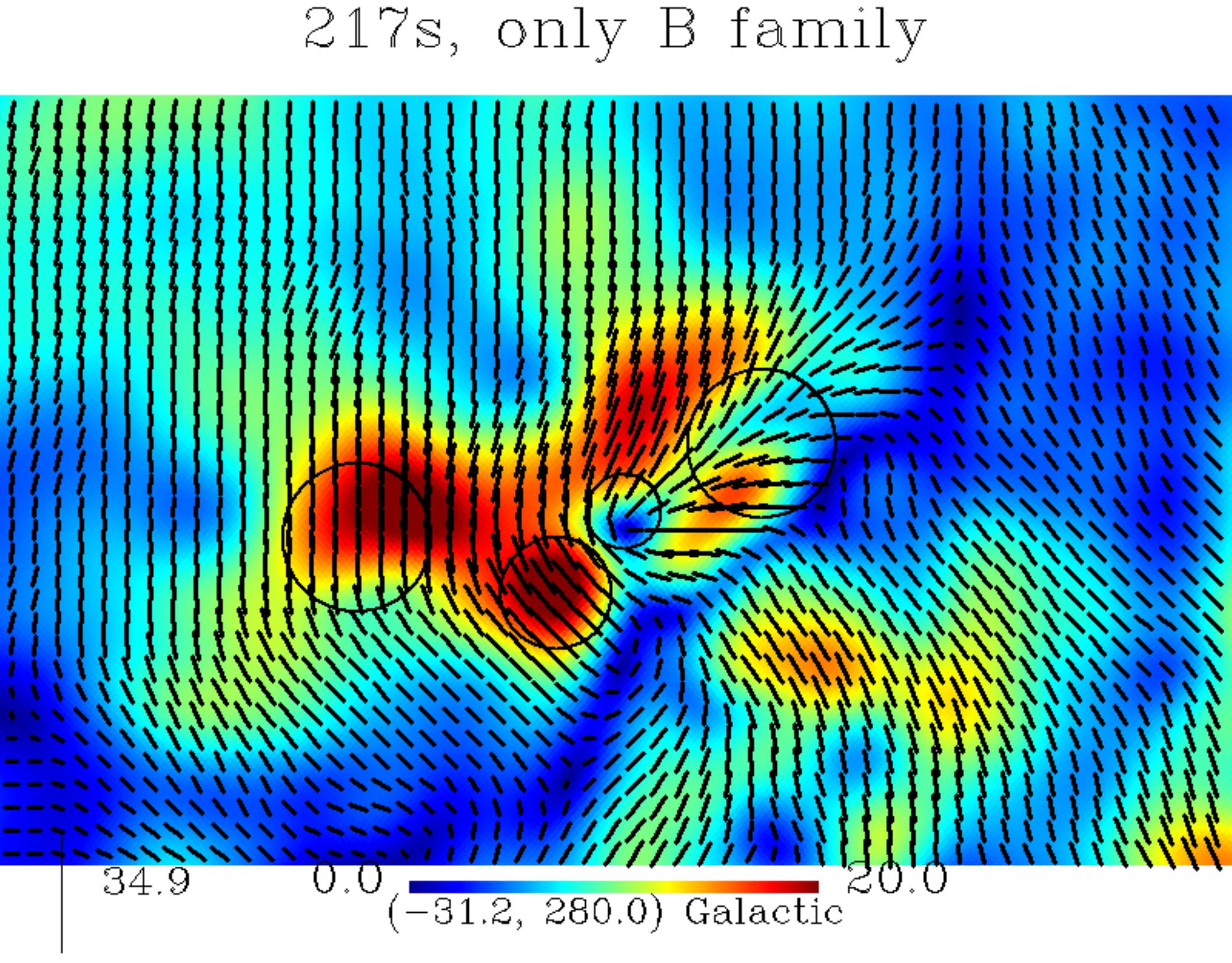}
  \includegraphics[width=0.24\textwidth]{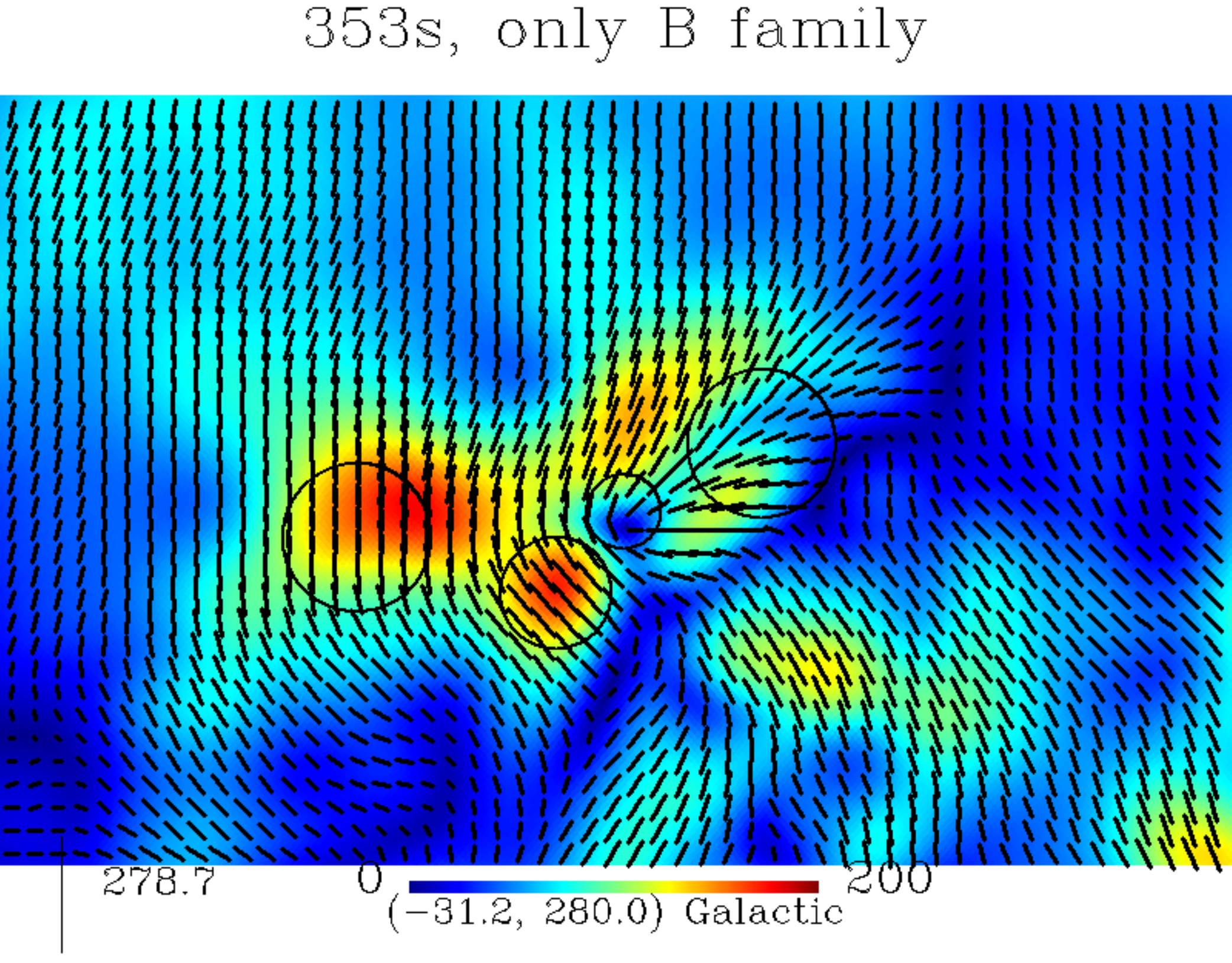}
  \caption{Similar to figure~\ref{fig:LMC example B} but focuses on the B-family
  ejecting structure in eight WMAP and Planck bands: \emph{Upper}, from
  left-to-right: the WMAP K-band (22.8 GHz) and Planck 30, 44 70 bands;
  \emph{Lower}, from left-to-right: the Planck 100, 143, 217 and 353 GHz bands.
  }
  \label{fig:LMC example B all bands}
\end{figure*}

\begin{figure*}[!htb]
  \centering
  \includegraphics[width=0.32\textwidth]{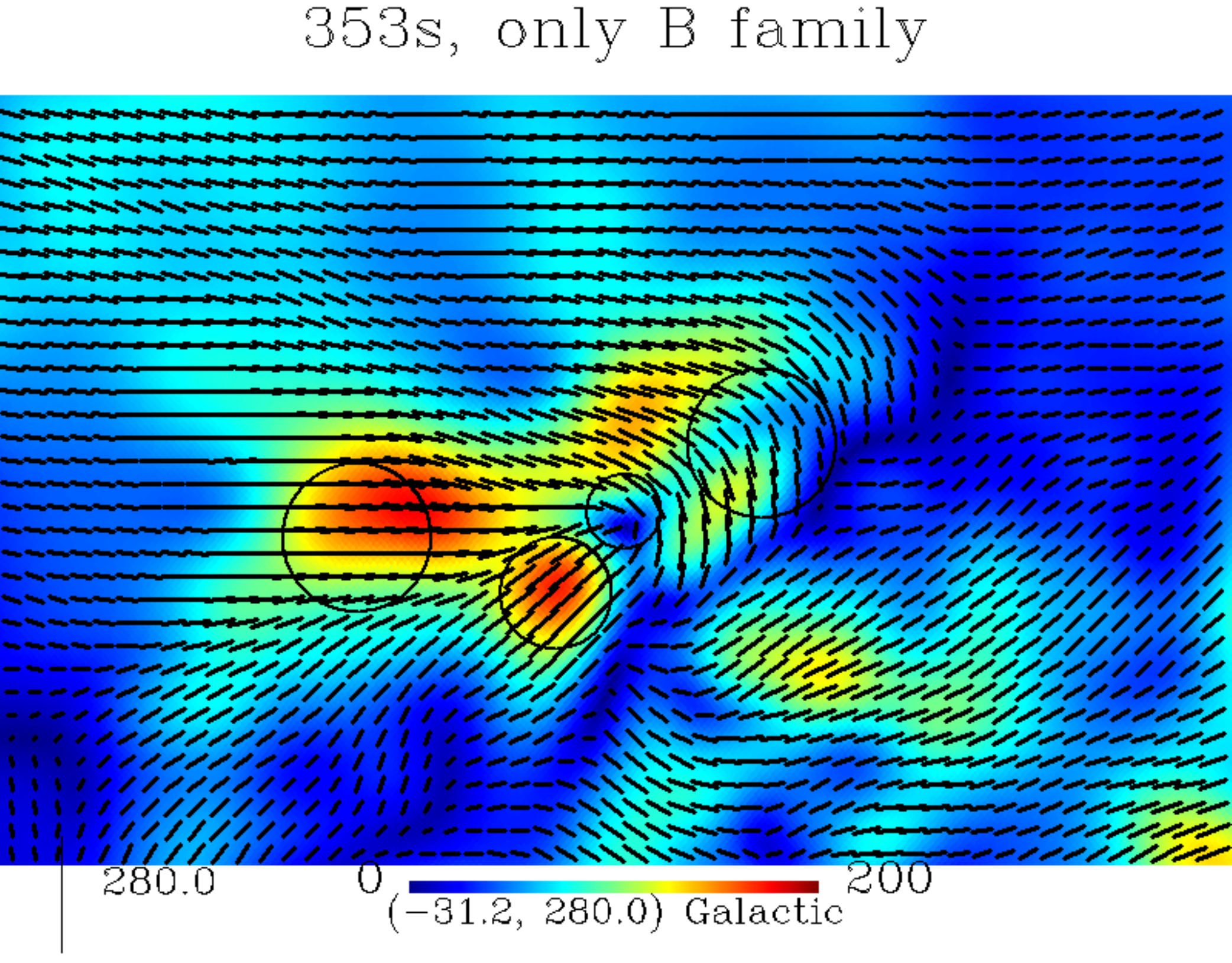}
  \includegraphics[width=0.32\textwidth]{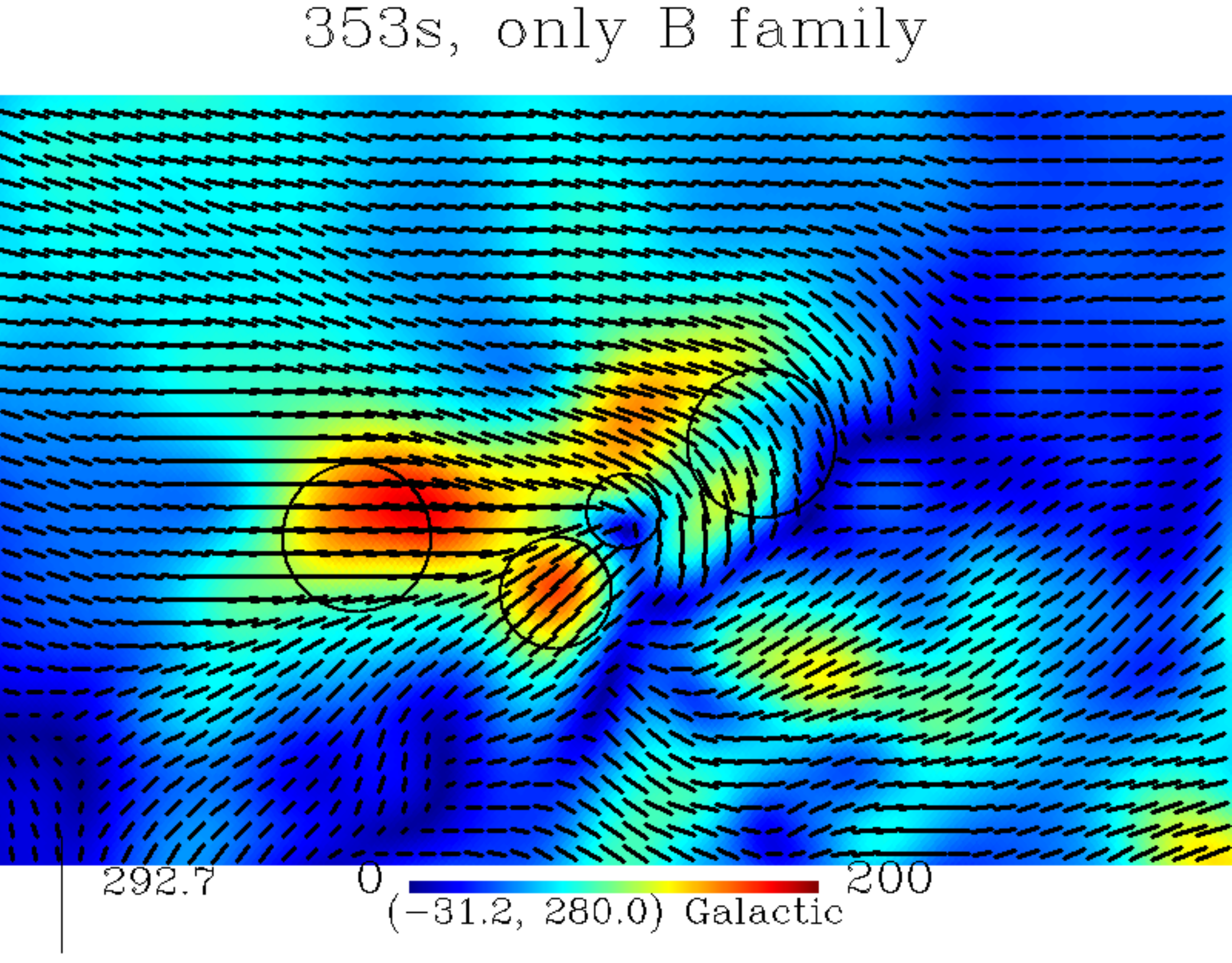}
  \includegraphics[width=0.32\textwidth]{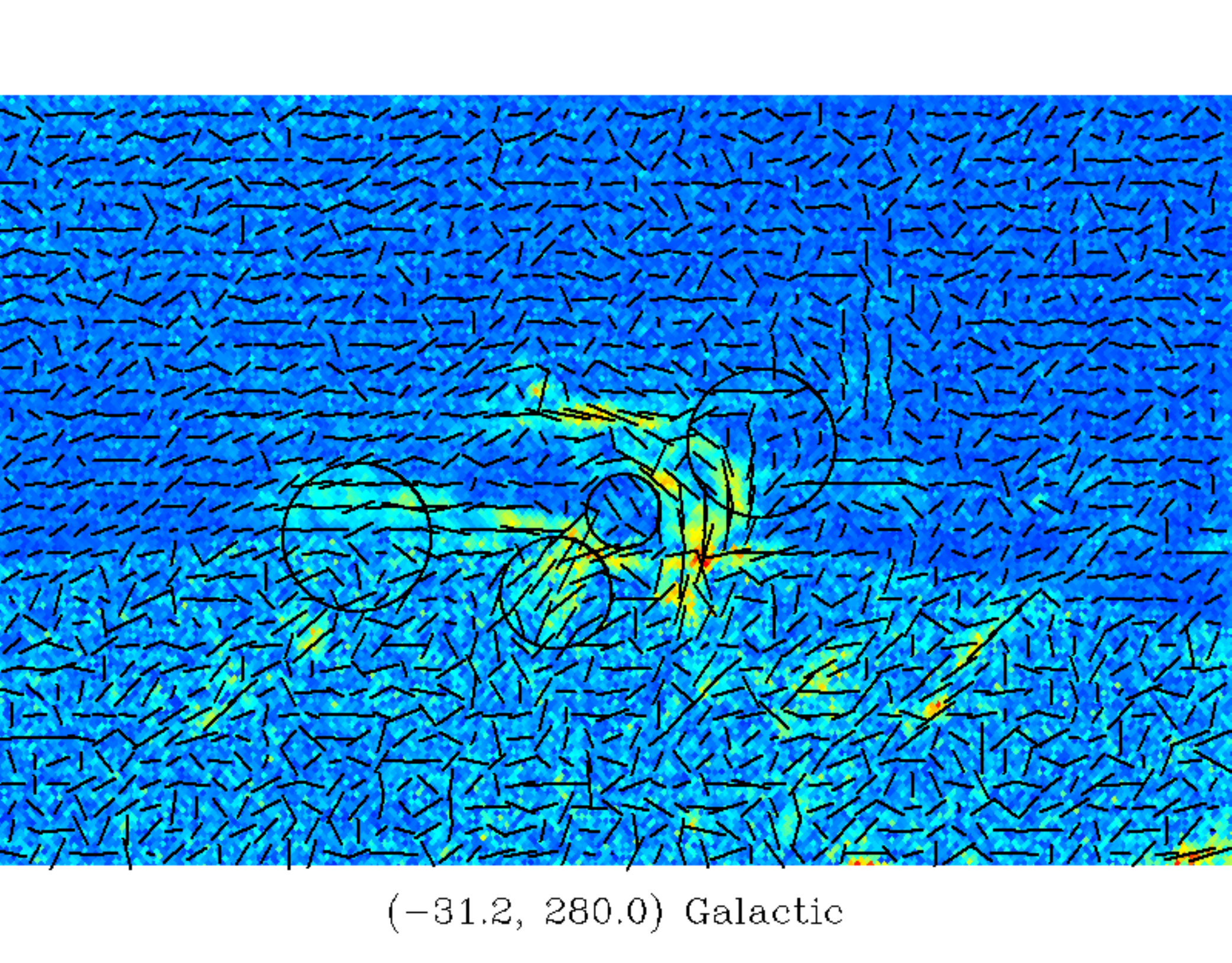}
  \caption{Test of figure~\ref{fig:LMC example B all bands} for: 1) the point
  source effect, 2) self-consistency and 3) rotation of polarization by
  $90\degree$. \emph{Left}: Similar to the 353 GHz results in
  figure~\ref{fig:LMC example B all bands} but rotate all polarizations by
  $90\degree$. \emph{Middle}: Similar to the left panel but exclude 3\% of the
  local region with strongest polarization intensities before any other
  operations, which shows no significant changes; thus the polarization patterns
  in figure~\ref{fig:LMC example B all bands} are nearly unaffected by point
  sources. \emph{Right}: The original 353 GHz polarization map (also rotate the
  polarizations by $90\degree$), no smoothing, no EB-separation. In this panel,
  we can see not only similar polarization pattern, but also alignment of the
  relatively hotter regions with the left panel. Which validates the B-family
  polarization patterns in figure~\ref{fig:LMC example B all bands} and strongly
  suggests there can be a foreground emission mechanism that prefers the B-mode.
  }
  \label{fig:test of LMC example B}
\end{figure*}

Immediately after figure~\ref{fig:LMC example B all bands}, we perform three
tests of the results: 1) Whether or not the results are affected by strong point
sources. 2) Can we see similar structures in the original 353 GHz map (no
smoothing, no EB-separation). 3) What is the structure look like when we rotate
all polarizations by $90\degree$. These test results are shown in
figure~\ref{fig:test of LMC example B}. This figures tells us that: 1) The
results in figures~\ref{fig:LMC example B all bands}--\ref{fig:test of LMC
example B} are unaffected by point sources. 2) Similar polarization pattern can
be seen even on the original polarization map (no smoothing, no EB-separation).
3) When we rotate all polarizations by $90\degree$, it is still a ejecting
structure but the direction is inverted. 4) The hot regions in B-family is
apparently aligned with the hot regions in the original polarization. In all,
the tests not only validates the results in figure~\ref{fig:LMC example B all
bands}, but also prefers the explanation that there is a foreground emission
mechanism that prefers the B-mode.

We also perform a test of the E-major ($P_E>P_B$) and B-major ($P_E<P_B$) pixels
by checking the fraction of such pixels in the regions of figures~\ref{fig:LMC
example B} and \ref{fig:LMC example E}; and on the original resolution 353 GHz
map. The fraction of E-major pixel is 43\% for the region in figure~\ref{fig:LMC
example B} and 33\% for the region in figure~\ref{fig:LMC example E}, so both
regions are dominated by the B-mode signals. The significance of such fractions
are also tested by checking the same fraction in other regions of the same shape
and size but centered at the $N_{side}=8$ pixels and located above Gal-latitudes
$|b|>25\degree$. Both tests show confidence levels of about $1-a=97\%$ that the
B-mode dominance is significant, as illustrated by figure~\ref{fig:EB major test
of LMC B and E}, which gives a joint confidence levels of about $1-a=99.9\%$.
\begin{figure*}[!htb]
  \centering
  \includegraphics[width=0.48\textwidth]{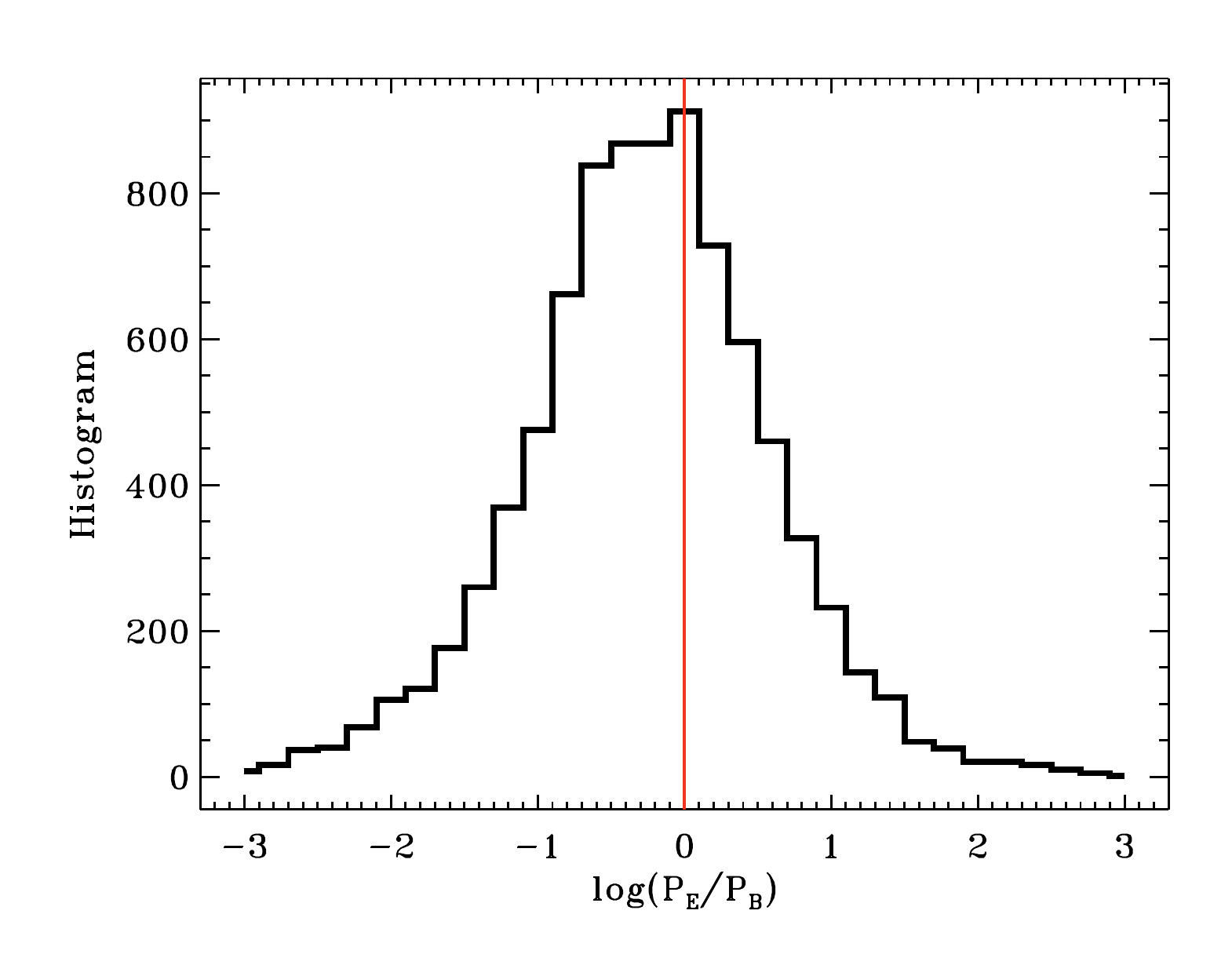}
  \includegraphics[width=0.48\textwidth]{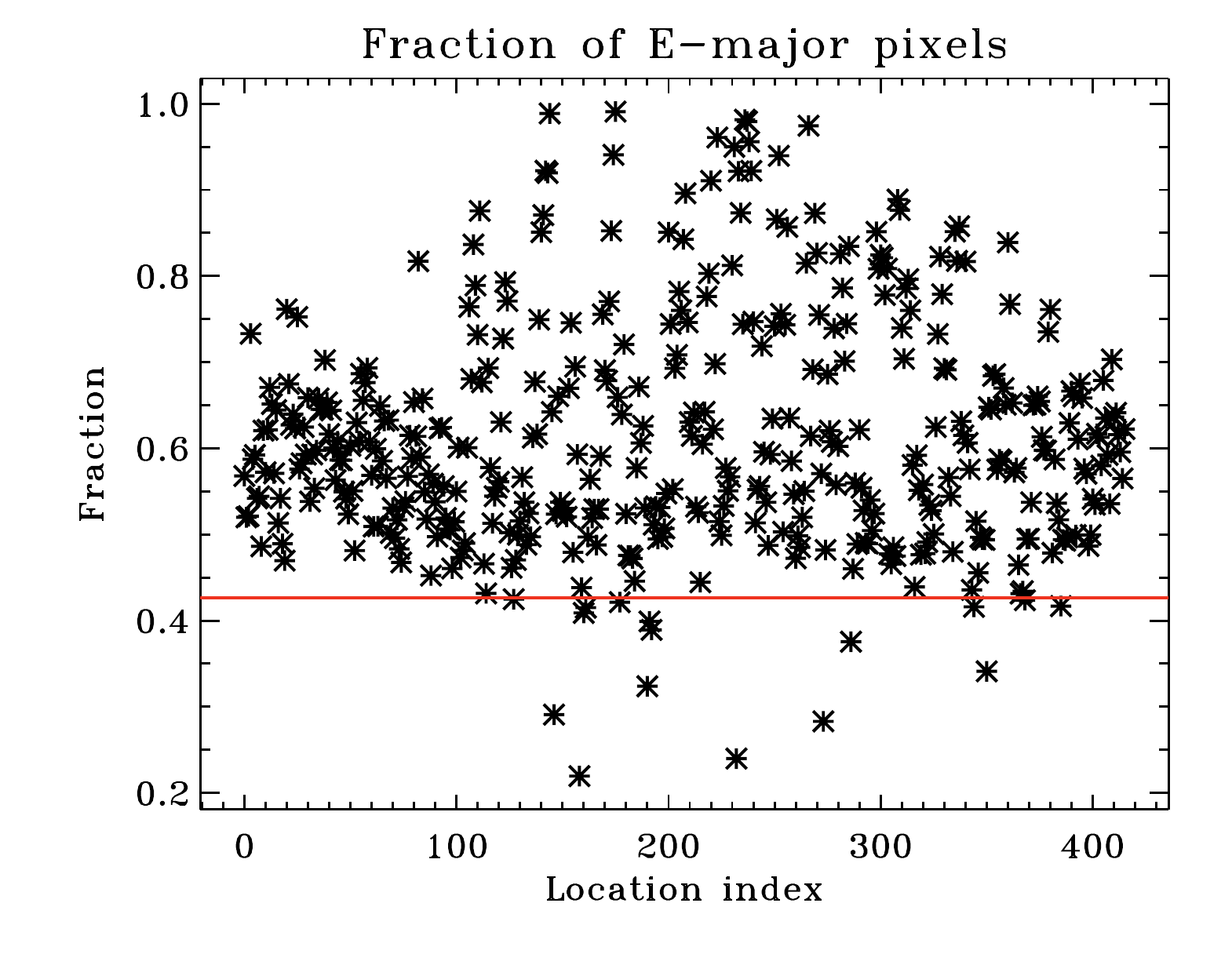}

  \includegraphics[width=0.48\textwidth]{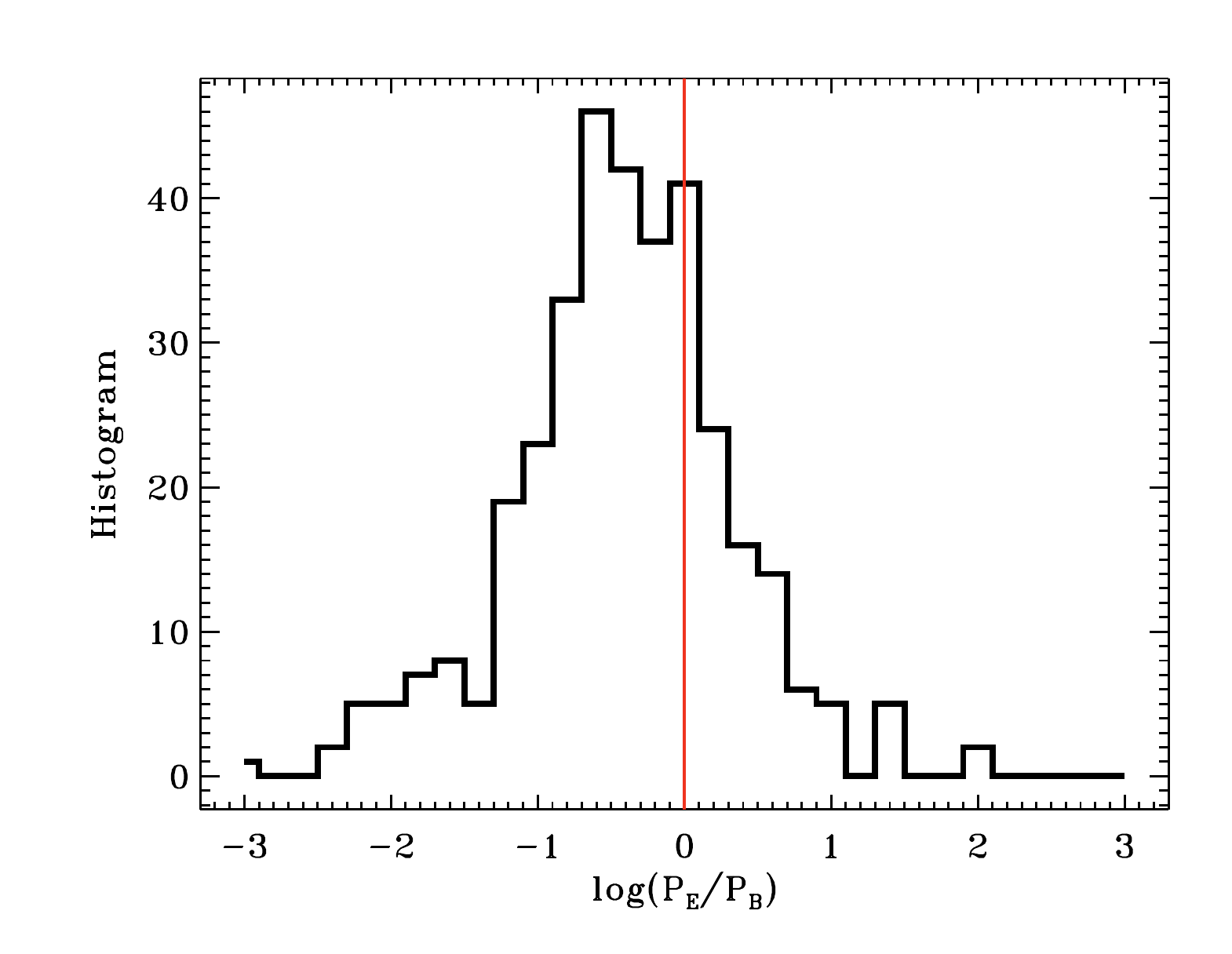}
  \includegraphics[width=0.48\textwidth]{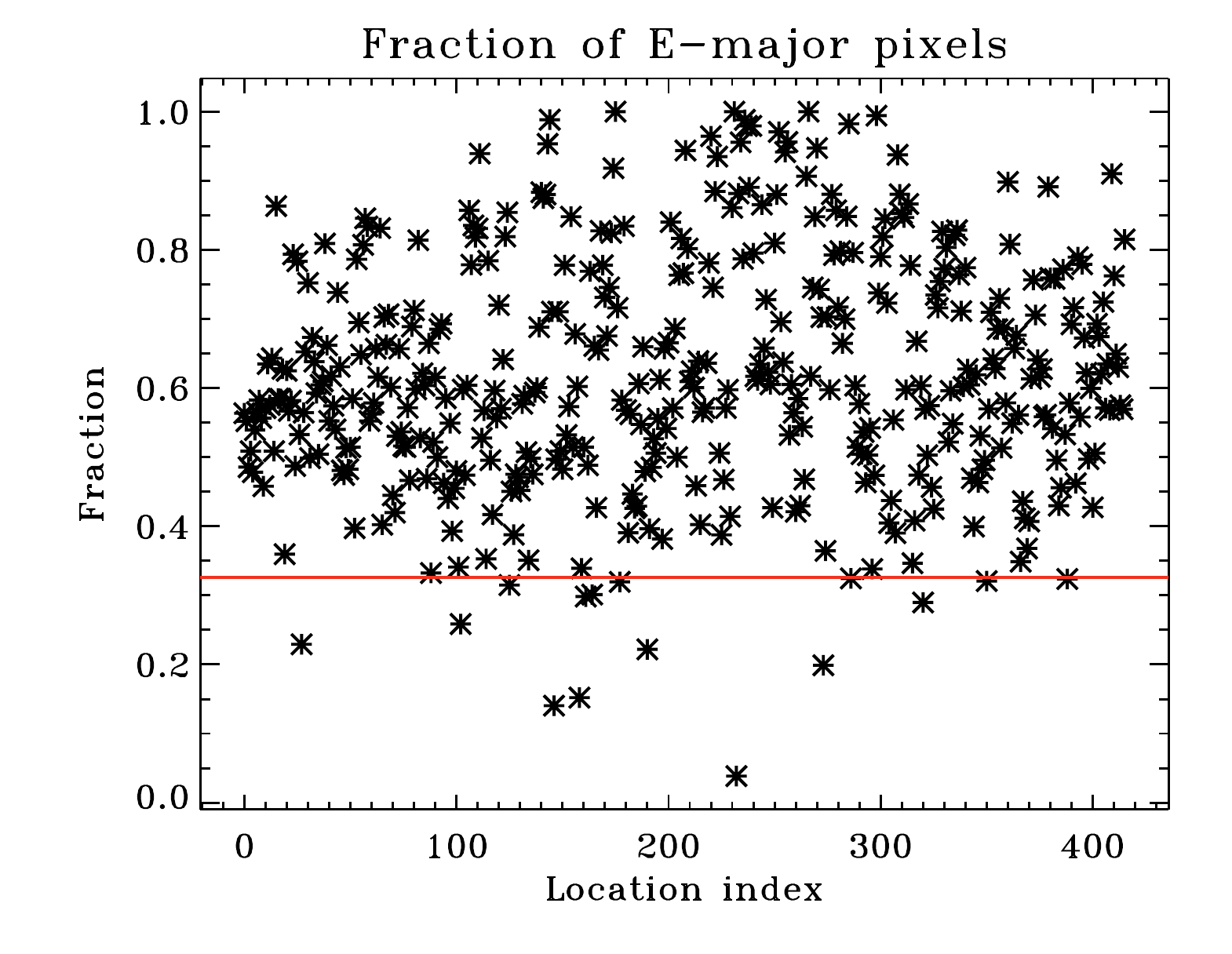}
  \caption{Test of the E- or B-major pixels in the regions of
  figure~\ref{fig:LMC example B} (upper) and figure~\ref{fig:LMC example E}
  (lower). \emph{Left}: The histogram of $\log(P_E/P_B)$ in the corresponding
  regions. \emph{Right}: The fraction of E-major pixel in the region (the red
  line) compared with other regions of the same size and on the same map,
  located above Gal-latitudes $|b|>25\degree$ and centered at the $N_{side}=8$
  pixels. }
  \label{fig:EB major test of LMC B and E}
\end{figure*}

\begin{figure*}[!htb]
  \centering
  \includegraphics[width=0.45\textwidth]{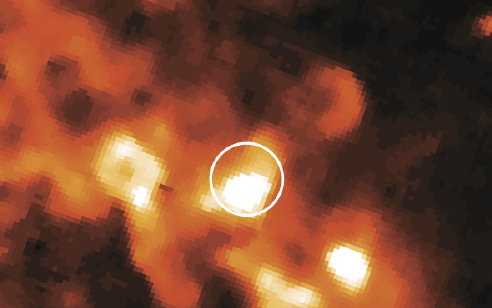}
  \includegraphics[width=0.45\textwidth]{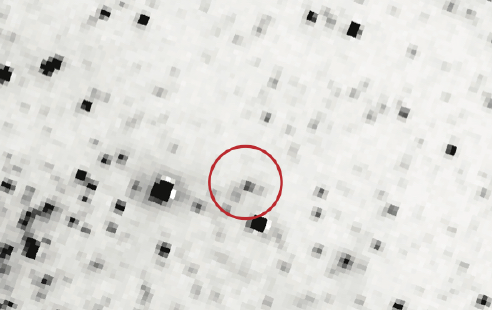}
  \caption{The foreground H$\alpha$ line emission (\emph{left}) and stellar
  continuum emission (\emph{right}) images of the northern region of LMC in
  Galactic coordinate system with the galactic north pole to the top of the
  figures. Both images cover the same area as that in Figure \ref{fig:LMC
  example E}. The circle with a 0.1$^{\circ}$ radius in each plot with marks the
  same corresponding location of the sky. The circle coincides with the emission
  region LH 120-N 64 \cite{2002AJ....123.2754W}. The data presented here are
  obtained from The Southern H-Alpha Sky Survey Atlas
  \citep{2001PASP..113.1326G}.}
  \label{fig:foreground_HII_stars_LMC_I}
\end{figure*}

\begin{figure*}[!htb]
  \centering
  \includegraphics[width=0.45\textwidth]{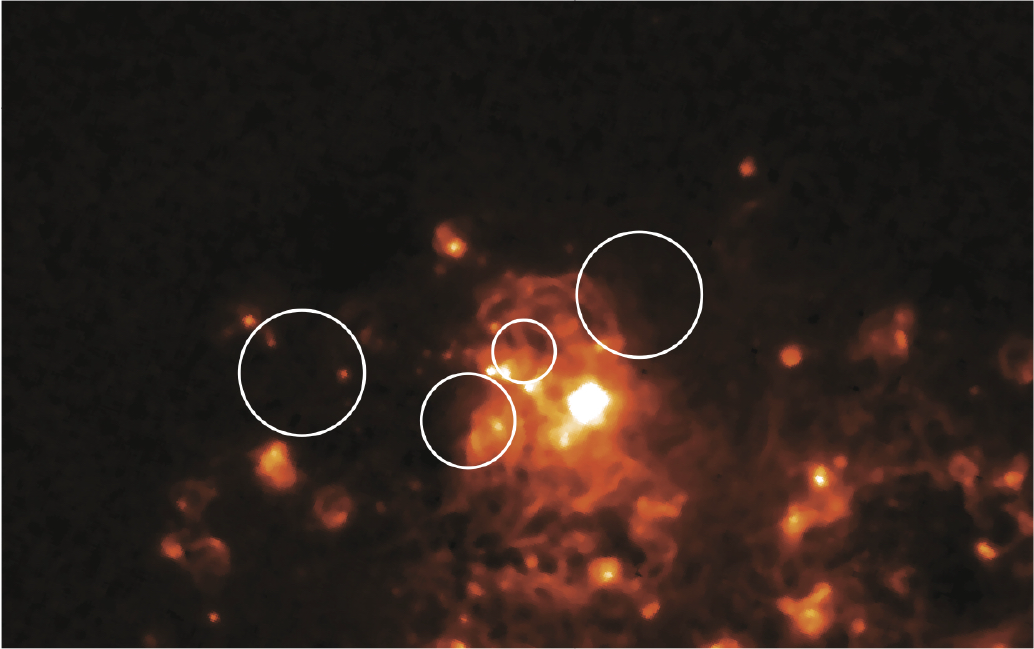}
  \includegraphics[width=0.45\textwidth]{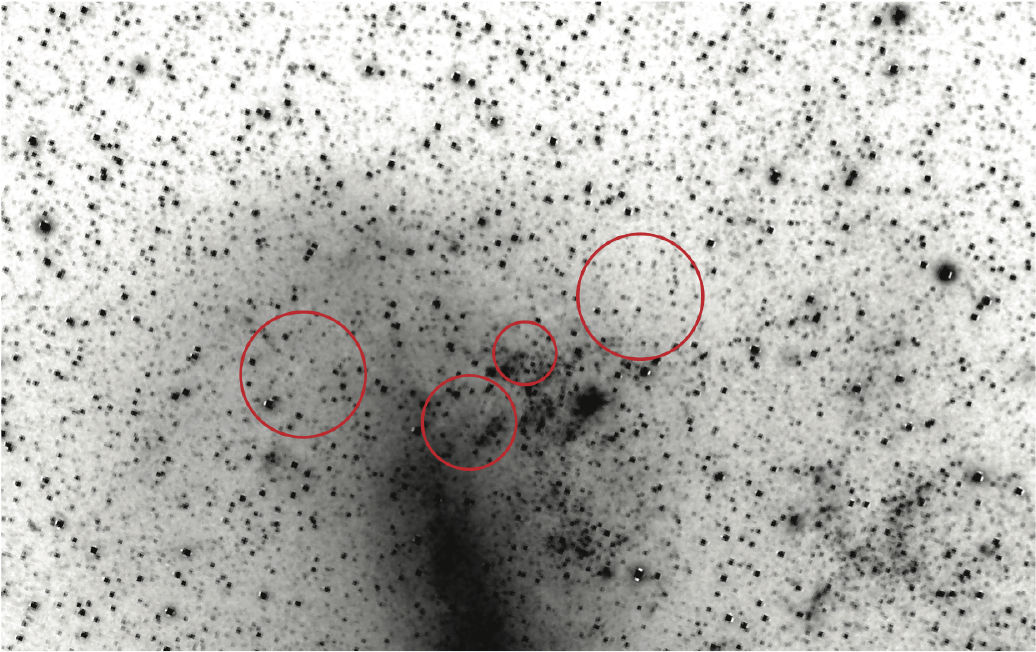}
  \caption{The foreground H$\alpha$ line emission (\emph{left}) and
  stellar continuum emission (\emph{right}) images of the south-east region of
  LMC in Galactic coordinate system with the galactic north pole to the top of
  the figures. Both images cover the same area as that in Figure \ref{fig:test
  of LMC example B} and the circles within mark the same corresponding locations
  of the sky. Several young star clusters \cite{2008MNRAS.389..678B} associated
  with nebulae NGC\,2080, NGC\,2085, and NGC\,2086 reside in between the two
  circles in the central region of the plot. The bright source in the H$\alpha$
  image is 30 Doradus, a giant H\,{\sc ii} region in LMC. The data presented
  here are obtained from The Southern H-Alpha Sky Survey Atlas
  \citep{2001PASP..113.1326G}.}
  \label{fig:foreground_HII_stars_LMC_II}
\end{figure*}

We further check the regions in figures~\ref{fig:LMC example E}
and~\ref{fig:test of LMC example B} using the foreground H$\alpha$ line emission
and stellar continuum emission, as shown in
figures~\ref{fig:foreground_HII_stars_LMC_I}
and~\ref{fig:foreground_HII_stars_LMC_II}. The results convincingly show that
these regions are associated with bright and active foreground regions, and
further prefers an explanation by foreground emission mechanism.

\begin{figure*}[!htb]
  \centering
  \includegraphics[width=0.32\textwidth]{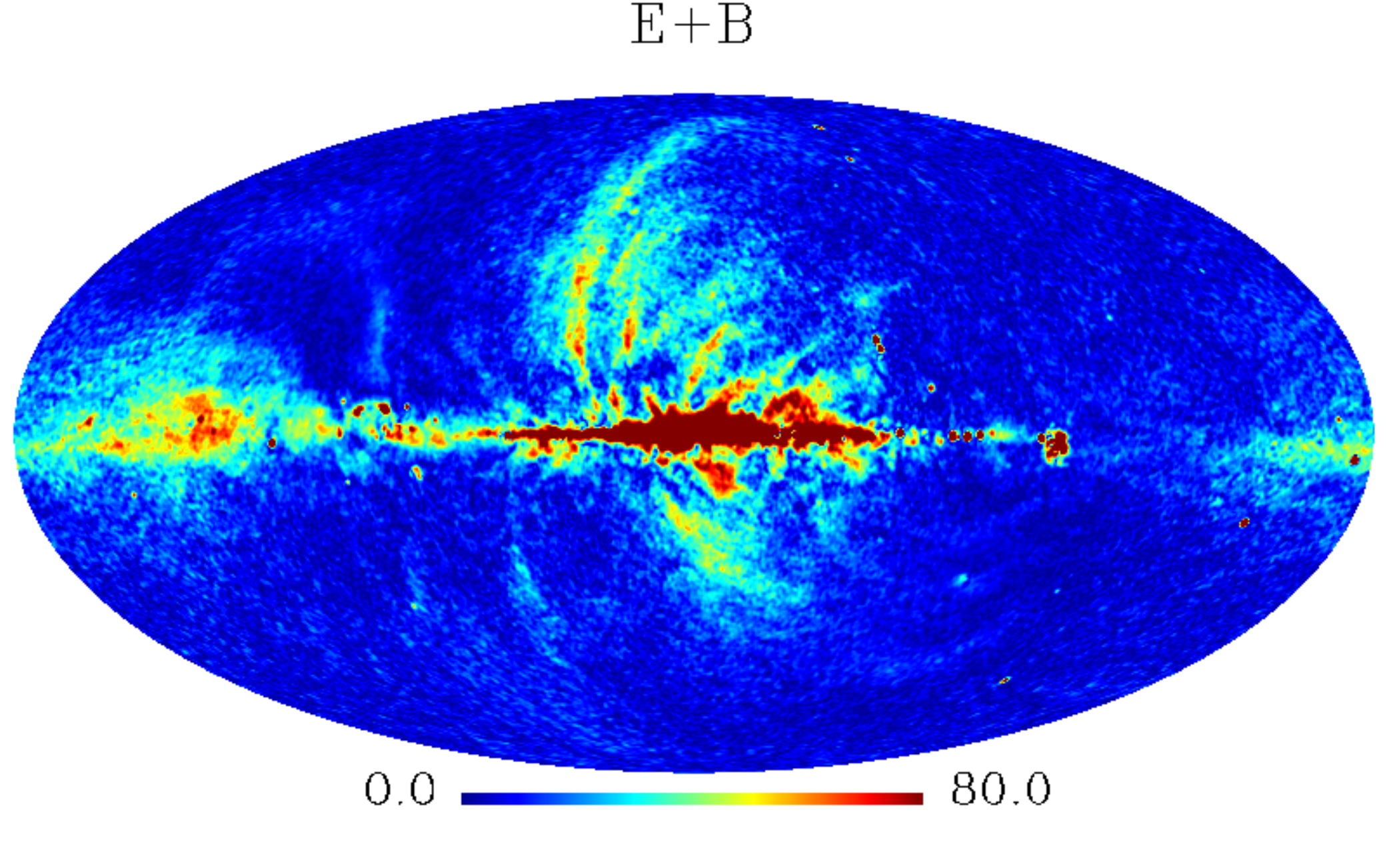}
  \includegraphics[width=0.32\textwidth]{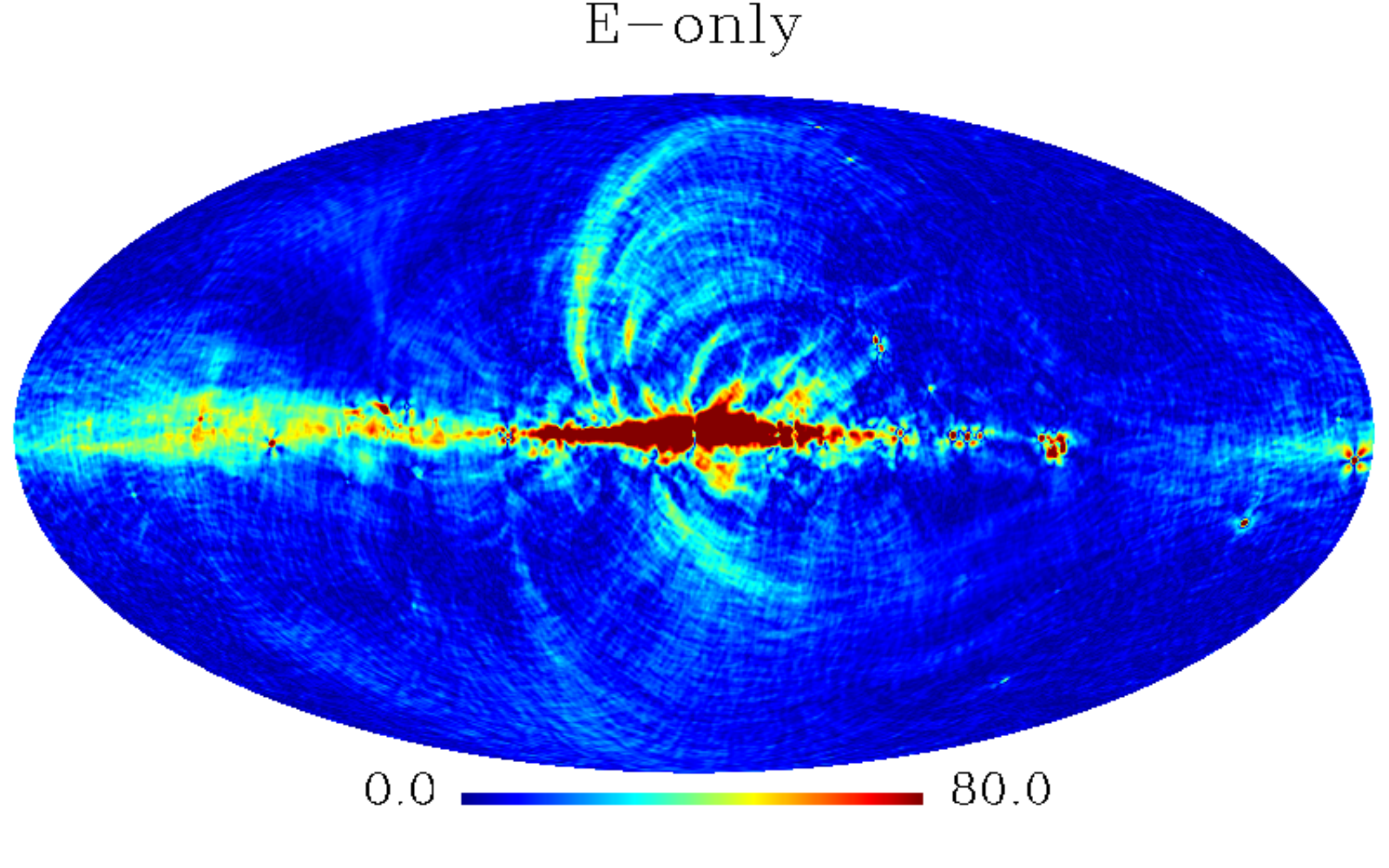}
  \includegraphics[width=0.32\textwidth]{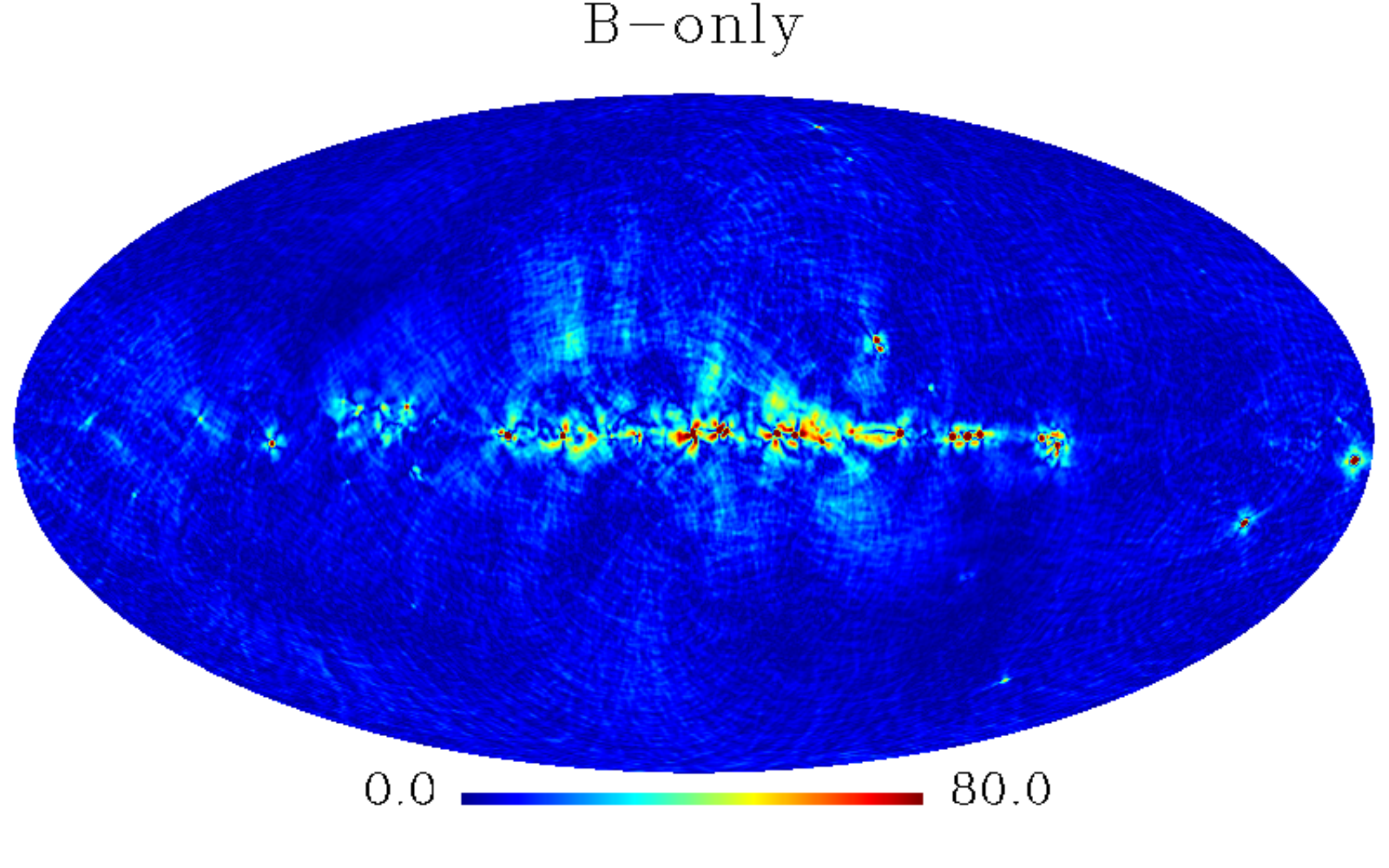}
  \caption{The fullsky Planck 30 GHz polarization mapat
  $N_{\mathrm{side}}=1024$ and smoothed to $1\degree$, for: the original
  polarization (left, including both the E and B families), the E-family
  (middle) and the B-family (right). Apparently, all loop-like structures appear
  only in the E family, which was discovered in~\cite{2018JCAP...05..059L} and
  explained in~\cite{2018A&A...617A..90L}.}
  \label{fig:loops}
\end{figure*}

Finally, we show the fullsky Planck 30 GHz polarized map with the original and
E, B family polarization in figure~\ref{fig:loops}. There are many loop
structures in the original polarization map, but with the E and B family
decomposition, we can easily see that all loop-like structures are only from the
E family, which was first discovered in~\cite{2018JCAP...05..059L} and whose
mechanism was explained in~\cite{2018A&A...617A..90L}. We also point out that in
figure 5 of a recent work~\cite{2022arXiv220310669N}, the mechanism of the Odd
Radio Circle is found to be the same as the one found in
in~\cite{2018A&A...617A..90L}.

\subsection{Toy model of the mechanism of B-family emissions}
\label{sub:toy model}

In this section, we provide a toy model as an attempt to explain the B-family
emissions shown in figures~\ref{fig:LMC example B} and~\ref{fig:LMC example E}.
The model is based on the model in~\citep{2018A&A...617A..90L} but gives an
analytic solution at small angles. The model is briefly illustrated as follows:

Assume we stays at point $\bm{O}=(0,0,0)$, and a compact center object (star,
supernova, blackhole, etc.) stays at $\bm{S}=(0,0,1)$. The line-of-sight is
along the spherical polar direction $(\theta,\varphi)$; thus, a test point at
distance $d$ along the line-of-sight has Cartesian coordinates
\begin{align}
	\bm{P}=(d\sin \theta\cos\varphi, d\sin \theta\sin \varphi,
d\cos \theta).
\end{align}
The vector from the compact object to the test point is
\begin{align}
	\bm{R} &= \bm{P}-\bm{S} = (d\sin \theta\cos\varphi, d\sin \theta\sin
\varphi, d\cos \theta-1)\\\nonumber
	|\bm{R}| &= r = \sqrt{1-2d\cos\theta+d^2}.
\end{align}
Then we use cross products to define the natural reference frames for
polarization~\citep{PhysRevD.55.1830} (normalized). The plus and minus axes for
the Q-Stokes parameters are $\bm{A}_{\pm}$:
\begin{align}
	\bm{A}_+ &= \frac{\bm{P}\times\bm{S}}{d\sin\theta} =
	(\sin\varphi,-\cos\varphi, 0) \\\nonumber
	\bm{A}_- &= \frac{\bm{P}\times\bm{A}_+}{d} =
	(\cos\theta\cos\varphi,\cos\theta\sin\varphi,-\sin\theta).
\end{align}
\red{As the first step}, we assume the backgroud magnetic field is smooth at
small scales, and its projection on $\bm{R}$ (radial component) is erased by the
shock wave or stellar wind. Then the initial and resulting magnetic fields are
\begin{align}\label{equ:bpbm}
	\bm{B}_0 &= b_0(\sin \theta'\cos\varphi', \sin \theta'\sin \varphi',
\cos \theta') \\\nonumber
	\bm{B} &= \bm{B}_0 - (\bm{B}_0\cdot\bm{R})\frac{\bm{R}}{r},
\end{align}
where $\bm{B}_0$ is the initial magnetic field, $(\theta',\varphi')$ is the
direction of $\bm{B}_0$ in the spherical polar coordinate system, and $\bm{B}$
is the resulting magnetic field. Then we compute the projection of the magnetic
field on the $\bm{A}_{\pm}$ axes as $B_{\pm}$:
\begin{align}
	B_+ &= \bm{B}\cdot\bm{A}_+ = \sin\theta'\sin (\varphi-\varphi') \\\nonumber
	B_- &= \bm{B}\cdot\bm{A}_- =
	\frac{-(d-\cos\theta)(d\cos\theta'\sin\theta +
	(1-d\cos\theta)\cos(\varphi-\varphi')\sin\theta'
	)}{1-2d\cos\theta+d^2}.
\end{align}
If the polarization is perpendicular to the magnetic field, then the Q-stokes
parameter is proportional to the power (square) difference: $Q
\propto B_-^2-B_+^2$, whereas for parallel polarizations the Q-stokes
parameter is proportional to $Q \propto B_+^2-B_-^2$. 

Because in the natural reference system, the Q-Stokes parameter corresponds to
the E-mode, the fraction of the E-mode in the total polarization can be
described by
\begin{align}
	f_E = \pm\frac{B_+^2-B_-^2}{B_+^2+B_-^2}\in[-1,1].
\end{align}
If $f_E$ is close to $\pm1$, then the polarization is dominated by the E-family.
If $f_E$ is close to zero, then the polarization is dominated by the B-family.

Now we introduce the small angle approximation: $\theta\ll 1$, which means the
line-of-sight is not far from the center compact object. Then
eq.~(\ref{equ:bpbm}) is simplified to:
\begin{align}
	B_+ &= \bm{B}\cdot\bm{A}_+ = \sin\theta'\sin (\varphi-\varphi') \\\nonumber
	B_- &= \bm{B}\cdot\bm{A}_- \approx \sin\theta'\cos(\varphi-\varphi').
\end{align}
Thus we have
\begin{align}
	f_E \approx \pm\frac{B_+^2-B_-^2}{B_+^2+B_-^2} = \pm\cos 2(\varphi-\varphi')
\end{align}
Therefore, if the following conditions are satisfied: 
\begin{enumerate}
	\item The background magnetic field does not change significantly at small
	scales.
	\item For some reason, the effect of shock wave or stellar wind is
	asymmetric. This includes at least two possibilities: a) The shockwave or
	stellar wind itself is asymmetric. b) The shock wave or stellar wind is
	symmetric, but the interstellar medium is asymmetric.
	\item The major direction of effect is roughly $45\degree$ apart from the
	direction of the smooth background magnetic field (both consider only the
	projection to the $xy$-plane), i.e., $\cos 2(\varphi-\varphi')\approx 0$.
\end{enumerate}
Then the polarization will be dominated by the B-family.

By checking the foreground H$\alpha$ line emission and stellar continuum
emission in figures~\ref{fig:foreground_HII_stars_LMC_I}
and~\ref{fig:foreground_HII_stars_LMC_II} and comparing them with the
polarization structures in figures~\ref{fig:LMC example B} and ~\ref{fig:LMC
example E}, we can see asymmetric bubble-like structures in the former, which is
qualitatively consistent with the B-family structure in the latter, especially
the direction of the bubble-like structures' shell. This fact indicates that the
above toy model is at least qualitatively reasonable.

\section{Discussion}
\label{sec:disscuss}

Earlier works \cite{2001PhRvD..64j3001Z, 2018JCAP...05..059L,
2018A&A...617A..90L, 2018arXiv180711940R} have established the E/B decomposition
as a non-local operation, which can be mathematically expressed in terms of an
integral convolution in pixel space, and the corresponding convolution kernels
have been calculated and visualized. In this work, we return to the E--B
decomposition theory, and, working in the space of the Stokes parameters $Q$ and
$U$, show how the operation can be formulated using quaternion multiplication.
The quaternion algebra naturally accommodates the harmonic-space E/B modes, and
the relations between all relevant quantities can be written compactly as
products of quaternion matrices. For this purpose, we depend on the different
types of quaternion conjugation, which are described in the appendix. Apart from
its mathematical concision and beauty, the quaternion representation of the E/B
decomposition can be developed into a general quaternionic eigenproblem, where
the different spins are related to the eigenvectors and the parities to the
eigenvalues.

As example of applications, we also study the Stoke-space E/B modes and
associated estimators in the polarized foregrounds analysis, which discovers
several interesting foreground structures in the E- and B-families; and give
reasonable explanations of their mechanisms (see
also~\cite{2018A&A...617A..90L}), which was also used recently to explain the
mechanism of the Odd Radio Circle~\cite{2022arXiv220310669N}.

\Ack{

This work is supported in part by the National Key R\&D Program of China
(2021YFC2203100, 2021YFC2203104) and the Anhui project Z010118169.

}

\appendix

\section{The system of quaternion multiplication}
\label{app:sys of quat mul}

A quaternion is formed by four real numbers or, equivalently, by two complex
numbers as follows:
\begin{align}
	\bm{q} = (a,b,c,d) = a+b\bm{i}+c\bm{j}+d\bm{k} = z_1 + z_2\bm{j},
\end{align}
where $z_1=a+b\bm{i}$ and $z_1=c+d\bm{i}$, and $\bm{i}$, $\bm{j}$,
$\bm{k}$ are three imaginary units that satisfy:
\begin{enumerate}
	\item $\bm{ii}=\bm{jj}=\bm{kk}=-1$.
	\item $\bm{ij}=\bm{k}$, $\bm{jk}=\bm{i}$ and $\bm{ki}=\bm{j}$.
	\item $\bm{ij}=-\bm{ji}$, $\bm{jk}=-\bm{kj}$ and $\bm{ki}=-\bm{ik}$.
\end{enumerate}
The elements attached to $\bm{i}$, $\bm{j}$, and $\bm{k}$ are called the
imaginary parts of a quaternion (or vector part), and the rest is called the
real or scalar part of a quaternion. With these rules, it is easy to see that
the multiplication of two quaternions $q_1=(a_1,b_1,c_1,d_1)$ and
$q_2=(a_2,b_2,c_2,d_2)$ is
\begin{align} \label{equ:quat_mul}
\bm{q}_1\bm{q}_2 =
& (a_1a_2 - b_1b_2 - c_1c_2 - d_1d_2) + 
\\ \nonumber
&(a_1b_2 + b_1a_2 + c_1d_2 - d_1c_2) \bm{i} +
\\ \nonumber
& (a_1c_2 - b_1d_2 + c_1a_2 + d_1b_2) \bm{j} +
\\ \nonumber
& (a_1d_2 + b_1c_2 - c_1b_2 + d_1a_2) \bm{k},
\end{align}
where the elements of the first quaternion are aligned vertically, and the
elements of the second quaternion are placed in a way shown in
figure~\ref{fig:quat mul}:
\begin{figure*}[!htb]
  \centering
  \includegraphics[width=0.48\textwidth]{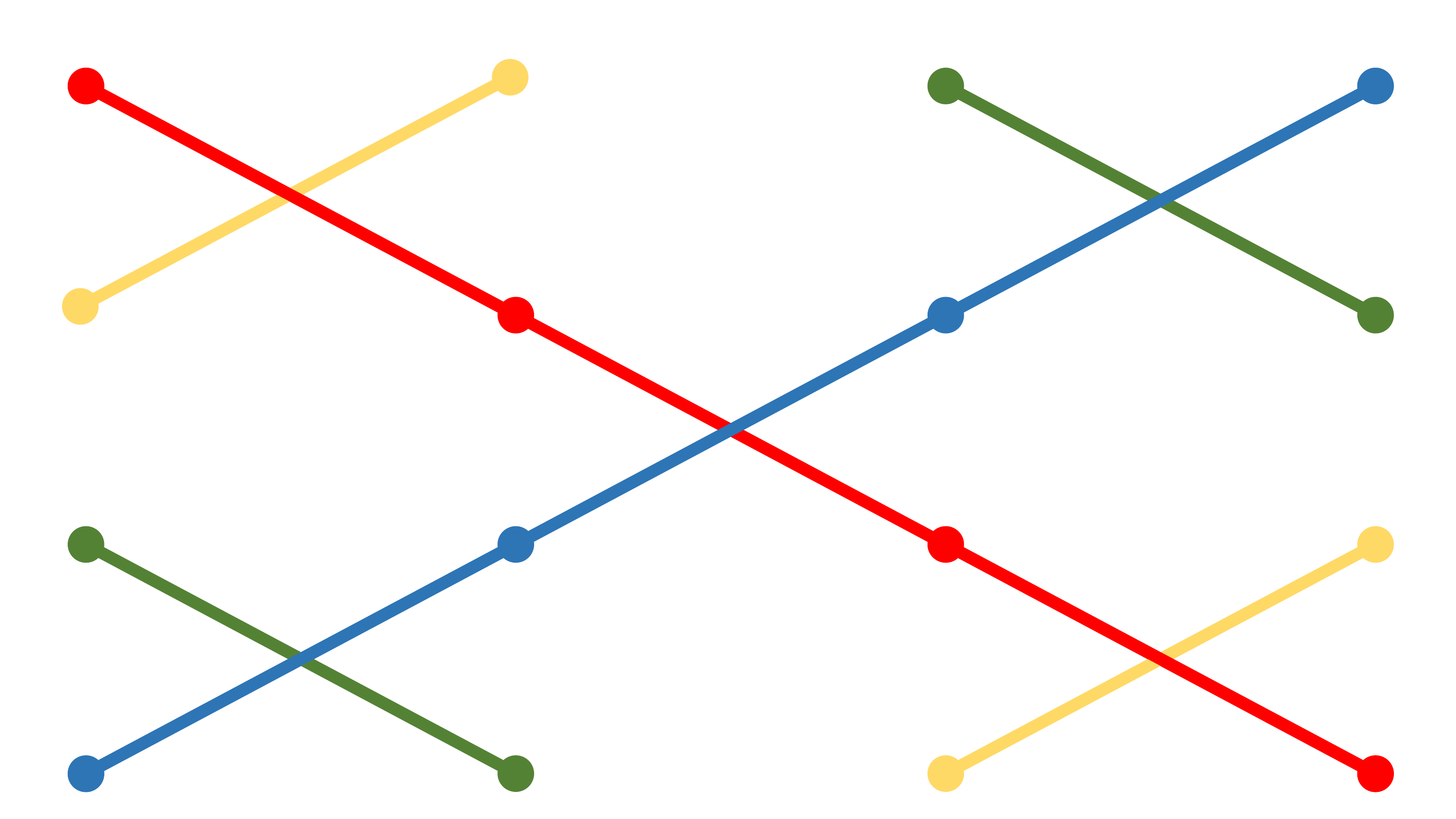}
  \caption{Placement of the second quaternion's elements in a quaternion
  multiplication.}
  \label{fig:quat mul}
\end{figure*}

\subsection{The system of quaternion conjugates}
\label{sub:quat conj}

Unlike the complex numbers that have only one conjugate, the quaternions have
seven different conjugates. Because this is seldom mentioned in literature, in
this work we define a symbol for these seven conjugates as $\bm{q}^{*_{x}}$,
where $x$ is a three-bit binary number with the bits corresponding to $\bm{i}$,
$\bm{j}$, and $\bm{k}$ respectively. If the value of one bit is equal to 1, then
the corresponding imaginary part will be inverted. For example,
$(a,b,c,d)^{*_{100}} = (a,-b,c,d)$. It is worth to mention that
$\bm{q}^{*_{111}}$ can also be shortened as $\bm{q}^{*}$, which is the most
widely used quaternion conjugate that inverts all imaginary parts
simultaneously.

Below we provide examples of how the seven quaternion conjugates are connected
to actual transforms, but note that they are apparently not the only
possibilities:
\begin{enumerate}
	\item Type-001 (Type-1, symbol: $\bm{q}^{*_{001}}$ or
	$\bm{q}^{*_{1}}$) conjugate: $z_1+z_2\bm{j} \longrightarrow
	z_1+z_2^*  \bm{j}$.
	\item Type-010 (Type-2, symbol: $\bm{q}^{*_{010}}$ or
	$\bm{q}^{*_{2}}$) conjugate: $z_1+z_2\bm{j} \longrightarrow
	z_1-z_2^*  \bm{j}$.
	\item Type-011 (Type-3, symbol: $\bm{q}^{*_{011}}$ or
	$\bm{q}^{*_{3}}$) conjugate: $z_1+z_2\bm{j} \longrightarrow z_1-z_2
	\bm{j}$.
	\item Type-100 (Type-4, symbol: $\bm{q}^{*_{100}}$ or
	$\bm{q}^{*_{4}}$) conjugate: $z_1+z_2\bm{j} \longrightarrow
	z_1^*+z_2  \bm{j}$.
	\item Type-101 (Type-5, symbol: $\bm{q}^{*_{101}}$ or
	$\bm{q}^{*_{5}}$) conjugate: $z_1+z_2\bm{j} \longrightarrow
	z_1^*+z_2^*\bm{j}$.
	\item Type-110 (Type-6, symbol: $\bm{q}^{*_{110}}$ or
	$\bm{q}^{*_{6}}$) conjugate: $z_1+z_2\bm{j} \longrightarrow
	z_1^*-z_2^*\bm{j}$.
	\item Type-111 (Type-7, symbol: $\bm{q}^{*_{111}}$ or
	$\bm{q}^{*_{7}}$) conjugate: $z_1+z_2\bm{j} \longrightarrow
	z_1^*-z_2  \bm{j}$.
\end{enumerate}
For convenience and as shown above, the quaternion conjugates can possibly be
shortened as $*_{x'}$, where $x'$ is the decimal value of the binary number $x$.
However, the binary symbol is recommended because it makes the following rule
more clear: Let $x_{1,2,3}$ be the values of the three digits of $x$, so they
are either 1 or 0; then by eq.~(\ref{equ:quat_mul}) it is easy to prove that
\begin{align}\label{equ:quat conj odd even}
	\begin{matrix}
		\mathrm{If} & x_1+x_2+x_3=\mathrm{Odd}  & \mathrm{then:} &
		(\bm{q}_1\bm{q}_2)^{*_{x}} =
		\bm{q}_2^{*_{x}}\bm{q}_1^{*_{x}},
		\\ 
		\mathrm{If} & x_1+x_2+x_3=\mathrm{Even} & \mathrm{then:} &
		(\bm{q}_1\bm{q}_2)^{*_{x}} =
		\bm{q}_1^{*_{x}}\bm{q}_2^{*_{x}}.
	\end{matrix}
\end{align}
When $x_1+x_2+x_3$ is odd/even, the corresponding conjugate is called an
odd/even conjugate. The above equation means, for the quaternions, odd and even
conjugates follow different combinations of the distr`'ibution and commutative
laws. It is also easy to see that the combination of odd conjugates can give
both odd and even conjugates, but the combination of even conjugates will
\emph{not} give an odd conjugate.

From eq.~(\ref{equ:quat conj odd even}), it is also easy to prove that, for an
odd conjugate, the corresponding imaginary part will be erased in the quaternion
multiplication $\bm{q}\bm{q}^{*_x}$, for example, when $\bm{q}=(a,b,c,d)$, we
have:
\begin{align}
	\bm{q}\bm{q}^{*_{100}} &= (a^2+b^2-c^2-d^2,\bm{0},2ac-2bd,2ad+2bc)
	\\ \nonumber
	\bm{q}\bm{q}^{*_{111}} &= (a^2+b^2+c^2+d^2,\bm{0},\bm{0},\bm{0}).
\end{align}	
However, for an even conjugate, the corresponding imaginary parts are erased not
in a direct multiplication, but in the combination $\bm{q}\bm{q}^{*_x} +
\bm{q}^{*_x}\bm{q}$, like:
\begin{align}
	&\bm{q}\bm{q}^{*_{110}} = (a^2+b^2+c^2-d^2,2cd,-2bd,2ad)
	\\ \nonumber
	&\bm{q}\bm{q}^{*_{110}}+\bm{q}^{*_{110}}\bm{q} =
	2(a^2+b^2+c^2-d^2,\bm{0},\bm{0},2ad)
\end{align}
Therefore, the quaternions have at least three independent parity states -- one
for each imaginary part. Each single conjugate ($x_1+x_2+x_3=1$) will help to
erase one parity state in the multiplication because we have $0=\pm0$ at the
same time.

\subsection{The system of quaternion matrix multiplication}
\label{sub:quat mat mul}

The algebra of quaternion matrix has been discussed in detail
by~\cite{1994..quat..mat}. However, the order problem in quaternion matrix
multiplication has not been studied, which will be discussed below.

Let $\bm{a}=(a_1,a_2,a_3,a_4)$, $\bm{b}=(b_1,b_2,b_3,b_4)$ and
$\bm{x}=(x_1,x_2,x_3,x_4)$ be quaternions, and $\bm{A} \equiv
\bm{a}_{ij}$, $\bm{B} \equiv \bm{b}_{jk}$, $\bm{X} \equiv \bm{a}_{k}$ be
two $n\times n$ square matrices and one $n\times 1$ column matrix of
quaternions. Then a naive definition of the matrix multiplication
$\bm{ABX}$ is
\begin{align}
	\bm{ABX} \xLongrightarrow[]{\mathrm{naive}} \sum_{jk}
	\bm{a}_{ij}\bm{b}_{jk}\bm{x}_{k}.
\end{align}
However, because quaternion multiplication is non-commutative, there are
actually six quaternion orders for the about matrix multiplication:
\begin{align}
	\begin{matrix}
		(\bm{ABX})_{012} \equiv \sum_{jk} \bm{a}_{ij}\bm{b}_{jk}\bm{x}_{k}, & 
		(\bm{ABX})_{021} \equiv \sum_{jk} \bm{a}_{ij}\bm{x}_{k}\bm{b}_{jk}, \\ 
		(\bm{ABX})_{102} \equiv \sum_{jk} \bm{b}_{jk}\bm{a}_{ij}\bm{x}_{k}, & 
		(\bm{ABX})_{120} \equiv \sum_{jk} \bm{b}_{jk}\bm{x}_{k}\bm{a}_{ij}, \\ 
		(\bm{ABX})_{201} \equiv \sum_{jk} \bm{x}_{k}\bm{a}_{ij}\bm{b}_{jk}, & 
		(\bm{ABX})_{210} \equiv \sum_{jk} \bm{x}_{k}\bm{b}_{jk}\bm{a}_{ij}. 
	\end{matrix}
\end{align}
Some of the orders can be implemented by changing the order of matrix
multiplication with a proper transposing; however, this is not always possible,
like order-120, because the rule of matrix multiplication does not allow a
column matrix $\bm{X}$ to be placed between two square matrices $\bm{A}$ and
$\bm{B}$. Therefore, in order to properly define the quaternion matrix
multiplication, we have to find out a way to separate the order of quaternion
multiplication from the order of matrix multiplication.

According to eq.~(\ref{equ:quat_mul}), it is easy to write the quaternion
multiplication of $\bm{a}$ and $\bm{x}$ in a matrix forms:
\begin{align}
	\bm{a}\bm{x} \equiv& 
	\begin{pmatrix*}[r]
		a_1 & -a_2 & -a_3 & -a_4 \\
		a_2 &  a_1 & -a_4 &  a_3 \\
		a_3 &  a_4 &  a_1 & -a_2 \\
		a_4 & -a_3 &  a_2 &  a_1 \\
	\end{pmatrix*} \cdot
	\begin{pmatrix*}[r]
		x_1 \\ x_2 \\ x_3 \\ x_4
	\end{pmatrix*} = \bm{a}_L\cdot\bm{x}_o
	\\ \nonumber
	\bm{x}\bm{a} \equiv&
	\begin{pmatrix*}[r]
		a_1 & -a_2 & -a_3 & -a_4 \\
		a_2 &  a_1 &  a_4 & -a_3 \\
		a_3 & -a_4 &  a_1 &  a_2 \\
		a_4 &  a_3 & -a_2 &  a_1 \\
	\end{pmatrix*} \cdot
	\begin{pmatrix*}[r]
		x_1 \\ x_2 \\ x_3 \\ x_4
	\end{pmatrix*} = \bm{a}_R\cdot\bm{x}_o,
\end{align}
where subscripts ``$_{L/R}$'' means to augment each quaternion to its left/right
$4\times4$ real matrix, and subscript ``$_o$'' means to augment each quaternion
to its $4\times1$ real column matrix form, and the augmentations work for both
quaternion and quaternion matrix. The above equations show how to write
quaternion multiplication with different orders with a fixed-order matrix
multiplication, which is exactly what we need. Now the six quaternion orders of
$\bm{ABX}$ can be written as
\begin{align}
	\begin{matrix*}[l]
		(\bm{ABX})_{012} \equiv  
		\sum_{jk} \bm{a}_{ij}\bm{b}_{jk}\bm{x}_{k}
		\equiv \bm{A}_L\bm{B}_L\bm{X}_o, & 
		(\bm{ABX})_{021} \equiv  
		\sum_{jk} \bm{a}_{ij}\bm{x}_{k}\bm{b}_{jk}
		\equiv \bm{A}_L\bm{B}_R\bm{X}_o, \\ 
		(\bm{ABX})_{102} \equiv  
		\sum_{jk} \bm{b}_{jk}\bm{a}_{ij}\bm{x}_{k}
		\equiv (\bm{A}_R\bm{B}_o)_L\bm{X}_o, & 
		(\bm{ABX})_{120} \equiv  
		\sum_{jk} \bm{b}_{jk}\bm{x}_{k}\bm{a}_{ij}
		\equiv \bm{A}_R\bm{B}_L\bm{X}_o, \\ 
		(\bm{ABX})_{201} \equiv  
		\sum_{jk} \bm{x}_{k}\bm{a}_{ij}\bm{b}_{jk}
		\equiv (\bm{A}_L\bm{B}_o)_R\bm{X}_o, & 
		(\bm{ABX})_{210} \equiv  
		\sum_{jk} \bm{x}_{k}\bm{b}_{jk}\bm{a}_{ij}
		\equiv (\bm{A}_R\bm{B}_o)_R\bm{X}_o. 
	\end{matrix*}
\end{align}
By this way, the orders of quaternion multiplication and matrix multiplication
are separated, and we are free to use any appropriate orders to solve a problem.

\section{Quick reference of the E and B family decomposition with quaternions}
\label{app:eb family quick reference}

Here we summarize the main results of the E and B family decomposition
for a quick reference:

Given the basic spin-2 spherical harmonic decomposition:
\begin{align}\label{equ:spin 2 sht app}
	a_{\pm2,\ell m} &= \int (Q \pm U\bm{i}) [_2 Y^*_{\ell
	m}] d\bm{n},
\end{align}
The E and B mode spherical harmonic coefficients are defined as
\begin{align}\label{equ: def of almEB app}
	a_{E,\ell m} &= \frac{a_{2,\ell m} + a_{-2,\ell m}}{2}
	\\ \nonumber
	a_{B,\ell m}\bm{i} &= \frac{a_{2,\ell m} - a_{-2,\ell m}}{2}.
\end{align}
By defining
\begin{align}\label{equ:define F app}
    F_{+,lm} &= \frac{1}{2} \left({}_{2}Y_{lm} + {}_{-2}Y_{lm} \right)
    \quad
    F_{-,lm} = \frac{1}{2} \left({}_{2}Y_{lm} - {}_{-2}Y_{lm} \right),
\end{align}
we get the E and B families as
\begin{align}\label{equ: QU eb solution E app}
	&\left\{
	\begin{matrix*}[r]
		Q_E = \sum_{\ell m} a_{E,\ell m} F_{+,\ell m} \\		
		U_E\bm{i} = \sum_{\ell m} a_{E,\ell m} F_{-,\ell m}
	\end{matrix*}
	\right.
	\\ \nonumber
	&\left\{
	\begin{matrix*}[l]
		Q_B = \sum_{\ell m} (a_{B,\ell m}\bm{i}) F_{-,\ell m} \\		
		U_B\bm{i} = \sum_{\ell m} (a_{B,\ell m}\bm{i}) F_{+,\ell m}
	\end{matrix*}
	\right. .
\end{align}
The harmonic domain summation of $F_{\pm}$ is
\begin{align}\label{equ:fpfm all app}
	\sum_{\ell m} F_{+,\ell m}(\bm{n})F^*_{+,\ell m}(\bm{n}') &=
	\mathrm{Re}(\mathcal{Y}_\mathrm{+}) ;\;\;
	\sum_{\ell m} F_{+,\ell m}(\bm{n})F^*_{-,\ell m}(\bm{n}') =
	\bm{i}\,\mathrm{Im}(\mathcal{Y}_\mathrm{-})
	\\ \nonumber
	\sum_{\ell m} F_{-,\ell m}(\bm{n})F^*_{-,\ell m}(\bm{n}') &=
	\mathrm{Re}(\mathcal{Y}_\mathrm{-}) ;\;\;
	\sum_{\ell m} F_{-,\ell m}(\bm{n})F^*_{+,\ell m}(\bm{n}') =
	\bm{i}\,\mathrm{Im}(\mathcal{Y}_\mathrm{+}),
\end{align}
where $\mathscr{F_\pm}$ are defined as
\begin{align}
	\mathscr{F}_+(\bm{n},\bm{n}') &= \sum_\ell \sqrt{\frac{2\ell+1}{4\pi}} \,
	\frac{_2Y_{\ell, -2}(\beta,\alpha) + {}_2Y_{\ell,
	2}(\beta,\alpha)}{2}
	\\ \nonumber
	\mathscr{F}_-(\bm{n},\bm{n}') &= \sum_\ell \sqrt{\frac{2\ell+1}{4\pi}} \,
	\frac{_2Y_{\ell, -2}(\beta,\alpha) - {}_2Y_{\ell,
	2}(\beta,\alpha)}{2}.
\end{align}
Then we define five quaternions straightforwardly:
\begin{align}\label{equ:quat form base app}
	\mathcal{F}_{\ell m} &= F_{+,\ell m} + F_{-,\ell m} \bm{j}
	\\\nonumber
	\mathcal{G} &= \mathscr{F}_+ + \mathscr{F}_- \bm{j}
	\\\nonumber
	\mathcal{D} &= P_E + P_B \bm{j}
	\\ \nonumber
	\mathcal{A}_{\ell m} &= a_{E,\ell m} + (a_{B,\ell m}\bm{i})\bm{j} =
	a_{E,\ell m} + a_{B,\ell m}\bm{k}
	\\\nonumber
	\mathcal{P} &= Q + (U\bm{i})\bm{j} = Q + U\bm{k},
\end{align}
where $\mathcal{F}_{\ell m}$ is the base function of EB-decomposition,
$\mathcal{G}$ is the pixel domain convolution kernel,  $\mathcal{D}$ is the
expected pixel domain EB-family, $\mathcal{A}_{\ell m}$ is the harmonic domain
EB-coefficient, and $\mathcal{P}$ is a two-element quaternion containing the
pixel domain $Q$ and $U$ Stokes parameters. The matrix forms of these
quaternions are expressed in bold letters. With the above definition, we have:
\begin{enumerate}
    \item The forward and backward transforms between $\bm{\mathcal{P}}$
   (containing the input $Q$, $U$) and $\bm{\mathcal{A}}$ (containing the
   resulting $a_{E,\ell m}$, $a_{B,\ell m}$) is given by the following
   quaternion matrix equations:
    \begin{align}
    	\bm{\mathcal{A}} &= \bm{\mathcal{F}}^{H_{101}}
    	\bm{\mathcal{P}}
    	\\ \nonumber
    	\bm{\mathcal{P}} &= \bm{\mathcal{F}}^{*_{010}}
    	\bm{\mathcal{A}},
    \end{align}
    where $^{H_{101}}$ and $^{*_{010}}$ are quaternion conjugates
    defined in Appendix~\ref{sub:quat conj}.
    
    \item The pixel-pixel domain decomposition of the E and B families is
    \begin{align}
    	\bm{\mathcal{D}} = \bm{\mathcal{G}} \bm{\mathcal{P}},
    \end{align}
    
    \item The forward and backward transforms between the pixel domain E and B
	families and $a_{E,\ell m}$, $a_{B,\ell m}$ are given in the following
	quaternion matrix form, where $\bm{\mathcal{Y}}_2={}_2Y_{\ell m}(\bm{n})$ is
	the matrix form of ${}_2Y_{\ell m}(\bm{n})$:
    \begin{align}
    	\bm{\mathcal{D}} &= \bm{\mathcal{Y}}_2 \bm{\mathcal{A}}
    	\\ \nonumber
    	\bm{\mathcal{A}} &= \bm{\mathcal{Y}}^{H}_2
    	\bm{\mathcal{D}},
    \end{align}
\end{enumerate}


%

\end{document}